EVOLUTIONARY BIOLOGY

# Deep learning on butterfly phenotypes tests evolution's oldest mathematical model

**Jennifer F. Hoyal Cuthill[1,2,3]*, Nicholas Guttenberg[1]*, Sophie Ledger[4], Robyn Crowther[4], Blanca Huertas[4]**

**Traditional anatomical analyses captured only a fraction of real phenomic information. Here, we apply deep learning to quantify total phenotypic similarity across 2468 butterfly photographs, covering 38 subspecies from the polymorphic mimicry complex of *Heliconius erato* and *Heliconius melpomene*. Euclidean phenotypic distances, calculated using a deep convolutional triplet network, demonstrate significant convergence between interspecies co-mimics. This quantitatively validates a key prediction of Müllerian mimicry theory, evolutionary biology's oldest mathematical model. Phenotypic neighbor-joining trees are significantly correlated with wing pattern gene phylogenies, demonstrating objective, phylogenetically informative phenome capture. Comparative analyses indicate frequency-dependent mutual convergence with coevolutionary exchange of wing pattern features. Therefore, phenotypic analysis supports reciprocal coevolution, predicted by classical mimicry theory but since disputed, and reveals mutual convergence as an intrinsic generator for the unexpected diversity of Müllerian mimicry. This demonstrates that deep learning can generate phenomic spatial embeddings, which enable quantitative tests of evolutionary hypotheses previously only testable subjectively.**

## INTRODUCTION

Analyses of biological phenotype have traditionally used only subjective (*1*), categorical descriptions, although variation in many traits is fundamentally quantitative (*2*, *3*). Consequently, evolutionary questions that have been the subject of speculation for more than 150 years (*4*, *5*) have never before been addressed definitively because of the lack of phenotypic trait quantification. More recent, quantitative approaches to phenotypic analysis have included geometric measurement, outline or landmark analysis (geometric morphometrics) (*6*, *7*), spectrophotometry (*3*), and image pixel comparisons. While such standard analyses have underpinned comparative anatomy, taxonomic designation (*3*), investigation of phenotypic evolution, and morphological phylogenetics (*2*), even the largest studies capture only a small fraction of real phenomic information, require laborious manual processing (*8*), may be sensitive to feature conjunction (e.g., pattern element duplication) or minor misalignment (*7*, *9*), can produce descriptive statistics that are nonreplicable (*1*) or internally inconsistent (*10*), and produce evolutionary trees with consistent topological differences to genomic phylogenies (*11*). For such reasons, real-world evolutionary systems, such as the classic case study of mimicry in *Heliconius* butterflies (*4*, *12*, *13*), have previously been beyond the reach of comprehensive phenomic analysis. In consequence, fundamental questions in evolutionary theory have remained outstanding, including (i) the statistical congruence between phenomic and genomic phylogenies (*11*); (ii) the statistical significance and quantitative extent, during mutualistic Müllerian mimicry, of phenotypic convergence, the principal prediction of evolutionary biology's first mathematical model (*4*, *13*, *14*); (iii and iv) the quantitative effects of hybridization (*15*, *16*) and biogeographic distance (*17*, *18*) on phenotypic convergence in mimicry; (v) the existence and quantitative extent of mutual phenotypic convergence [and therefore strict coevolution (*19*)] in mimicry (*4*, *13*); and (vi) the effects of mimetic convergence on the coevolution of phenotypic novelty.

New advances in machine learning using deep neural networks (deep learning) (*20*) enable automated quantification of total similarity across large and diverse data samples, in Euclidean spaces of moderate dimensionality (*21*, *22*). Within biology, however, deep learning has been applied primarily to image classification tasks (*20*). Here, we apply deep learning using a convolutional triplet neural network (*21*, *22*) to additionally quantify total visible phenotypic similarity among dorsal and ventral photographs of 38 subspecies of Neotropical butterfly from the species *Heliconius erato* and *Heliconius melpomene*. The analysis includes 1234 butterfly specimens, widely sampling the total geographic range (fig. S2) and extensive wing pattern polymorphism within each species (fig. S4). This enables the first comprehensive, quantitative analysis of Müllerian mimicry across this classic (*12*, *23*), but controversial (*13*) coevolutionary system. This provides quantitative answers to major outstanding questions in evolutionary theory, as well as the most phenotypically inclusive investigation, to date, of evolutionary convergence (*24*, *25*) and a fully automated, objective data-capture method for phenotypic (morphological) phylogenetics.

## RESULTS

### Deep learning network output and accuracy

To quantify phenotypic distances between *Heliconius* butterflies, a deep convolutional neural network (fig. S1) was trained to classify photographs of *Heliconius* butterflies by subspecies, with 1500 of 2468 total images used for network training and the remainder for testing. The training method (*21*, *22*) used triplets of images, each replicate showing the network two images sampled from the same subspecies and one sampled from a different subspecies. The deep convolutional neural network (which we name ButterflyNet) learnt both to correctly classify images by subspecies and to calculate an internally consistent set of Euclidean distances between input images, with those of the same subspecies closer together and those of different subspecies further apart. The trained network achieved 86% accuracy in classifying test images by subspecies (compared to a chance value of 5%). Crucially, the network also calculated coordinates of all 2468 images within a phenotypic spatial embedding, with dimensionality set to 64 (table S3). In doing so, the network identifies and uses a subset of information from the input images that is sufficient to discriminate the taxonomic groups with the reported accuracy. A support vector classifier (SVC) trained on the spatial embeddings similarly achieved 87% subspecies classification accuracy, confirming that the spatial embedding preserves information sufficient for successful classification.

[1]Earth-Life Science Institute, Tokyo Institute of Technology, Tokyo 152-8550, Japan. [2]Department of Earth Sciences, University of Cambridge, Cambridge CB2 3EQ, UK. [3]Institute of Analytics and Data Science and School of Life Sciences, University of Essex, Wivenhoe Park, Colchester CO4 3SQ, UK. [4]Department of Entomology, Natural History Museum, Cromwell Road, London SW7 5BD, UK.
*Corresponding author. Email: j.hoyal.cuthill@elsi.jp (J.F.H.C.); ngutten@gmail.com (N.G.)

### Phylogenetic results

Neighbor-joining trees, reconstructed from phenotypic distances between subspecies (Fig. 1 and fig. S6), group interspecies mimics rather than separating the species, consistent with evolutionary convergence, as discussed below (although species are successfully separated by specific embedding axes; Fig. 2F). However, recovered intraspecific relationships, in particular, show significant similarities to independent, published phylogenies based on gene sequences (for an overlapping subset of 25 subspecies) (Fig. 1) (*26*, *27*). For both species (separately and combined), the phenotypic trees are significantly more similar to color pattern gene trees than are randomly generated trees (Mann-Whitney pairwise comparisons, $P \leq 2.59 \times 10^{-256}$; table S10). This demonstrates that the phenotypic distances generated by deep learning contain significant phylogenetic information relative to a null hypothesis of random similarity. On average, intraspecific trees of phenotypic distance are significantly more similar to trees reconstructed from genes associated with wing color pattern (*27*) than they are to neutral gene loci ("housekeeping genes") or to a control sample of random phylogenetic topologies (Mann-Whitney pairwise comparisons: *H. erato*, $P \leq 7.128 \times 10^{-35}$; *H. melpomene*, $P \leq 1.645 \times 10^{-13}$; and all subspecies, $P \leq 0.0066$; Fig. 1 and table S10). For *H. erato* (Fig. 1E) phenotypic phylogenies are more similar to color gene phylogenies than color gene phylogenies are to neutral gene phylogenies. Statistics repeated after exclusion of hybrid specimens (table S2) confirm these results (table S11). We can note that the differences between gene histories are expected because of standard population genetic processes and a greater difference between genetic partitions is apparent in *H. erato* (Fig. 1E), which tends to have larger population sizes and carries more genetic variation (*26*, *28*, *29*). Overall, these results show that, despite a strong signal from mimicry between the species, the phenotypic spatial embedding has successfully recovered additional phylogenetic information within the image dataset [originating from the combination of lineal descent and historical hybridization (*26*, *28*)].

### Quantification of phenotypic similarity

Across the complete dataset, specimens of the same subspecies [exhaustively sampled from the Natural History Museum (NHM) collection] are clustered in a visualization of the first two principal component axes constructed from the phenotypic spatial embedding (Fig. 2A). Similarly, statistical analyses of average pairwise Euclidean distances (table S4) calculated directly from the original embedding coordinates (table S3) show that the most phenotypically similar butterfly images are those of the same subspecies (Kruskal-Wallis, $P = 1.043 \times 10^{-28}$; $H = 128.9$; Mann-Whitney pairwise, $P \leq 5.495 \times 10^{-7}$; Fig. 3A, left). Alongside the high subspecies classification accuracy (≥86%), this demonstrates that named subspecies of *H. erato* and *H. melpomene* (*30*) are objectively distinguishable, despite phenotypic variability among individuals and incomplete reproductive isolation between subspecies within each species.

### Testing phenotypic convergence

Pairwise comparisons across subspecies (Fig. 3A, middle right) show that traditionally hypothesized sympatric co-mimics (Figs. 1 and 2, fig. S3, and table S1) (*12*) are significantly more similar to each other than are other subspecies pairs (Mann-Whitney pairwise, $P = 5.50 \times 10^{-7}$). Comparisons performed separately for each species confirm these results (Kruskal-Wallis, $P = 1.48 \times 10^{-23}$; $H = 113.2$; Mann-Whitney pairwise comparisons mimicry versus identity or other, $P \leq 4.16 \times 10^{-4}$; identity versus other, $P \leq 6.43 \times 10^{-11}$; interspecies identity and other, nonsignificant, respectively $P = 0.9532$, $P = 0.3285$; Fig. 3B). This highlights the remarkable level of adaptive phenotypic evolution by these species, in which mimetic phenotypic similarity between the species (Figs. 2B and 3B, middle) is greater than the subspecies similarity within them (Figs. 2E and 3B, right). Across the 38 sampled subspecies, six distinct phenotypic clusters, identified by both hierarchical clustering (Fig. 2C and Table 1) and neighbor joining (fig. S6), include interspecies co-mimics.

### Hybridization

Statistical comparisons of phenotypic distance with hybrids excluded confirm that co-mimics (Fig. 1, fig. S3, and table S1) (*12*) are significantly more similar to each other than are other subspecies pairs, with a reduced average distance between the co-mimics and increased statistical significance compared to the complete dataset (hybrids excluded Mann-Whitney pairwise, $P = 2.33 \times 10^{-8}$; average distance mimics = 0.897, $n$ mimics = 15; complete dataset: average distance mimics = 1.027, $n$ mimics = 19).

### Biogeography

Butterfly individuals sampled from subspecies that are traditionally hypothesized co-mimics are also confirmed to be geographically closer, on average, than those from other subspecies pairs (Kruskal-Wallis, $P = 4.23 \times 10^{-16}$; $H = 70.8$; Mann-Whitney, $P \leq 0.0066$; pairwise, $P = 0.0041$; fig. S5). This indicates that geographic proximity has promoted interspecies convergence in mimicry (*17*, *18*), between *H. erato* and *H. melpomene*, across their Neotropical range.

### Testing mutual convergence

The extent of mutual convergence in mimicry is illustrated by a case study (Fig. 4) from three comparative analyses (covering 12 of the studied subspecies), all of which are supportive of reciprocal coevolutionary influence (fig. S7). Comparisons of the average locations of these subspecies in phenotypic space (Fig. 4) indicate statistically significant mutual convergence for this case study (with *H. erato cyrbia* closer to *H. melpomene cythera* than is *H. erato venus* and *H. melpomene cythera* closer to *H. erato cyrbia* than is *H. melpomene vulcanus*; *H. erato* mean distance from conspecific = 0.26; Mann-Whitney, $P = 1.0195 \times 10^{-15}$; *H. melpomene* mean distance = 0.41; $P = 5.1718 \times 10^{-31}$). The calculated extent of convergence by *H. melpomene* is 1.6 times that by *H. erato* (1.4 with hybrids excluded; fig. S7). Of the three comparative analyses (fig. S7), two indicate mutual convergence (Fig. 4 and fig. S7, A and B), and one indicates divergence by *H. erato* outside the geographic range of *H. melpomene*.

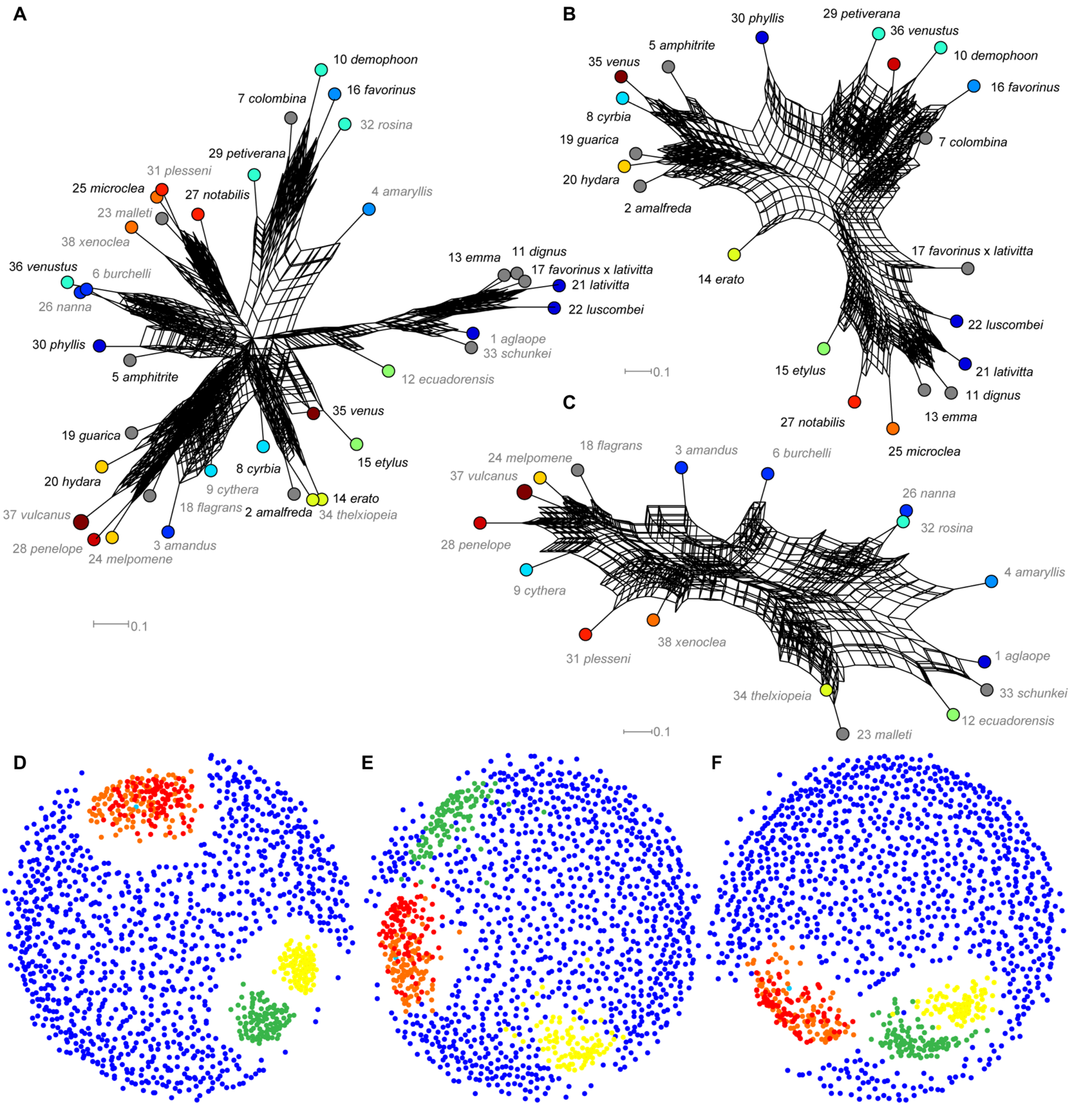

**Fig. 1. Phylogenetic relationships between subspecies of *H. erato* and *H. melpomene*.** (**A** to **C**) Neighbor-joining consensus networks of phenotypic distance (eight sampled embedding axes, 100 replicates). (**D** to **F**) Tree space visualizations for phylogenies based on phenotype (orange and red, 32 embedding axes; red, hybrids excluded; cyan, 64 embedding axes), color pattern genes (green) (*27*), neutral genes (yellow), and randomized topologies (blue, 1000 replicates). (A and D) All subspecies. (B and E) Only *H. erato*. (C and F) Only *H. melpomene*. (A to C) Node label color indicates species (black, *H. erato* and gray, *H. melpomene*). Node colors show mimicry groups (tables S1 and S9). Node numbers show subspecies numbers (table S1).

## Convergence on novel patterns

This case study (Fig. 4) also shows that, for both *H. erato* and *H. melpomene*, a pattern feature shared by the focal co-mimics is predominant in one of their conspecifics, but not the other, supporting mutual transfer of pattern features between these lineages. Specifically, *H. erato cyrbia* and *H. melpomene cythera* both have strongly blue, iridescent wings, a characteristic typical of *H. erato venus* but not of *H. melpomene vulcanus* [which is more black, particularly on the hindwing, e.g., individuals closest to the mean location for the subspecies (Fig. 4); see also type descriptions (*31*)]. Conversely, phenotypically average *H. erato cyrbia* and *H. melpomene cythera* have comparatively narrow red forewing bands, a feature more typical of *H. melpomene vulcanus* than of *H. erato venus* [average individuals (Fig. 4); type descriptions (*31*)]. This represents a newly demonstrated mechanism for the coevolution of novel phenotypes, as discussed below.

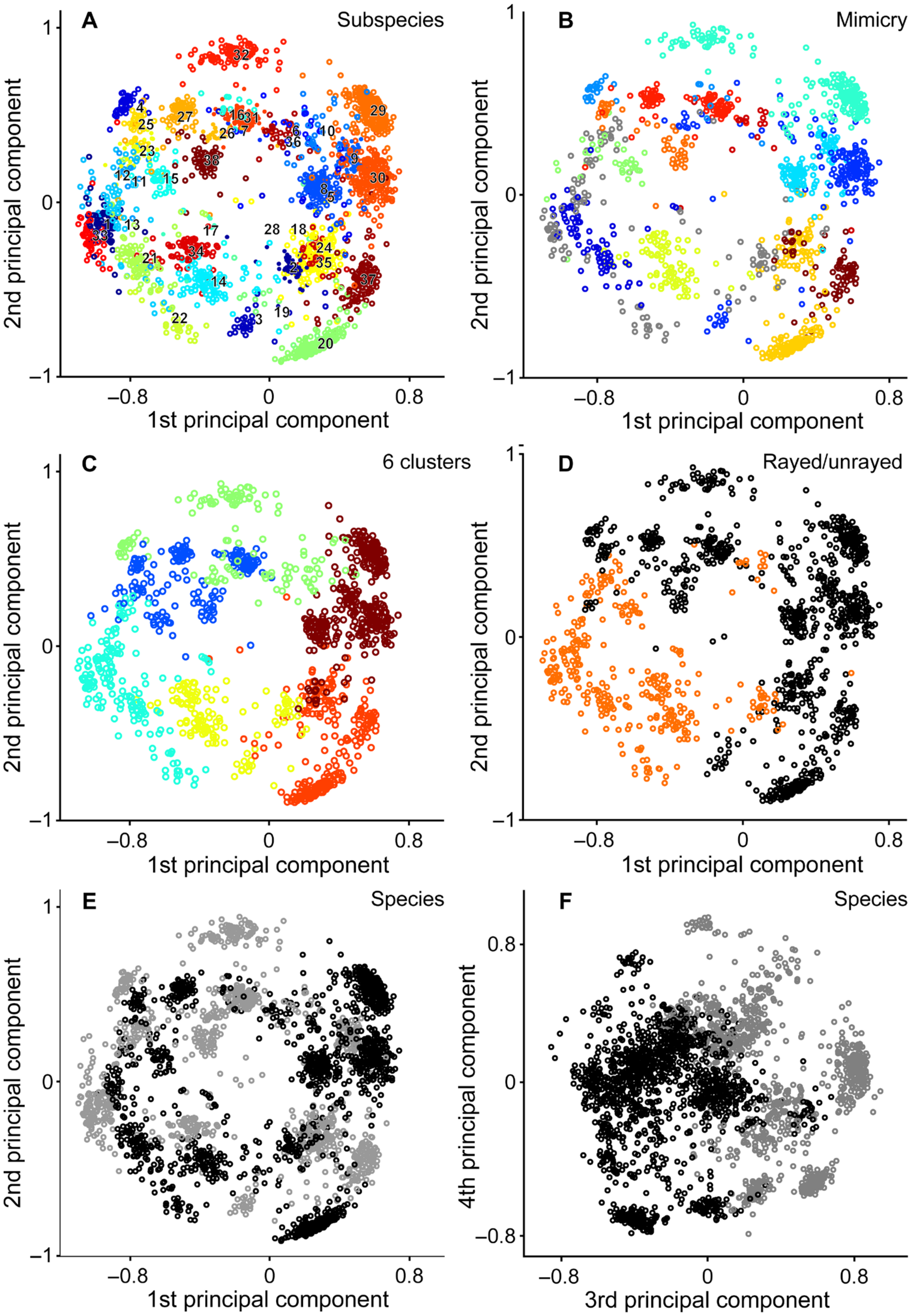

**Fig. 2. Principal component visualization of phenotypic variation among *Heliconius* butterflies.** Principal component scores calculated for 2468 images of butterfly species *H. erato* and *H. melpomene* on the basis of image coordinates in a 64-dimensional phenotypic space, generated using a deep convolutional triplet network. Cumulative variance explained by displayed principal component axes: 1, 28%; 2, 50%; 3, 68%; and 4, 81%. (**A**) Butterfly subspecies 1 to 38 (Fig. 1 and table S7). (**B**) Twelve traditionally hypothesized (*12*) mimicry complexes (tables S1 and S9) of *H. erato* and *H. melpomene* subspecies (gray circles indicate nonmimics, not included in any of these mimicry complexes). (**C**) Six hierarchical clusters. (**D**) Two broad classes of type pattern for each subspecies (table S8), with orange rays (orange circles) or without rays (black circles). (**E** and **F**) Species, *H. erato* (black circles) and *H. melpomene* (gray circles).

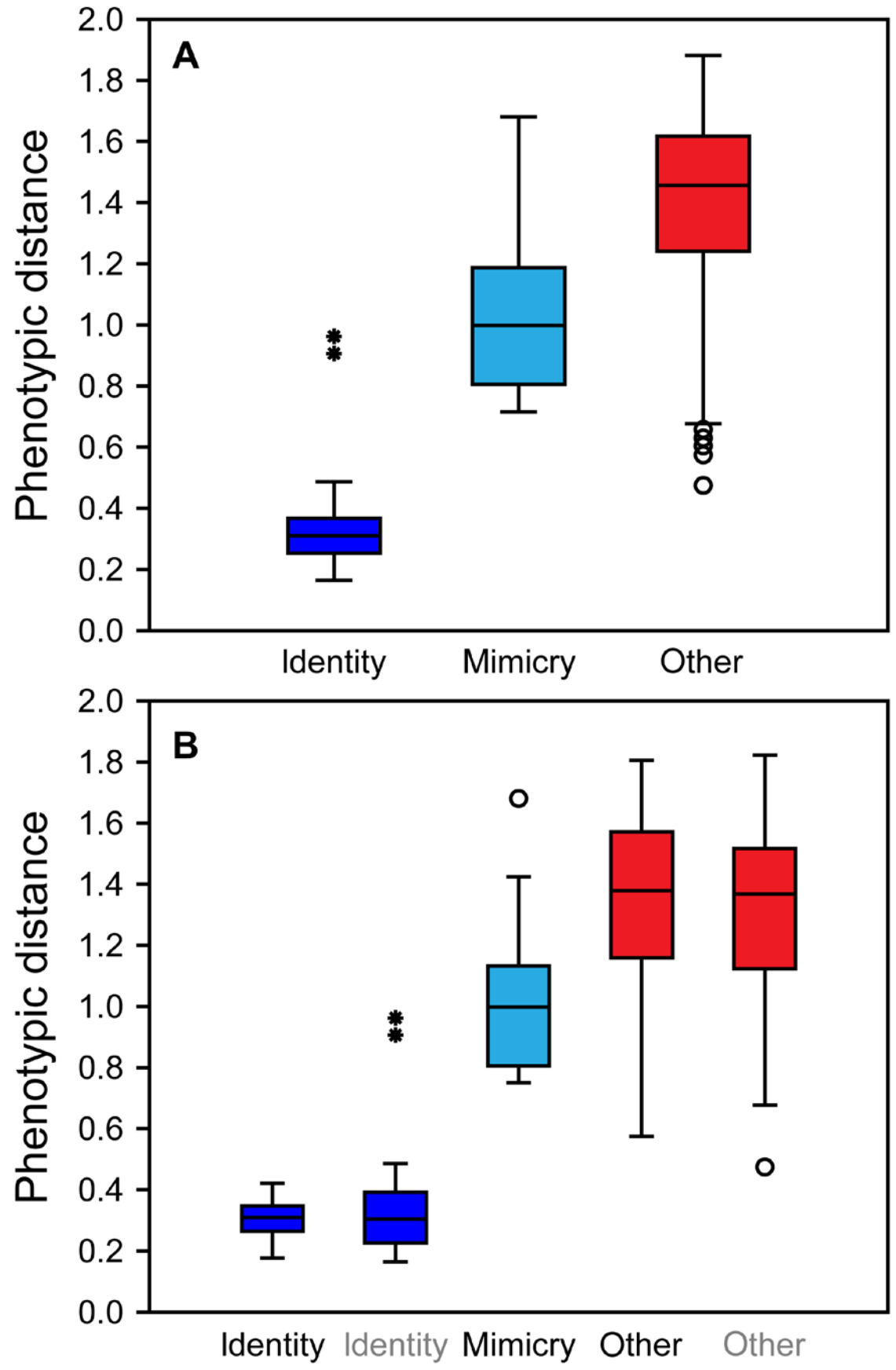

**Fig. 3. Average pairwise Euclidean phenotypic distances between subspecies of *H. erato* and *H. melpomene*.** (**A**) Box plot of mean pairwise phenotypic distances (table S4) within subspecies (identity), between co-mimic subspecies (mimicry), and between all other subspecies (other). Sample sizes are 38, 19, and 684 subspecies pairs, respectively. (**B**) Separated species, *H. erato* (black labels, identity and other) and *H. melpomene* (gray) and interspecies co-mimics (mimicry). Sample sizes are 21, 17, 15, 209, and 133 pairs. Boxes show 25 to 75% quartiles; horizontal lines, medians; whiskers, inner fence within 1.5 × box height; circles and asterisks, outliers, respectively, within or beyond 3 × box height.

## DISCUSSION

### Quantification of phenotypic similarity using deep learning

The spatial embedding constructed by the deep convolutional neural network captures and systematizes the phenotypic variation among *Heliconius* butterfly subspecies, from relatively subtle differences in the size, shape, number, position and color of wing pattern features (Fig. 2A, Table 1, and fig. S4) to broad divisions between major patterns groups (Fig. 2, C and D, and Table 1). This provides a comprehensive quantification of visible phenotypic similarity and an objective test of taxonomic delimitation (*3*), difficult or impossible to achieve with traditional morphometric methods (*6*, *7*) because of the scale and complexity of phenotypic variation.

The availability of genetic distances for butterfly subspecies studied here provides a ground truth against which the phenotypic distances, generated using the deep convolutional triplet network, can be tested. This ground truth is closely related to the natural selection pressures on wing pattern phenotype that these butterflies experienced during their evolutionary history, especially for genes specifically associated with wing pattern phenotype (*27*). The statistically significant similarity between the phenotypic neighbor-joining trees generated using deep learning and independently reconstructed pattern gene phylogenies (*26*) therefore confirms the utility of this spatial embedding method (*21*, *22*) for evolutionary analysis. This demonstrates that deep learning can achieve informative quantification of biological phenome samples at microevolutionary (intraspecies) to macroevolutionary (interspecies) scales. Our finding that deep learning can recover significant phylogenetic information from phenome samples also validates what is, to our knowledge, the first fully automated, objective method for the construction of phenotypic (or "morphological") phylogenies. This is of particular significance since subjective morphological phylogenies and taxonomic relationships (traditionally based on morphology) have been shown to exhibit statistical nonreplicability between researchers (*1*) and consistent topological biases (*11*).

### Phenotypic convergence in Müllerian mimicry

Traditional hypotheses of mimicry were based on subjective qualitative assessments of phenotypic similarity between subspecies with overlapping geographic ranges (*12*). The significantly lower phenotypic distances between co-mimic subspecies of *H. erato* and *H. melpomene* relative to nonmimics (Fig. 3B) now quantitatively demonstrate evolutionary convergence (*24*, *25*) in visible phenotype during the evolution of mimicry, considering as input all phenotypic information from the 2468 dorsal and ventral butterfly photographs. Historical, qualitative discussions of convergence have also focused primarily on taxonomic type patterns (based on a small number of type specimens selected as representative examples for each subspecies) (*31*). Our analysis of specimens exhaustively sampled from the Natural History Museum collection demonstrates statistically significant convergence across this comprehensive sample of wild phenotypic variation. This indicates that, even with the inclusion of individual variation and natural hybrids (table S2), evolutionary convergence in visual mimicry exerts a dominant statistical signal.

Hybridization occurs naturally between some intraspecific subspecies (and with some other *Heliconius* species, although *H. erato* and *H. melpomene* do not interbreed) both at hybrid zones where adjacent geographic ranges meet and, sometimes, far from these range boundaries. Hybrids may deviate perceptibly from a locally common wing pattern, in which case preferential predation can provide purifying selection that stabilizes mimicry (*17*). Hybridization is also increasingly recognized as a source of wing pattern variation (in addition to standard mutation) that can promote reproductive isolation and ecological speciation (*15*). Consequently, hybridization forms an important component of this mimicry system, with potentially complex effects on mimicry evolution. Statistical comparisons of phenotypic distance with hybrids excluded demonstrate that an even stronger signal of mimicry is apparent than for the complete dataset. These findings confirm theoretical suggestions that the general effect of wild hybridization between *Heliconius* subspecies is to dilute the strength of phenotypic mimicry (*16*), although mimicry remains the dominant phenotypic signal even with hybrids included.

The two species, *H. erato* and *H. melpomene*, have independently evolved similar wing patterns multiple times, with a hierarchy of convergent pattern detail (Fig. 2, B to D, and Table 1). Across the 38 subspecies, there are six distinct and convergent phenotypic clusters (Fig. 2C, Table 1, and fig. S6), which include interspecies co-mimics. These convergent phenotypic clusters include, for example, the orange "rayed" patterns, nonrayed pattern types containing red and black "postman"

**Table 1. Hierarchical clusters of subspecies from *H. erato* and *H. melpomene* calculated from the phenotypic spatial embedding.** Cluster membership is based on the modal specimen value for each subspecies after exclusion of hybrid specimens (table S2). Subspecies are illustrated by the dorsal photograph closest to the subspecies principal component analysis centroid (corresponding to Fig. 2A). *N* indicates subspecies number (table S1). Cluster numbers are colored by value to highlight divisions.

| | | | | N clusters | | | | | | | | | | | | | | | | | | | | | | | | | | | | | | | | | |
|---|---|---|---|---|---|---|---|---|---|---|---|---|---|---|---|---|---|---|---|---|---|---|---|---|---|---|---|---|---|---|---|---|---|---|---|---|---|
| **Species** | **Subspecies** | **N** | **Image** | **2** | **3** | **4** | **5** | **6** | **7** | **8** | **9** | **10** | **11** | **12** | **13** | **14** | **15** | **16** | **17** | **18** | **19** | **20** | **21** | **22** | **23** | **24** | **25** | **26** | **27** | **28** | **29** | **30** | **31** | **32** | **33** | **34** | **35** |
| *erato* | *venustus* | 36 | 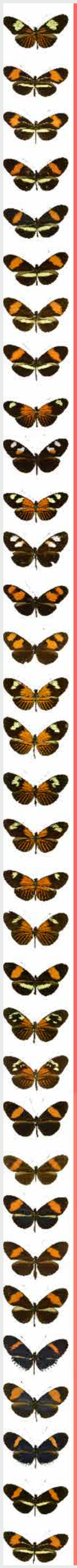 | 1 | 3 | 1 | 3 | 1 | 3 | 4 | 6 | 8 | 1 | 3 | 5 | 7 | 9 | 11 | 13 | 14 | 15 | 17 | 18 | 19 | 20 | 21 | 1 | 3 | 5 | 7 | 9 | 10 | 12 | 14 | 16 | 18 | 20 |
| *melpomene* | *nanna* | 26 | | 1 | 3 | 1 | 3 | 1 | 3 | 4 | 6 | 8 | 1 | 3 | 5 | 7 | 9 | 11 | 13 | 14 | 15 | 17 | 18 | 19 | 20 | 21 | 2 | 4 | 6 | 8 | 10 | 11 | 1 | 3 | 5 | 7 | 9 |
| *melpomene* | *burchelli* | 6 | | 1 | 3 | 1 | 3 | 1 | 3 | 4 | 6 | 8 | 1 | 3 | 5 | 7 | 9 | 11 | 13 | 14 | 15 | 17 | 18 | 19 | 20 | 21 | 2 | 4 | 6 | 8 | 10 | 11 | 2 | 4 | 6 | 8 | 10 |
| *erato* | *demophoon* | 10 | | 1 | 3 | 1 | 3 | 1 | 3 | 4 | 6 | 8 | 2 | 4 | 6 | 8 | 10 | 12 | 1 | 3 | 5 | 7 | 9 | 11 | 13 | 14 | 16 | 18 | 19 | 21 | 1 | 3 | 5 | 7 | 9 | 11 | 13 |
| *erato* | *colombina* | 7 | | 1 | 3 | 1 | 3 | 1 | 3 | 4 | 6 | 8 | 2 | 4 | 6 | 8 | 10 | 12 | 1 | 3 | 5 | 7 | 9 | 11 | 13 | 14 | 16 | 18 | 19 | 21 | 2 | 4 | 6 | 8 | 10 | 12 | 14 |
| *erato* | *favorinus* | 16 | | 1 | 3 | 1 | 3 | 1 | 3 | 4 | 6 | 8 | 2 | 4 | 6 | 8 | 10 | 12 | 1 | 3 | 5 | 7 | 9 | 11 | 13 | 14 | 16 | 18 | 19 | 21 | 2 | 4 | 6 | 8 | 10 | 12 | 14 |
| *melpomene* | *amaryllis* | 4 | | 1 | 3 | 1 | 3 | 1 | 3 | 4 | 6 | 8 | 2 | 4 | 6 | 8 | 10 | 12 | 2 | 4 | 1 | 3 | 5 | 7 | 9 | 10 | 12 | 14 | 16 | 18 | 20 | 21 | 22 | 24 | 26 | 28 | 29 |
| *melpomene* | *rosina* | 32 | | 1 | 3 | 1 | 3 | 1 | 3 | 4 | 6 | 8 | 2 | 4 | 6 | 8 | 10 | 12 | 2 | 4 | 2 | 4 | 6 | 8 | 10 | 11 | 13 | 15 | 17 | 19 | 21 | 22 | 23 | 25 | 27 | 29 | 30 |
| *erato* | *etylus* | 15 | | 1 | 3 | 1 | 3 | 2 | 4 | 5 | 7 | 1 | 3 | 5 | 7 | 9 | 11 | 13 | 14 | 15 | 16 | 1 | 3 | 5 | 7 | 9 | 11 | 13 | 15 | 17 | 19 | 20 | 21 | 23 | 25 | 27 | 1 |
| *erato* | *notabilis* | 27 | | 1 | 3 | 1 | 3 | 2 | 4 | 5 | 7 | 1 | 3 | 5 | 7 | 9 | 11 | 13 | 14 | 15 | 16 | 2 | 4 | 6 | 8 | 1 | 3 | 5 | 7 | 9 | 11 | 12 | 13 | 15 | 17 | 19 | 21 |
| *melpomene* | *malleti* | 23 | | 1 | 3 | 1 | 3 | 2 | 4 | 5 | 7 | 1 | 3 | 5 | 7 | 9 | 11 | 13 | 14 | 15 | 16 | 2 | 4 | 6 | 8 | 2 | 4 | 6 | 8 | 10 | 12 | 13 | 14 | 16 | 18 | 20 | 22 |
| *melpomene* | *plesseni* | 31 | | 1 | 3 | 1 | 3 | 2 | 4 | 5 | 7 | 2 | 4 | 1 | 3 | 5 | 7 | 9 | 11 | 12 | 13 | 15 | 16 | 17 | 19 | 20 | 22 | 23 | 24 | 26 | 27 | 28 | 29 | 31 | 32 | 33 | 34 |
| *erato* | *microclea* | 25 | | 1 | 3 | 1 | 3 | 2 | 4 | 5 | 7 | 2 | 4 | 2 | 4 | 6 | 8 | 10 | 12 | 13 | 14 | 16 | 17 | 18 | 1 | 3 | 5 | 7 | 9 | 11 | 13 | 14 | 15 | 17 | 19 | 21 | 23 |
| *melpomene* | *xenoclea* | 38 | | 1 | 3 | 1 | 3 | 2 | 4 | 5 | 7 | 2 | 4 | 2 | 4 | 6 | 8 | 10 | 12 | 13 | 14 | 16 | 17 | 18 | 2 | 4 | 6 | 8 | 10 | 12 | 14 | 15 | 16 | 18 | 20 | 22 | 24 |
| *melpomene* | *aglaope* | 1 | | 1 | 3 | 2 | 1 | 3 | 5 | 6 | 8 | 9 | 10 | 11 | 13 | 14 | 1 | 3 | 5 | 7 | 8 | 10 | 12 | 14 | 16 | 17 | 19 | 21 | 22 | 24 | 25 | 26 | 27 | 29 | 30 | 1 | 3 |
| *melpomene* | *schunkei* | 33 | | 1 | 3 | 2 | 1 | 3 | 5 | 6 | 8 | 9 | 10 | 11 | 13 | 14 | 1 | 3 | 5 | 7 | 8 | 10 | 12 | 14 | 16 | 17 | 19 | 21 | 22 | 24 | 25 | 26 | 27 | 29 | 30 | 2 | 4 |
| *erato* | *luscombei* | 22 | | 1 | 3 | 2 | 1 | 3 | 5 | 6 | 8 | 9 | 10 | 11 | 13 | 14 | 2 | 4 | 6 | 1 | 3 | 5 | 7 | 9 | 11 | 12 | 14 | 16 | 18 | 20 | 22 | 23 | 24 | 26 | 1 | 3 | 5 |
| *erato* | *lativitta* | 21 | | 1 | 3 | 2 | 1 | 3 | 5 | 6 | 8 | 9 | 10 | 11 | 13 | 14 | 2 | 4 | 6 | 1 | 3 | 5 | 7 | 9 | 11 | 12 | 14 | 16 | 18 | 20 | 22 | 23 | 24 | 26 | 2 | 4 | 6 |
| *melpomene* | *ecuadorensis* | 12 | | 1 | 3 | 2 | 1 | 3 | 5 | 6 | 8 | 9 | 10 | 11 | 13 | 14 | 2 | 4 | 6 | 2 | 4 | 6 | 8 | 10 | 12 | 13 | 15 | 17 | 1 | 3 | 5 | 7 | 9 | 11 | 13 | 15 | 17 |
| *erato* | *emma* | 13 | | 1 | 3 | 2 | 1 | 3 | 5 | 6 | 8 | 9 | 10 | 11 | 13 | 14 | 2 | 4 | 6 | 2 | 4 | 6 | 8 | 10 | 12 | 13 | 15 | 17 | 2 | 4 | 6 | 1 | 3 | 5 | 7 | 9 | 11 |
| *erato* | *dignus* | 11 | | 1 | 3 | 2 | 1 | 3 | 5 | 6 | 8 | 9 | 10 | 11 | 13 | 14 | 2 | 4 | 6 | 2 | 4 | 6 | 8 | 10 | 12 | 13 | 15 | 17 | 2 | 4 | 6 | 2 | 4 | 6 | 8 | 10 | 12 |
| *erato* | *erato* | 14 | | 1 | 3 | 2 | 2 | 4 | 6 | 7 | 9 | 10 | 11 | 12 | 1 | 3 | 5 | 7 | 9 | 10 | 11 | 13 | 15 | 16 | 18 | 19 | 21 | 2 | 4 | 6 | 8 | 9 | 11 | 13 | 15 | 17 | 19 |
| *melpomene* | *thelxiopeia* | 34 | | 1 | 3 | 2 | 2 | 4 | 6 | 7 | 9 | 10 | 11 | 12 | 2 | 4 | 6 | 8 | 10 | 11 | 12 | 14 | 2 | 4 | 6 | 8 | 10 | 12 | 14 | 16 | 18 | 19 | 20 | 22 | 24 | 26 | 28 |
| *erato* | *guarica* | 19 | | 2 | 1 | 3 | 4 | 5 | 7 | 8 | 1 | 3 | 5 | 6 | 8 | 10 | 12 | 14 | 15 | 16 | 17 | 18 | 19 | 20 | 21 | 22 | 23 | 24 | 25 | 27 | 28 | 29 | 30 | 1 | 3 | 5 | 7 |
| *erato* | *hydara* | 20 | | 2 | 1 | 3 | 4 | 5 | 7 | 8 | 1 | 3 | 5 | 6 | 8 | 10 | 12 | 14 | 15 | 16 | 17 | 18 | 19 | 20 | 21 | 22 | 23 | 24 | 25 | 27 | 28 | 29 | 30 | 2 | 4 | 6 | 8 |
| *melpomene* | *vulcanus* | 37 | | 2 | 1 | 3 | 4 | 5 | 7 | 8 | 2 | 4 | 6 | 7 | 9 | 11 | 13 | 1 | 3 | 5 | 6 | 8 | 10 | 12 | 14 | 15 | 17 | 19 | 20 | 22 | 23 | 24 | 25 | 27 | 28 | 30 | 31 |
| *melpomene* | *flagrans* | 18 | | 2 | 1 | 3 | 4 | 5 | 7 | 8 | 2 | 4 | 6 | 7 | 9 | 11 | 13 | 2 | 4 | 6 | 7 | 9 | 11 | 13 | 15 | 16 | 18 | 20 | 21 | 23 | 24 | 25 | 26 | 28 | 29 | 31 | 32 |
| *melpomene* | *melpomene* | 24 | | 2 | 1 | 3 | 4 | 5 | 7 | 8 | 2 | 4 | 6 | 7 | 9 | 11 | 13 | 2 | 4 | 6 | 7 | 9 | 11 | 13 | 15 | 16 | 18 | 20 | 21 | 23 | 24 | 25 | 26 | 28 | 29 | 31 | 32 |
| *melpomene* | *cythera* | 9 | | 2 | 2 | 4 | 5 | 6 | 1 | 3 | 5 | 7 | 9 | 10 | 12 | 1 | 3 | 5 | 7 | 8 | 9 | 11 | 13 | 15 | 17 | 18 | 20 | 22 | 23 | 25 | 26 | 27 | 28 | 30 | 31 | 32 | 33 |
| *erato* | *venus* | 35 | | 2 | 2 | 4 | 5 | 6 | 1 | 3 | 5 | 7 | 9 | 10 | 12 | 2 | 4 | 6 | 8 | 9 | 10 | 12 | 14 | 1 | 3 | 5 | 7 | 9 | 11 | 13 | 15 | 16 | 17 | 19 | 21 | 23 | 25 |
| *erato* | *cyrbia* | 8 | | 2 | 2 | 4 | 5 | 6 | 1 | 3 | 5 | 7 | 9 | 10 | 12 | 2 | 4 | 6 | 8 | 9 | 10 | 12 | 14 | 2 | 4 | 6 | 8 | 10 | 12 | 14 | 16 | 17 | 18 | 20 | 22 | 24 | 26 |
| *erato* | *petiverana* | 29 | | 2 | 2 | 4 | 5 | 6 | 2 | 1 | 3 | 5 | 7 | 8 | 10 | 12 | 14 | 15 | 16 | 17 | 18 | 19 | 20 | 21 | 22 | 23 | 24 | 25 | 26 | 28 | 29 | 30 | 31 | 32 | 33 | 34 | 35 |
| *erato* | *phyllis* | 30 | | 2 | 2 | 4 | 5 | 6 | 2 | 2 | 4 | 6 | 8 | 9 | 11 | 13 | 15 | 16 | 17 | 18 | 19 | 20 | 21 | 22 | 23 | 24 | 25 | 26 | 27 | 2 | 4 | 6 | 8 | 10 | 12 | 14 | 16 |

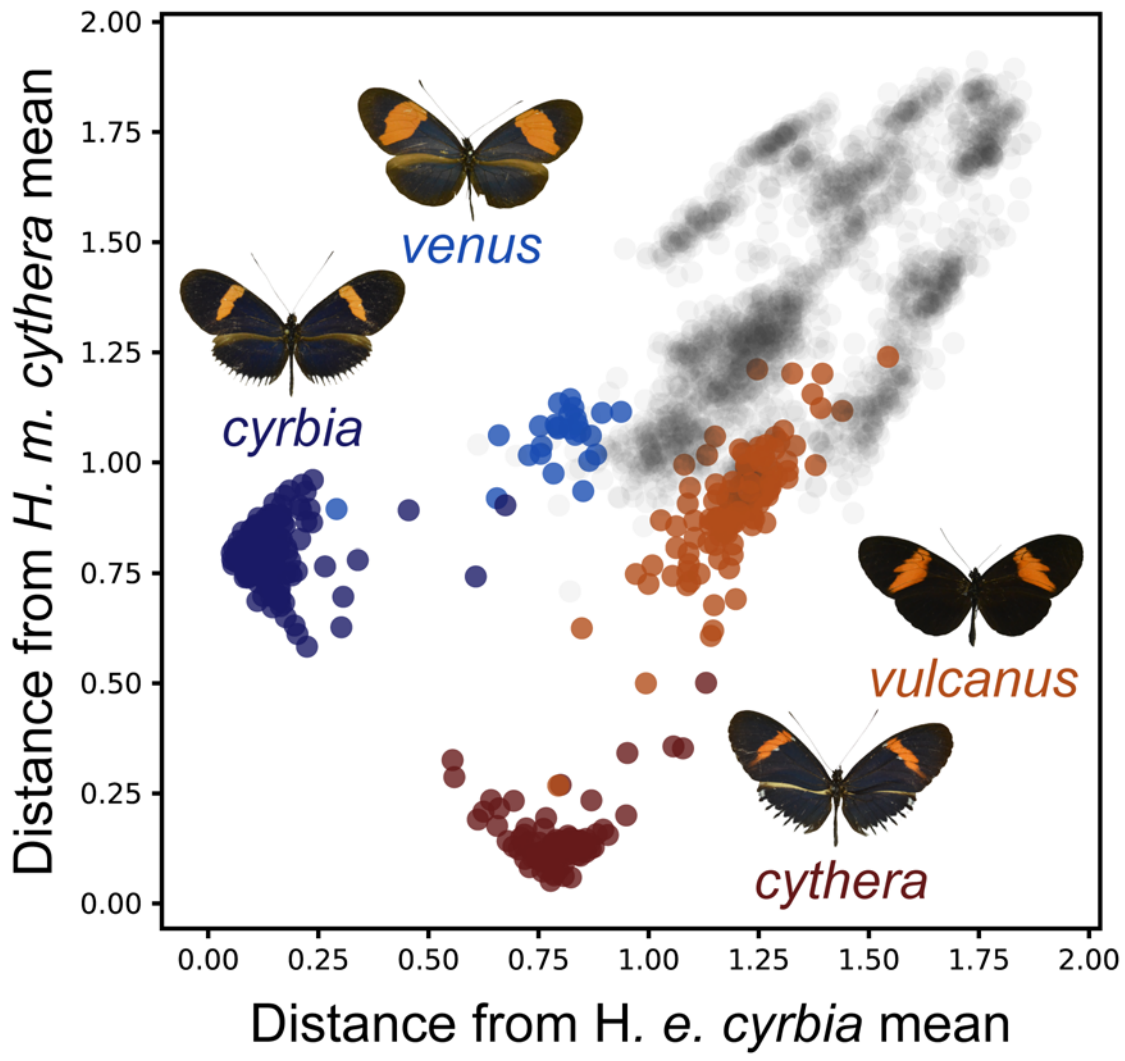

**Fig. 4. Comparative analysis of the extent of phenotypic convergence in mimicry.** Case study from comparative analyses of 12 subspecies (fig. S7). The locations of two focal co-mimics (*H. erato cyrbia*, dark blue circles; *H. melpomene cythera*, dark red circles) in phenotypic space are compared alongside their nearest conspecifics (*H. erato venus*, light blue circles; *H. melpomene vulcanus*, light red circles). Subspecies are illustrated by dorsal photographs of the butterfly closest to the mean location for the subspecies. Gray circles indicate images of other subspecies in the dataset. Axes show the squared distance from the mean location of the focal co-mimic, summed across all 64 spatial embedding axes. Distance between subspecies means on the *y* axis, *H. erato venus* − *H. erato cyrbia* = 0.26. Distance between subspecies means on the *x* axis, *H. melpomene vulcanus* − *H. melpomene cythera* = 0.41.

patterns (Fig. 2D), with or without a yellow hindwing band (hierarchical clustering with three clusters) and with red versus yellow forewing pattern features (four clusters), as well as patterns with a yellow forewing feature of increased complexity (five clusters).

This comprehensive, quantitative demonstration of phenotypic convergence meets the principal prediction of Müller's 1879 (*4*) model for the evolution of mimicry in mutually protected species (Müllerian mimics), such as unpalatable *Heliconius* butterflies (*12*). In this, the first mathematical model of Darwinian relative fitness (*4*, *14*), Müller predicted that two equally unpalatable, co-occurring populations will come to resemble each other because both benefit by sharing the cost of predator avoidance learning, with relative fitness benefits proportional to the inverse square of abundance [with additional abundance effects suggested by later reformulations (*14*)] (*4*). Relative abundance has been estimated respectively at 1.3:1 to 2.4:1 for *H. erato* and *H. melpomene* overall (*28*, *29*) [predicting respective benefits of 1:1.7 to 1:5.8 (*4*)] but is subject to marked local fluctuations, in which momentary relative abundances of the species may be equal or reversed relative to the overall trend (*32*).

## Mutual convergence and strict coevolution

The historically controversial (*13*) extent of mutual convergence in mimicry is illustrated by a detailed case study (Fig. 4) from three comparative analyses (covering 12 subspecies), all of which support reciprocal influence (fig. S7), which is the definitive feature of strict coevolution (*19*). Focal co-mimics *H. erato cyrbia* and *H. melpomene cythera* (Fig. 4) share a white hindwing marginal fringe and strong blue iridescence of the dorsal wing surface also present in their phenotypically closest conspecifics, *H. erato venus* and *H. melpomene vulcanus* [conspecifics also recovered as sister groups in independent gene phylogenies (*26*)]. Both pattern features are likely to have been secondarily derived (rather than ancestral within each species) on the basis of gene phylogenies, biogeographic distribution, and phylogeographic reconstruction (*12*, *16*, *26*, *33*). Geographic ranges (fig. S2) and phylogeographic reconstruction (*26*) also suggest that *H. erato cyrbia* and *H. melpomene cythera* are at an extreme of a west Andean subradiation (with conspecifics adjacent, north-east). This supports pattern derivation from the forms of *H. erato venus* and *H. melpomene vulcanus* to those of *H. erato cyrbia* and *H. melpomene cythera* (rather than vice versa). This shared biogeographic history is also compatible with the potential for strict interspecies coevolution, since reciprocal evolutionary influence between two taxa requires that they co-occur in both space and time (*26*). With regard to timing, while early phylogenetic studies cast doubt on the extent of contemporaneous diversification by *H. erato* and *H. melpomene* (*34*), subsequent genomic analyses (taking into account population sizes) have reconstructed their diversification over closely overlapping time ranges (*28*, *29*).

Comparisons of the average locations of these subspecies in phenotypic space (Fig. 4) show statistically significant mutual convergence. The implied extent of convergence by *H. melpomene* is 1.6 times that by *H. erato*, in line with the general frequency-dependent fitness benefits expected in Müllerian mimicry (*4*) between these species (with hybrids excluded, the value is 1.4; fig. S7). Of the three comparative analyses (fig. S7), two indicate mutual convergence (Fig. 4 and fig. S7, A and B), and one indicates divergence by *H. erato* where this subspecies leaves the geographic range and coevolutionary influence of *H. melpomene* [in Central-North American *H. erato petiverana*, fig. S7C; see also discussion in (*23*)]. For comparison, reversal of evolutionary polarities (swapped conspecific foci) would imply two cases of mutual convergence (fig. S7, A and B) and one of divergence by *H. melpomene* (fig. S7C).

Interpreted with regard to the extent of reciprocal influence in wing pattern evolution, which has been controversial and difficult to test (*13*, *35*), mutual convergence (Figs. 3 and 4 and fig. S7, A and B) and maintenance of wing pattern similarity in sympatry (e.g., fig. S7C) suggest that *H. erato* and *H. melpomene* have influenced each other (albeit to varying extents), with each species acting as both model and mimic to some degree, meeting the essential condition of strict reciprocal (*13*, *19*) coevolution. This is in sympathy with classical Müllerian mimicry theory (*4*) while in contrast to theoretical alternatives such as entirely one-sided advergence (Fig. 5) (*13*) or divergence of a sympatric model during parasitic, quasi-Batesian mimicry (*36*).

## Coevolution of phenotypic novelty

The focal case study of *H. erato cyrbia*, *H. melpomene cythera*, and their closest conspecifics (Fig. 4) also indicates mutual coevolutionary transfer of pattern features between interspecific lineages (Fig. 4; conceptual diagram in Fig. 5). This represents a class of phenomic recombination, generating novel phenotypic combinations, but acting via predator-mediated selection on phenotype (and underlying genotype) without direct gene exchange. To our knowledge, this mechanism for the generation of evolutionary novelty represents a newly identified coevolutionary phenomenon. However, coevolutionary pattern recombination is also a logical consequence of the mutual convergence predicted by classical Müllerian mimicry theory (*4*, *23*, *37*) and exemplifies the definitive feature of coevolution: coordination of evolutionary change in genetically distinct populations (*38*).

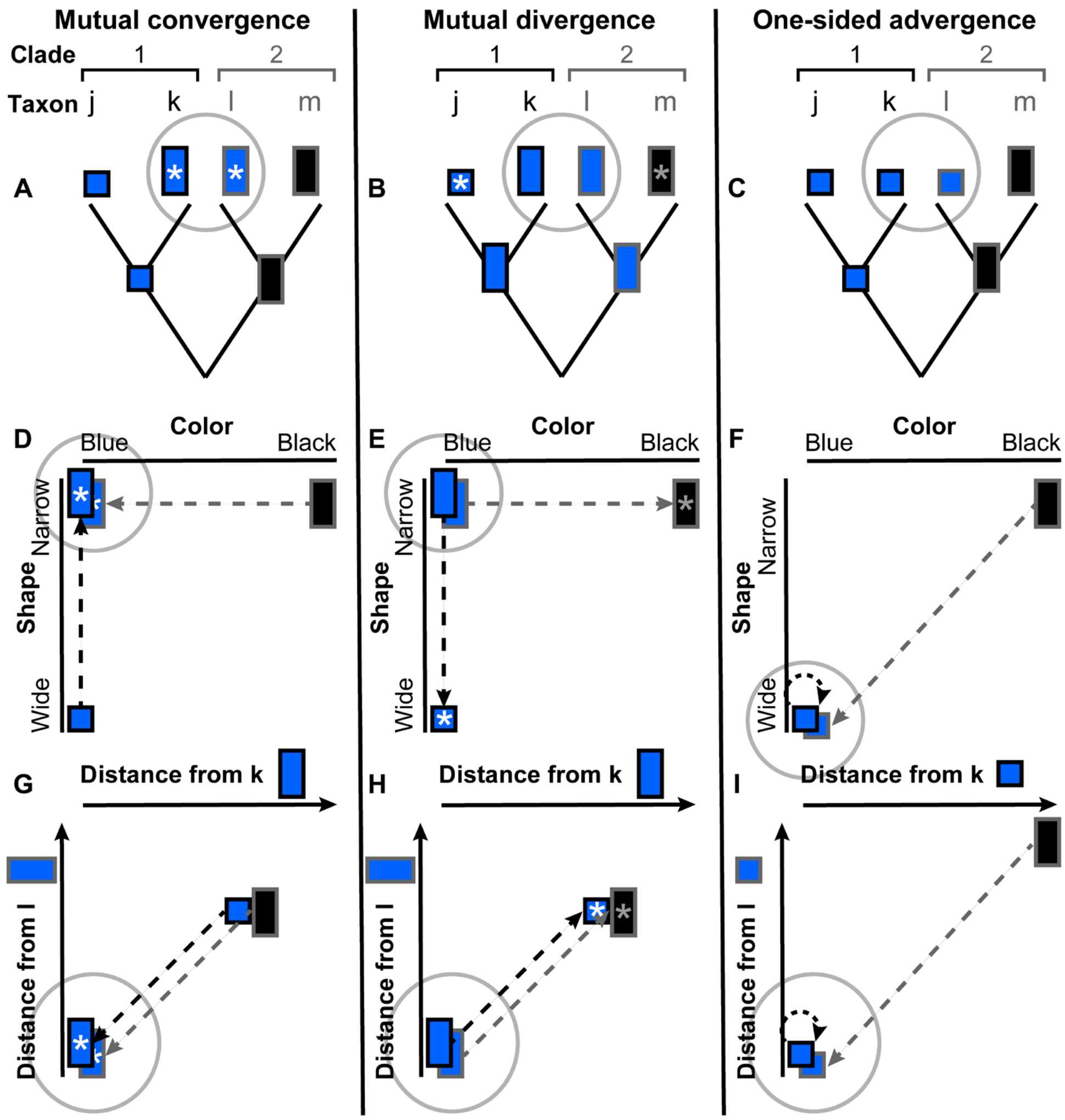

**Fig. 5. Conceptual diagrams illustrating the evolutionary alternatives of mutual convergence, mutual divergence, and one-sided convergence ("advergence").** Dashed arrows indicate the direction of evolutionary change. Left (**A**, **D**, and **G**): Mutual convergence in focal taxa (focal taxa, gray circles) with reciprocal transfer of pattern features (e.g., forewing band shape versus wing color) between two clades (1 and 2, respectively black versus gray outlines). Middle (**B**, **E**, and **H**): Reversed polarity with mutual divergence from the focal taxa. Right (**C**, **F**, and **I**): Advergence by one clade onto another (*13*). Asterisks indicate new derived patterns (feature combinations). When expressed in terms of the phenotypic distance from the focal taxa (G to I), mutual convergence (G) is characterized by a decreasing distance along the arrow of evolutionary change in both clades. Mutual divergence (H) is characterized by a increasing distance in both clades. Advergence (*13*) is characterized by a decreasing distance (and a greater distance traveled) in one clade (I).

Mutual convergence, of the type identified here (Fig. 4), can be considered equivalent to reciprocal advergence (where, in this usage, advergence means one-sided mimicry evolution (*13*)). One-sided advergence involves evolution in one lineage to match a phenotypic feature already present in its mimicry model (Fig. 5). This represents a coevolutionary information transfer from one lineage to another. Consequently, mutual convergence (i.e., reciprocal advergence) in two distinct pattern features entails reciprocal information exchange, with the potential to create a novel pattern, which did not previously exist in any lineage (Figs. 4 and 5). For example, in two lineages (labeled 1 and 2) that mutually converge (Figs. 4 and 5A) across two pattern features, forewing band shape (wide or narrow) and wing color (black or blue), the ancestral patterns (here inferred as wide, blue in lineage 1 and narrow, black in lineage 2) are then supplemented by a newly derived combination (here narrow, blue in both lineages). Thus, lineage 1 has taken on the ancestral band shape of lineage 2, while lineage 2 has picked up the ancestral wing color of lineage 1, to generate a new wing pattern.

This reveals that Müllerian mimicry can generate new phenotypes by combining pattern features from different lineages (Figs. 4 and 5 and fig. S7). While it is natural to expect that divergence will generate new phenotypic traits (Fig. 5, B, E, and H), the demonstration that evolutionary convergence can also generate new phenotypes is more unexpected (Figs. 4 and 5, A, D, and G). An intuitive expectation was that mimicry would drive convergence on a single-wing pattern (*13*) and classical discussion of coevolution (*4*, *12*) did not itself explain the remarkable phenotypic diversity observed among Müllerian mimics (*13*). Consequently, proposed diversification mechanisms have often focused on external factors such as microhabitat adaptation (*13*), variable predator abundance (*18*), and isolation in glacial refugia (*12*), as well as effects from additional butterfly species (*12*) and stochastic population genetics (*13*, *18*) (although all have been controversial).

Here, direct quantitative analysis of phenotype reveals that mutual convergence can itself increase phenotypic diversity.

## MATERIALS AND METHODS

### Image acquisition and preprocessing

Butterfly specimens (1269) from the species *H. erato* and *H. melpomene* were photographed at the NHM London (specimen numbers and image filenames, table S2), using consistent photographic protocols and lighting. Photographs were screened for poor image quality giving a final dataset of 1234 butterflies and 2468 photographs, including a dorsal and ventral photograph of each butterfly. For deep learning, images were then cropped and resized to a height of 64 pixels (maintaining the original image aspect ratio and padded to 140 pixels wide). The photographic dataset used for deep learning is provided in the Dryad Data Repository with filenames corresponding to joined data in table S2.

Taxonomic and locality data were recorded from NHM *Heliconius* butterfly specimen labels (table S2). Subspecies taxonomy follows reference (*30*). The complete photographic dataset covers 37 named subspecies and one-labeled cross: 21 subspecies from *H. erato* and 17 from *H. melpomene*. Specimens of these subspecies were sampled exhaustively from the NHM collection, within the limits of the data collection period. The complete photographic dataset (fig. S4, Dryad Data Repository) covers both specimens closely representative of subspecies descriptions (*30*, *31*) (including available holotypes, syntypes, and paratypes; table S2) as well as other, naturally varying, individuals. These variants include some likely hybrid specimens showing varying levels of phenotypic admixture from other subspecies (see additional taxonomic information, table S2). Inclusion of all available specimens in machine learning covered a very broad range of the phenotypic diversity within these species, providing the deep learning network with all available information from which to learn phenotypic features correlated with subspecies identification. Two sets of statistical analyses were then conducted, one set including all 1269 photographed butterfly specimens and the second set excluding potential hybrid specimens to give a reduced dataset of 815 specimens and 1630 photographs (fig. S4 and table S2).

### Deep learning

To quantify phenotypic distances between *Heliconius* butterflies, a deep convolutional neural network was trained to classify photographs of *Heliconius* butterflies by subspecies, with 1500 of the 2468 total images used for network training and the remainder for testing. The training method (*21*, *22*) used triplets of images, each replicate showing the network two images sampled from the same subspecies and one sampled from a different subspecies. Image classification and spatial embedding were performed using a 15-layer deep learning network (Supplementary Computer Code), which we name ButterflyNet (fig. S1). This makes use of a triplet embedding loss function (*21*, *22*) to train a network to organize its inputs (images) in a space such that proximity in that space (Euclidean distance) is highly correlated with identity (in this case subspecies). The learned embedding was then passed through an additional small network to perform direct categorical subspecies classification. Overall, the total network optimizes the sum of the triplet loss and the categorical cross entropy (eqs. S1 and S2). The computer code used for machine learning is provided as a Python script (Supplementary Computer Code), which makes use of the PyTorch, Scikit-l arn, and Adam packages (for further details, see Supplementary Methods).

After the network was trained on 1500 images randomly sampled from the 2468 images in the dataset, network testing was performed on the remainder (968 images). Testing presents the trained network with new images, which it has not encountered before. The network then classifies the new images by subspecies, image classifications are compared to the known subspecies identities, and the overall accuracy of test classifications is reported. Additional testing was performed using an SVC trained on the embeddings from the main network (ButterflyNet) to determine the accuracy of classification of specimens to subspecies based on their locations in the phenotypic spatial embedding.

### Quantification of phenotypic similarity

Pairwise Euclidean phenotypic distances between all images were calculated from the coordinates of all 2468 images within a spatial embedding with 64 dimensions, generated using the network (table S3). These distances were then used to calculate the average pairwise Euclidean phenotypic distances between all subspecies (table S4). These were then used in statistical comparisons between sets of unique unordered subspecies pairs (Fig. 3; for further details, see Supplementary Methods). Average phenotypic distances for subspecies sets and principal component scores were calculated using MATLAB scripts. Nonparametric statistical analyses were conducted using the program PAST 3, after Shapiro-Wilk's tests indicated that distances for some subspecies sets were non-normally distributed (using an $\alpha$ value of 0.05). These analyses included Kruskal-Wallis tests for equal medians and, where this overall test was significant, subsequent Mann-Whitney pairwise comparisons of statistical distributions between groups.

### Phylogenetic analyses

Neighbor-joining trees for subspecies were constructed on the basis of phenotypic distances using MATLAB scripts (sampling either all 64 embedding axes with 1 replicate or subsamples of 8 or 32 axes with 100 replicates). Neighbor joining is a simple and fast algorithm for phylogenetic reconstruction, which reconstructs relationships based on phenetic (overall) similarity (*39*) e.g., Euclidean phenotypic distance, as applied here. Neighbor joining does not require an a priori mechanistic model for the evolutionary process in question (e.g., evolution of butterfly wing pattern phenotype), in contrast to maximum likelihood models of DNA substitution, for example. This method is therefore suitable for this first phylogenetic study of phenomic distances calculated using deep learning on butterfly photographs. The correlation was then statistically compared between the resultant phenotypic neighbor-joining trees and genetic phylogenies reconstructed with Bayesian methods (which incorporate DNA substitution models) (*27*). To test the phylogenetic informativeness of the phenotypic distances against such independent data sources, sets of neighbor-joining phenotypic trees (of either all subspecies, *H. erato* only, or *H. melpomene* only) were compared against random tree topologies and phylogenies (*26*) reconstructed from published gene sequences (*27*) from gene loci (sampled from a different, smaller set of 127 butterfly individuals), which were either associated with *Heliconius* wing color pattern (*optix*, *bves*, *kinesin*, *GPCR*, and *VanGogh*) (*27*) or were neutral markers (*mt COI-COII*, *SUMO*, *Suz12*, *2654*, and *CAT*). Pairwise distances between trees from the different sets were calculated using the Robinson-Foulds (symmetric distance) metric in the program PAUP and statistically compared using nonparametric Mann-Whitney tests in the program PAST (after Shapiro-Wilk's tests indicated non-normal distributions). Tree space visualizations of tree similarity were produced, based on the Robinson-Foulds distance, using the tree set visualization package in the program Mesquite. Consensus networks were constructed to visualize all splits (taxon partitions) implied among sets of trees using the program SplitsTree 4.

## Testing evolutionary convergence

Quantitative tests of evolutionary convergence were applied using an operational definition of convergence essentially the same as that for discrete traits (independent derivation of the same morphological state, e.g., within two species), with quantitative analysis used to test relative phenotypic similarity across the complete dataset (*24*, *25*) and the number of quantitatively distinct phenotypic states, e.g., clusters (for further details, see Supplementary Methods).

To further explore the extent of reciprocal convergence in mimicry between *H. erato* and *H. melpomene*, detailed comparative analyses (*40*) were then conducted using 12 selected subspecies (fig. S7). First, sets of four subspecies were identified, with each set consisting of two pairs of interspecies co-mimics in which conspecifics were nearest neighbors in the phenotypic spatial embedding, permitting phenotypic sister-group comparisons (fig. S7, A and B). Focal co-mimics, with pattern features that are potentially derived, rather than ancestral, were identified for each comparative analysis based on all available independent information from gene phylogenies, biogeographic distribution (fig. S2), and phylogeographic reconstruction (*12*, *16*, *26*, *33*). For comparison, the analyses were then repeated with reversed polarity. In each comparative analysis, the position in phenotypic space of each of the focal subspecies was compared to that of their nearest conspecific [a type of sister-group comparison (*25*)]. Mann-Whitney tests for equal medians tested whether conspecifics differed significantly in their distance from the focal co-mimic of the other species (after Shapiro-Wilk's tests indicated that some subspecies values were non-normally distributed).

## SUPPLEMENTARY MATERIALS

Supplementary material for this article is available at http://advances.sciencemag.org/cgi/content/full/5/8/eaaw4967/DC1
Supplementary Methods
Fig. S1. Diagram of the architecture of the deep learning network ButterflyNet used in this study.
Fig. S2. Geographic localities for sampled butterfly specimens from the polymorphic mimicry complex of *H. erato* and *H. melpomene*.
Fig. S3. Heatmap showing mean pairwise phenotypic and geographic distances between 38 subspecies of *H. erato* (black labels) and *H. melpomene* (gray labels).
Fig. S4. Collections of specimen photographs used in this study, grouped by subspecies.
Fig. S5. Average pairwise Euclidean geographic distances between subspecies of *H. erato* and *H. melpomene*.
Fig. S6. Neighbor-joining trees of phenotypic distance between subspecies of *H. erato* and *H. melpomene*.
Fig. S7. Comparative analyses of the extent of phenotypic convergence in mimicry.
Fig. S8. Principal component visualization of *Heliconius* butterflies.
Table S1. Traditionally hypothesized co-mimic subspecies of *H. erato* and *H. melpomene*.
Table S2. Taxonomic and locality data recorded for historical specimens of *H. erato* and *H. melpomene* held in the collections of the NHM London.
Table S3. Coordinates of butterfly images on 64 axes of a Euclidean phenotypic space constructed using a deep convolutional network with triplet training.
Table S4. Mean pairwise Euclidean phenotypic distances between subspecies from *H. erato* and *H. melpomene*.
Table S5. Mean pairwise squared Euclidean phenotypic distances between subspecies from *H. erato* and *H. melpomene*.
Table S6. Mean pairwise Euclidean geographic distances between subspecies from *H. erato* and *H. melpomene*.
Table S7. Number of sampled butterfly individuals for each subspecies.
Table S8. Broad pattern class of the type specimen of each subspecies.
Table S9. Traditionally hypothesized mimicry complexes of *H. erato* and *H. melpomene* subspecies.
Table S10. Statistical comparisons of pairwise Robinson-Foulds distances between sets of phylogenetic trees.
Table S11. Statistical comparisons with hybrids excluded of pairwise Robinson-Foulds distances between sets of phylogenetic trees.
Supplementary Computer Code in ipynb Format
Supplementary Computer Code in PDF Format
Reference (*41*)

**Acknowledgments:** Data compilation received additional assistance from volunteer research assistant S. Sjosten. We thank J. Mallet and S. Conway Morris for discussion of the manuscript as well as J. Turner and two anonymous referees for highly constructive reviews. **Funding:** Funding was received from an ELSI Origins Network (EON) Research Fellowship (J.F.H.C.) supported by a grant from the John Templeton Foundation. **Author contributions:** This study was originally designed by J.F.H.C. and N.G. S.L. and R.C. collected data, as supervised by B.H. and J.F.H.C. N.G. and J.F.H.C. wrote the computer code. J.F.H.C. performed statistical analyses. J.F.H.C. wrote the manuscript with input from all authors. **Competing interests:** The authors declare that they have no competing interests. **Data and materials availability:** All data needed to evaluate the conclusions in the paper are present in the paper and/or the Supplementary Materials. Additional photographic data related to this paper are available on the Dryad data repository, DOI: 10.5061/dryad.2hp1978.

Submitted 3 January 2019
Accepted 8 July 2019
Published 14 August 2019
10.1126/sciadv.aaw4967

**Citation:** J. F. Hoyal Cuthill, N. Guttenberg, S. Ledger, R. Crowther, B. Huertas, Deep learning on butterfly phenotypes tests evolution's oldest mathematical model. *Sci. Adv.* **5**, eaaw4967 (2019).

# Deep learning on butterfly phenotypes tests evolution's oldest mathematical model

Jennifer F. Hoyal Cuthill, Nicholas Guttenberg, Sophie Ledger, Robyn Crowther and Blanca Huertas

# Supplementary Materials for

## Deep learning on butterfly phenotypes tests evolution's oldest mathematical model

Jennifer F. Hoyal Cuthill*, Nicholas Guttenberg*, Sophie Ledger, Robyn Crowther, Blanca Huertas

*Corresponding author. Email: j.hoyal.cuthill@elsi.jp (J.F.H.C.); ngutten@gmail.com (N.G.)

**The PDF file includes:**

Supplementary Methods
Fig. S1. Diagram of the architecture of the deep learning network ButterflyNet used in this study.
Fig. S2. Geographic localities for sampled butterfly specimens from the polymorphic mimicry complex of *H. erato* and *H. melpomene*.
Fig. S3. Heatmap showing mean pairwise phenotypic and geographic distances between 38 subspecies of *H. erato* (black labels) and *H. melpomene* (gray labels).
Fig. S4. Collections of specimen photographs used in this study, grouped by subspecies.
Fig. S5. Average pairwise Euclidean geographic distances between subspecies of *H. erato* and *H. melpomene*.
Fig. S6. Neighbor-joining trees of phenotypic distance between subspecies of *H. erato* and *H. melpomene*.
Fig. S7. Comparative analyses of the extent of phenotypic convergence in mimicry.
Fig. S8. Principal component visualization of *Heliconius* butterflies.
Reference (*41*)

**Other Supplementary Material for this manuscript includes the following:**

(available at advances.sciencemag.org/cgi/content/full/5/8/eaaw4967/DC1)

Table S1 (Microsoft Excel format). Traditionally hypothesized co-mimic subspecies of *H. erato* and *H. melpomene*.
Table S2 (Microsoft Excel format). Taxonomic and locality data recorded for historical specimens of *H. erato* and *H. melpomene* held in the collections of the NHM London.
Table S3 (Microsoft Excel format). Coordinates of butterfly images on 64 axes of a Euclidean phenotypic space constructed using a deep convolutional network with triplet training.
Table S4 (Microsoft Excel format). Mean pairwise Euclidean phenotypic distances between subspecies from *H. erato* and *H. melpomene*.

Table S5 (Microsoft Excel format). Mean pairwise squared Euclidean phenotypic distances between subspecies from *H. erato* and *H. melpomene*.
Table S6 (Microsoft Excel format). Mean pairwise Euclidean geographic distances between subspecies from *H. erato* and *H. melpomene*.
Table S7 (Microsoft Excel format). Number of sampled butterfly individuals for each subspecies.
Table S8 (Microsoft Excel format). Broad pattern class of the type specimen of each subspecies.
Table S9 (Microsoft Excel format). Traditionally hypothesized mimicry complexes of *H. erato* and *H. melpomene* subspecies.
Table S10 (Microsoft Excel format). Statistical comparisons of pairwise Robinson-Foulds distances between sets of phylogenetic trees.
Table S11 (Microsoft Excel format). Statistical comparisons with hybrids excluded of pairwise Robinson-Foulds distances between sets of phylogenetic trees.
Supplementary Computer Code in ipynb Format
Supplementary Computer Code in PDF Format

## Supplementary Methods

### Image acquisition and pre-processing

1269 butterfly specimens from the species *Heliconius erato* and *H. melpomene* were photographed at the Natural History Museum London (NHM) by two full-time research assistants over a two month period in 2016 (specimen numbers and image filenames, Table S2). Data collection costs were £4207 ($5807) for the complete dataset, averaging £3.31 ($4.57) per specimen and £1.66 ($2.29) per photograph. A consistent photographic setup was used throughout with a Nikon digital camera, ring light, light-grey foam background, cm to 0.5 mm rulers, and colour analysis chart constructed from a Spyder Checkr 24 colour card. Photographs were then screened for poor image quality or specimen wing pattern damage, giving a final dataset of 1234 butterflies and 2468 photographs, including a dorsal and ventral photograph of each butterfly. Photographs were cropped to include only the butterfly specimen and a 10 pixel border by image segmentation using MatLab scripts. For deep learning, images were then re-sized to a consistent, low-resolution height of 64 pixels (maintaining the original image aspect ratio and padded to 140 pixels wide). Moderate image compression improves the performance of deep learning by reducing the number of pixels that must be compared between images. The photographic dataset used for deep learning is provided in the Dryad Data Repository with filenames corresponding to joined data in Table S2: doi:10.5061/dryad.2hp1978.

### Taxonomy, locality data and specimen selection

Taxonomic and locality data were recorded from NHM *Heliconius* butterfly specimen labels (Table S2). Capture localities were then matched to approximate latitudes and longitudes (Table S2) based on NHM records. Subspecies taxonomy follows reference (*30*). The complete photographic dataset covers thirty-seven named subspecies and one labelled cross: 21 subspecies from *H. erato* and 17 from *H. melpomene*. Specimens of these subspecies were sampled exhaustively from the NHM collection. All available specimens of *H. erato* and *H. melpomene* were selected, within the limits of the data collection period, moving systematically through collection drawers. The complete photographic dataset (fig. S4, Dryad Data Repository: doi:10.5061/dryad.2hp1978) covers both specimens closely representative of subspecies descriptions (*30*) (*31*) (including available holotypes, syntypes, and paratypes, Table S2) as well as other, naturally varying, individuals. These variants include some likely hybrid specimens showing varying levels of phenotypic admixture from other subspecies (see additional taxonomic information, Table S2). Inclusion of all available specimens in machine learning covers a very broad range of the phenotypic diversity within these species, providing the deep learning network with all available information from which to learn phenotypic features correlated with subspecies identification (see deep learning methods, below). The extent to which the named subspecies (*30*) are objectively distinguishable was then explicitly tested based on classification accuracy during network testing (Deep learning methods, below). Locations for all sampled specimens in a phenotypic spatial embedding (Table S3) were calculated (Deep learning methods, below) permitting further analysis of phenotypic distances between any subset of specimens. Two sets of statistical analyses were then conducted, one set including all 1269 photographed butterfly specimens and the second set excluding potential hybrid specimens to give a reduced dataset of 815 specimens and 1630 photographs (Table S2, fig. S4).

The extent of phenotypic similarity between subspecies, including Müllerian co-mimics, was explicitly tested using statistical analyses of phenotypic distances generated by the deep learning network (see statistical analyses of phenotypic distance, below). All sampled specimens were included in the main statistical analyses. Specimen selection was therefore independent of hypotheses of mimicry (e.g. as referring only to standard phenotypes of taxonomic type specimens (*31*)). This facilitated conservative tests of the extent of mimicry among wild-caught specimens, without any potential bias from specimen exclusion on our part. Supplementary statistical analyses on the reduced dataset (with hybrids excluded) then enabled further testing, independent of any effect from a preponderance of hybrids, which may potentially be overrepresented in museum collections relative to the wild.

Numbers of butterfly individuals sampled from each subspecies (Table S7) were variable, reflecting differences in abundance within the Natural History Museum collection. The average number of sampled butterfly individuals per subspecies was 32, the maximum number was 130 (for *H. erato petiverana*) and the minimum number was 1 (for *H. melpomene penelope*). Of 1234 total specimens in the dataset, 60% were from *H. erato* and 40% were from *H. melpomene*. The average number of sampled butterfly individuals per subspecies for *H. erato* was 35, for *H. melpomene* this was 29. Twenty-seven of the thirty-eight included subspecies were in one of twelve traditionally hypothesised (*12*) mimicry complexes (Tables S1, S9).

**Deep learning**

Image classification and spatial embedding were performed using a 15 layer deep learning network (Supplementary Computer Code), which we name ButterflyNet (figure 1). This makes use of a triplet embedding loss function (*21*) (*22*) to train a network to organise its inputs (images) in a space such that proximity in that space is highly correlated with identity (in this case subspecies). The network is trained on triplets of butterfly images, with each triplet containing two images sampled from the same subspecies and one image sampled from a different subspecies. The specific meaning of proximity is given by the distance function used in the loss function (the optimisation objective). In this study, the distance function was Euclidean distance. The learned embedding was then passed through an additional small network to perform direct categorical subspecies classification. Overall, the total network optimises the sum of the triplet loss and the categorical cross entropy

$$triplet\ loss = E[||z_A - z_{A'}||_2^2 - ||z_A - z_B||_2^2] \tag{1}$$

Where $E$ is the expectation over the dataset, $\|z_A - z_{A'}\|_2$ is the Euclidean distance, $z_A$ is the spatial embedding of butterfly $A$, butterflies $A$ and $A'$ are sampled from the same subspecies and butterfly $B$ is sampled from a different subspecies

$$categorical\ cross\ entropy = E[-\log(p(y))] \tag{2}$$

Where $p(y)$ is the probability assigned by the network of a given butterfly belonging to its named subspecies $y$.

The computer code used for machine learning is provided as a Python script (Supplementary Computer Code) in ipynb format which can be viewed using a text editor and run using the Jupyter Notebook App (http://ipython.org). The script makes use of the PyTorch, Scikit-learn and Adam packages.

***Network training***

Network training used the Adam optimizer with a learning rate of $10^{-4}$. The method of training is as follows. Each batch is composed of sets of image triplets sampled from the training data. To generate a triplet, first the majority label is chosen, and a pair of images sharing that label are randomly selected. Then a third image is sampled under the constraint that its label does not match the majority label (but is otherwise distributed according to the distribution of labels in the training data). The batch composition is chosen such that each majority label is selected an equal number of times to compose the batch. Images are subjected to augmentation by a uniform random translation in the range of [-3,3] pixels on x and y. We trained for a total of 30000 batches, each composed of 99 image triplets. For the first 1000 batches, the loss function was constructed as an equally weighted sum of the classification and triplet losses, with L2 regularization applied with a coefficient $6 \times 10^{-5}$. After 1000 batches, the weighting applied to the classification loss was reduced to 0.1 relative to the triplet loss. This helps for training during the initial transient phase in which the global structure of the embedding is first emerging by providing a stronger constraint on the embedding space (that it must contain sufficient information to predict the butterfly classes). These choices were determined as part of a hyperparameter search to optimize test classification accuracy as part of the process of development of the model (covering comparisons of regularization strategies, choice of activation function between ReLU and ELU, and general architecture dimensions and training process), but that search was not exhaustive and does not include some recent innovations such as ResNet-type architectures, nor does it include more extensive data augmentation strategies. Refinements of the model based on these avenues of exploration are left as possibilities for future work, and would likely be able to increase the classification accuracy further. In terms of structural considerations, in order for the embedding space to respect global topological relationships between the butterfly subspecies, it is necessary to have some relative ambiguity in classification so that positioning of clusters with respect to each-other has measurable consequences to the classifier loss. As such, if the network architecture is improved, it may be necessary at that time to either add additional data or refine the problem to maintain a fixed level of difficulty if the utility of the embedding is to be preserved.

***Network testing***

After the network was trained on 1500 images randomly sampled from the 2468 images in the dataset, network testing was performed on the remainder (968 images). Testing presents the trained network with new images, which it has not encountered before. The network then classifies the new images by subspecies, image classifications are compared to the known subspecies identities and the overall accuracy of test classifications is reported. The accuracy of classification expected simply by chance for this dataset was 5%. Additional testing was performed using a support vector classifier (SVC) trained on the embeddings from the main network (ButterflyNet). This tested the accuracy of classification of specimens to subspecies based on their locations in the phenotypic spatial embedding. This therefore tests the extent to which the spatial embedding locations generated by ButterflyNet are predictive of subspecies identity (e.g. relative to the original image data).

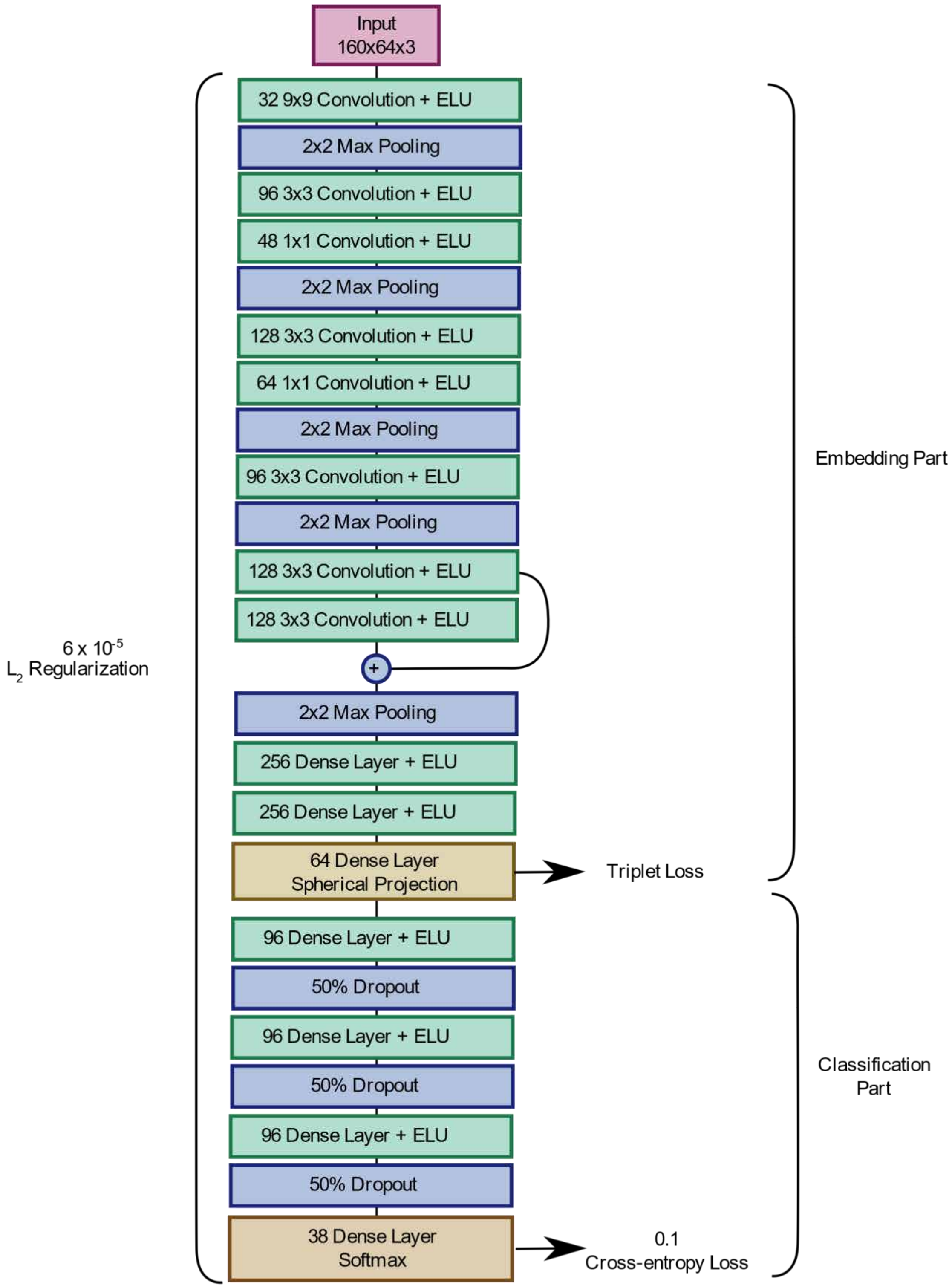

**Fig. S1. Diagram of the architecture of the deep learning network ButterflyNet used in this study.** Green boxes indicate intermediate network layers that perform matrix multiplications (image convolutions or dense, all-to-all, mappings). Blue layers perform operations on the output from previous layers (e.g. pooling or dropout operations). Brown layers indicate dense layers which produce a main output of the analysis (spatial embedding or image classification).

**Statistical analyses of phenotypic distance**

Pairwise Euclidean phenotypic distances between all images were calculated from the coordinates of all 2468 images within a spatial embedding with 64 dimensions, generated using the network (Table S3). These distances were then used to calculate the average pairwise Euclidean phenotypic distances between all subspecies (Table S4). Squared Euclidean distances were also calculated for heat-map visualisation (fig. S3, Table S5). Principal component analysis of the image coordinates was additionally used to visualise the principal component scores in the space of principal component axes 1 and 2 as well as 3 and 4 (Fig. 2). The Euclidean phenotypic distances between all subspecies (Table S4) were used in comparisons between different sets of unique unordered subspecies pairs (Fig. 3). The first comparisons (Fig. 3a) included identity pairs (diagonal of Table S4; average distances between images within each subspecies), pairs of subspecies traditionally hypothesised (*12*) to mimic each other (Table S1) and all other subspecies pairs (neither co-mimics nor identity pairs). The second comparisons (Fig. 3b) separated the two species to compare identity pairs from *H. erato* versus *H. melpomene*, other pairs (neither co-mimics nor identity pairs) where both members were of *H. erato* versus *H. melpomene*, and pairs of co-mimics in which one subspecies was of *H. erato* and the other of *H. melpomene*. Average phenotypic distances for subspecies sets and principal component scores were calculated using MatLab scripts. Nonparametric statistical analyses (robust to different sample sizes and non-normally distributed data) were conducted using the program Past 3, after Shapiro-Wilk's tests indicated that distances for some subspecies sets were non-normally distributed (using an alpha value of 0.05). These analyses included Kruskal-Wallis tests for equal medians and, where this overall test was significant, subsequent Mann-Whitney pairwise comparisons of statistical distributions between groups.

**Testing evolutionary convergence**

Relatively few examples exist of quantitatively demonstrated evolutionary convergence (*24*) (*25*). When considering categorical (discrete) traits, convergence can be defined as the repeated derivation of the same trait (e.g. phylogenetic character state) in two or more lineages (e.g. phylogenetic clades). For quantitative (continuous) traits, evolutionary convergence has been defined as an increase in similarity between two lineages (considering some specified axis or axes of variation) relative to their ancestral states (*25*). For consistency with categorical definitions, the broad phenomenon of evolutionary convergence may be considered to include some types of 'parallel' evolution (*24*) (*25*), such as parallel vectors of change to the same (or similar) trait values. Usually, ancestral states are unknown *a priori* (e.g. they must be estimated from contemporaneous taxa or non-contemporaries of uncertain ancestor-descent status). Consequently, tests of convergence in continuous traits have previously undertaken quantitative analyses of taxa that are qualitatively or functionally similar, and so potentially convergent, relative to respective sister-groups (immediate relatives) (*40*) (*25*) or to broader sets of close relatives (*24*). Where the putatively convergent taxa are quantitatively more similar to each other than are their relatives, this has been taken as support for convergence (*24*) (*25*). However, without additional information, especially on the temporal direction (polarity) of evolutionary change, it can be difficult to distinguish putative convergence from alternative patterns such as divergence by dissimilar sister taxa.

The studied case of *H. erato* and *H. melpomene* overcomes some such difficulties due to the sheer number of polymorphic mimicry types. For two compared clades with *n* cross-clade co-mimic pairs, each of which has a distinct pattern feature (or feature set), at most one of these distinct feature sets could potentially represent a shared ancestral state. All *n*-1, other co-mimic features must have been independently derived within each clade (since the minimum number of evolutionary derivations for a phylogenetic character on a given tree is the number of distinct character states minus one (*41*)). A test of evolutionary convergence can then be applied in which the operational definition of convergence is essentially that for discrete traits (independent derivation of the same state e.g. in two clades) and quantitative analysis is employed to test the relative similarity of the traits in question and the number of quantitatively distinct trait states (e.g. clusters).

**Comparative analyses of phenotypic convergence**

To further explore the extent of reciprocal convergence in mimicry between *H. erato* and *H. melpomene*, comparative analyses (*40*) were conducted using twelve selected subspecies (fig. S7). First, two sets of subspecies were identified (each set including four subspecies), with each set consisting of two pairs of interspecies co-mimics in which conspecifics are nearest neighbours in the phenotypic spatial embedding, permitting phenotypic sister-group comparisons (fig. S7a-b). The ancestral pattern types and order of pattern evolution within *H. erato* and *H. melpomene* are not known with certainty. However, focal-co-mimics, with pattern features that are potentially derived, rather than ancestral, were identified for each comparative analysis based on all available independent information from gene phylogenies (*26*) (*27*), biogeographic distribution (fig. S2) and phylogeographic reconstruction (*26*). This additional information aids assessment of the most likely polarity of pattern evolution (e.g. directed towards the focal taxa). From a cladistic perspective, this process is equivalent to assessing the most likely phenotypic states at the hypothetical ancestral node for two considered taxa. For comparison, the analyses were then repeated with reversed polarity. Two focal subspecies and their nearest conspecifics (fig. S7c) were selected based on previous discussion of the influence of *H. melpomene* on *H. erato* (e.g. *H. erato petiverana* (*23*)), which has been historically controversial (*13*) (*23*). The position in phenotypic space of each of the focal subspecies was then compared to that of their nearest conspecific (a type of sister-group comparison (*40*) (*25*)). Compared distances were the squared distance from the mean location of the focal co-mimic, summed across all 64 spatial embedding axes, calculated using a Python script (Supplementary Computer Code). Expressing the locations in phenotypic space in terms of distance from two focal taxa (fig. 5 g-j) enables two-dimensional visualisation of the distances among compared taxa across any number of phenotypic axes. This also facilitates tests of mutual convergence in which convergence is characterised by decreasing distance between one focal taxon and another, relative to a conspecific, and divergence is conversely characterised by increasing distance. In each comparative analysis, Mann Whitney tests for equal medians tested whether conspecifics differed significantly in their distance from the focal co-mimic of the other species (after Shapiro Wilk's tests indicated that some subspecies values were non-normally distributed).

**Phylogenetic analyses**

Neighbour joining trees for subspecies were constructed based on phenotypic distances (e.g. Table S4) using MatLab scripts. In order to visualise the phylogenetic agreement versus conflict among different axes of the phenotypic spatial embedding (Table S3), neighbour joining trees were constructed based on repeated sub-sampling of the axes (sampling either all 64 axes with 1 replicate, or subsamples of 8 or 32 axes with 100 replicates). Consensus networks were constructed to visualise all splits (taxon partitions) implied among sets of trees using the program SplitsTree 4. To test the phylogenetic informativeness of the phenotypic distances against independent data sources, sets of neighbour joining phenotypic trees (of either all subspecies, *H. erato* only, or *H. melpomene* only) were compared against phylogenies reconstructed from published gene sequences (*27*) as well as random tree topologies. The subspecies coverage and individual samples sizes of our analysis exceed those typically used in current gene sequencing studies. However, published phylogenies (*26*) based on multi-locus gene sequences (*27*) were available that included 25 of the 38 studied subspecies (13 *H. erato*, 12 *H. melpomene*), from gene loci (sampled from a different, smaller set of 127 butterfly individuals) which were either associated with *Heliconiu*s wing colour pattern (*27*) (*optix*, *bves*, *kinesin*, *GPCR*, *VanGogh*) or were neutral markers (*mt COI-COII*, *SUMO*, *Suz12*, *2654* and *CAT*). For each gene set (pattern versus neutral loci), 100 trees were sampled from the output of previously published Bayesian phylogenetic analyses (sampling the MCMC chain after burn-in) (*26*). One thousand equiprobable, random tree topologies were generated for each taxon set using the program Mesquite. Pairwise distances between trees from the different sets were calculated using the Robinson Foulds (symmetric distance) metric in the program PAUP. Robinson Foulds distances across different tree sets were statistically compared using nonparametric Mann-Whitney tests in the program PAST (after Shapiro-Wilk's tests indicated non-normal distributions). Tree-space visualisations were produced, based on the Robinson Foulds distance, using the TSV package in Mesquite.

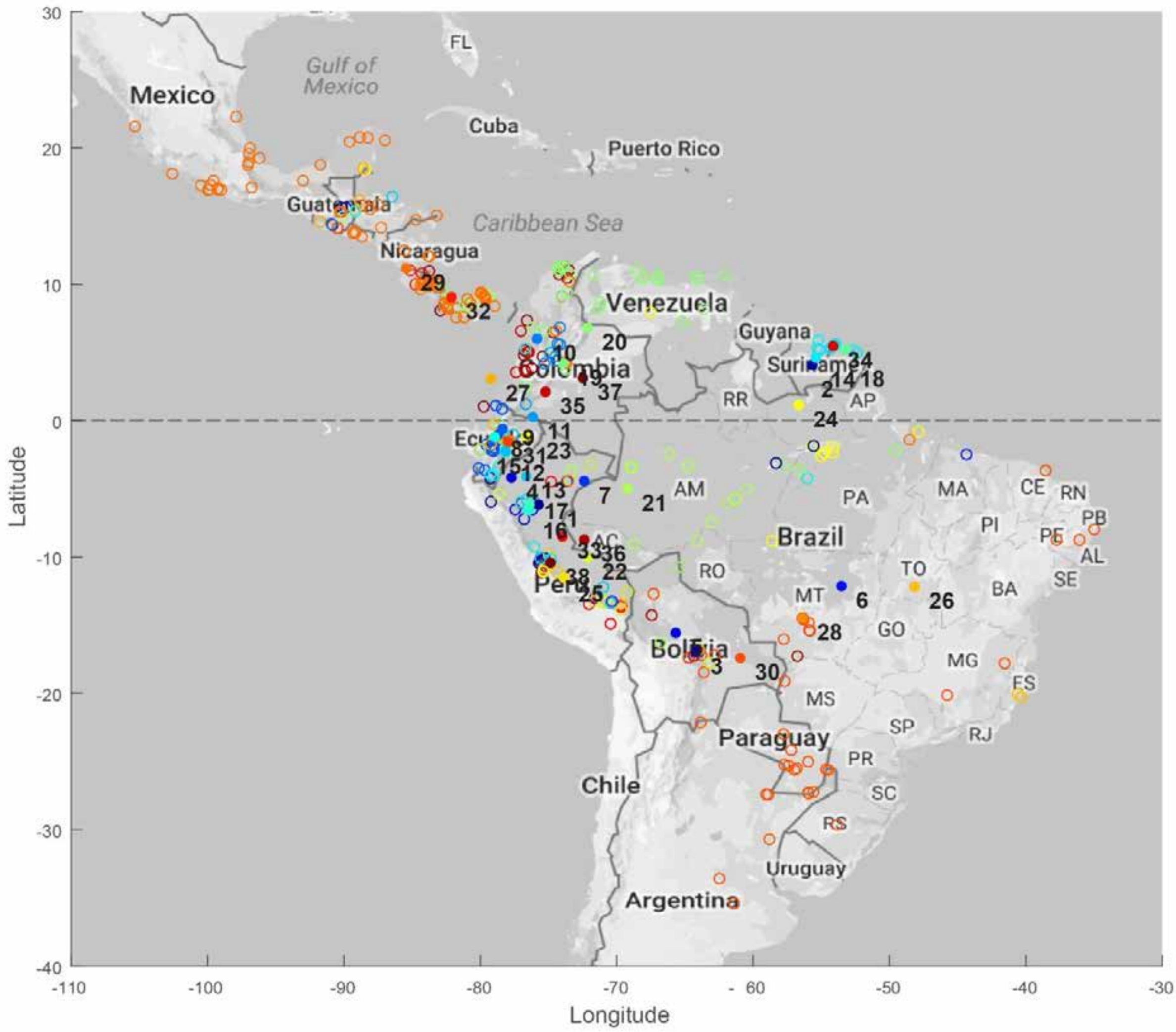

**Fig. S2. Geographic localities for sampled butterfly specimens from the polymorphic mimicry complex of *H. erato* and *H. melpomene*.** Open circles indicate approximate capture localities for historical butterfly specimens held in the Natural History Museum (NHM), London (Table S2). Based on specimen labels, capture localities were identifiable for 94% of 1234 butterflies in the dataset. Filled circles show the mean location for each subspecies. Numbers and circle colours indicate the subspecies number (Table S7).

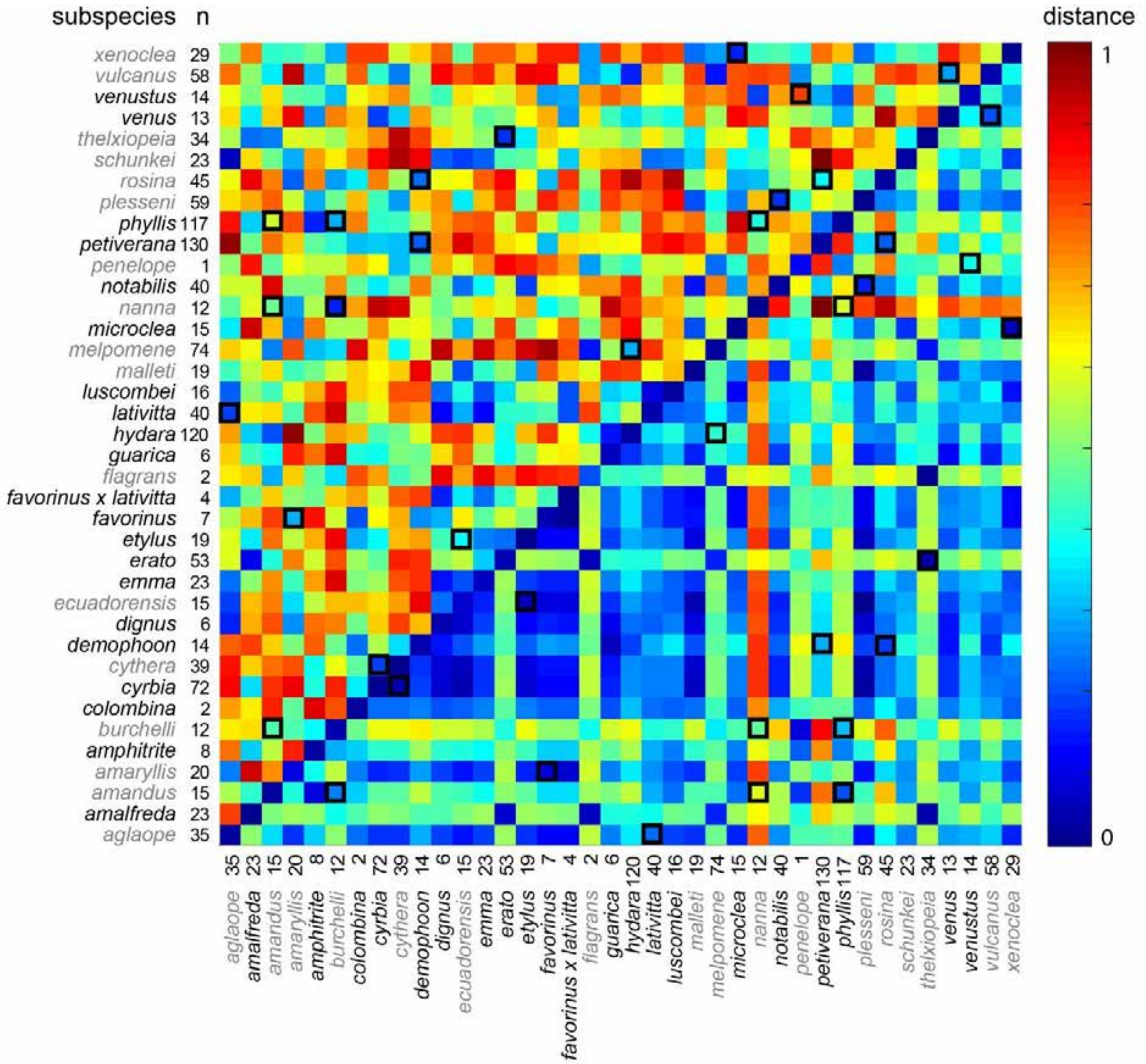

**Fig. S3. Heatmap showing mean pairwise phenotypic and geographic distances between 38 subspecies of *H. erato* (black labels) and *H. melpomene* (gray labels).** Upper matrix (including diagonal) shows mean pairwise distances for subspecies calculated from the 64-dimensional phenotypic embedding generated using a deep convolutional triplet network across 2468 butterfly images. Lower matrix shows mean pairwise geographic distance between butterfly specimens. Key shows correspondence between heat-map colours and distance, rescaled to vary between zero and one from original values (upper, squared Euclidean phenotypic distance and lower, Euclidean geographic distance, Tables S5-S6), from blue (most similar) to red (least similar). Black borders on squares indicate traditionally hypothesised co-mimics (*12*). Numbers adjacent to subspecies names show the number of butterfly specimens in the image database.

1 *Heliconius melpomene aglaope*

(A) Specimens identified as valid subspecies or accepted synonym

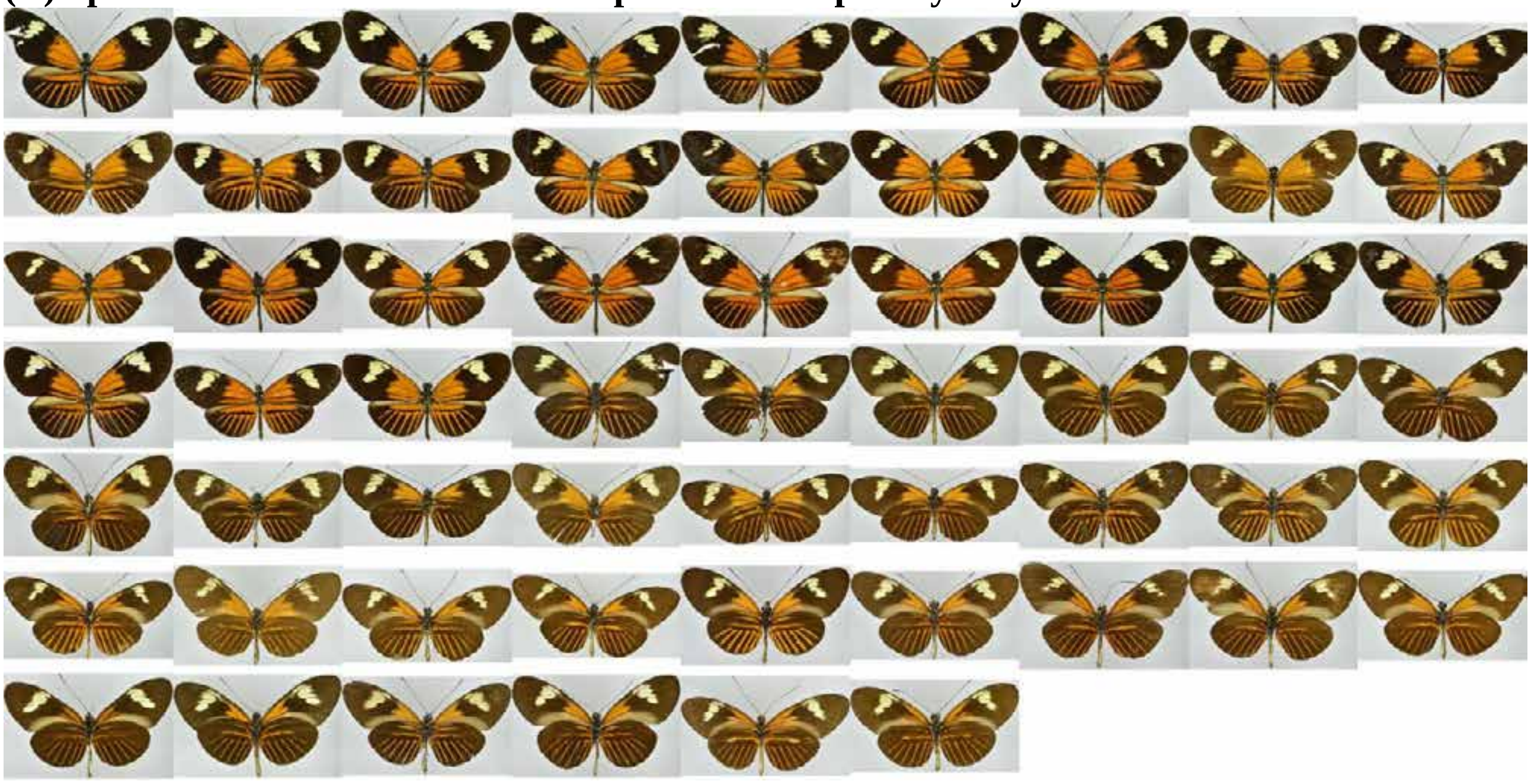

(B) Specimens identified as hybrids

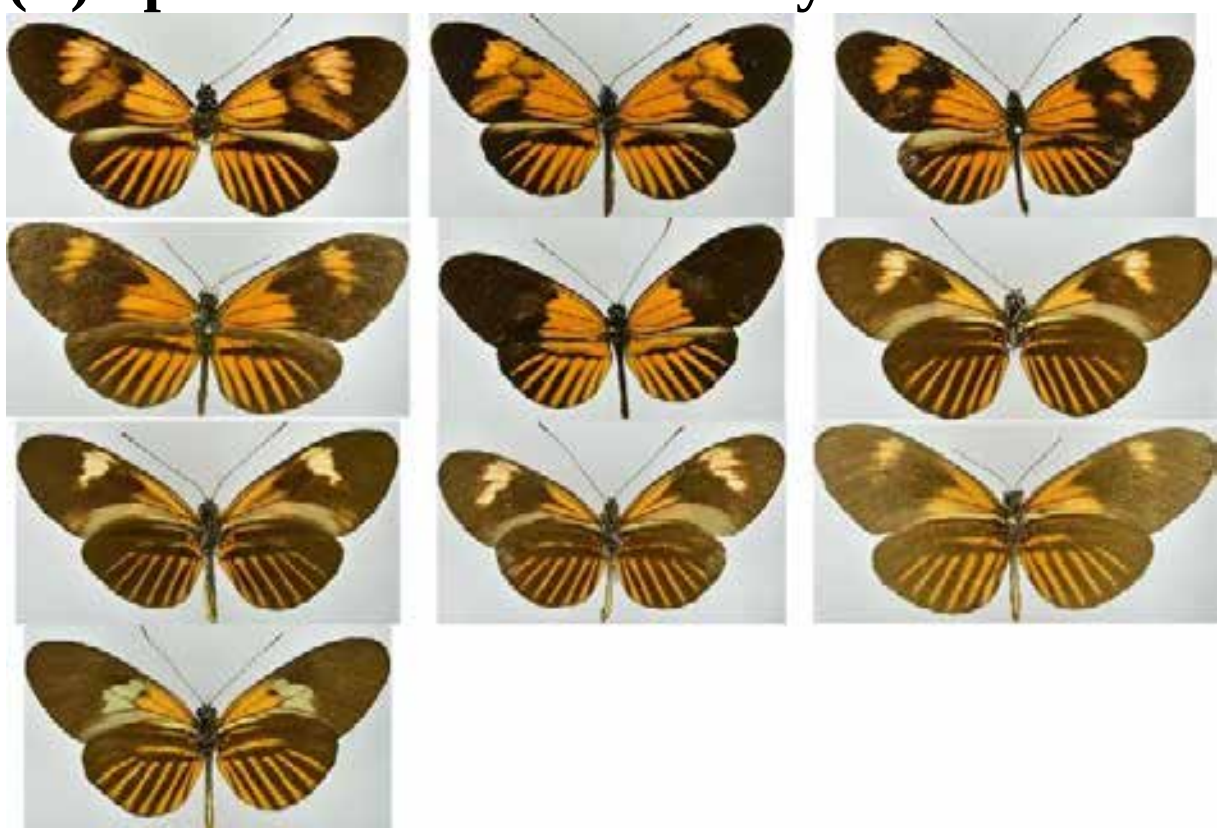

**2** ***Heliconius erato amalfreda***

**(B) Specimens identified as hybrids**

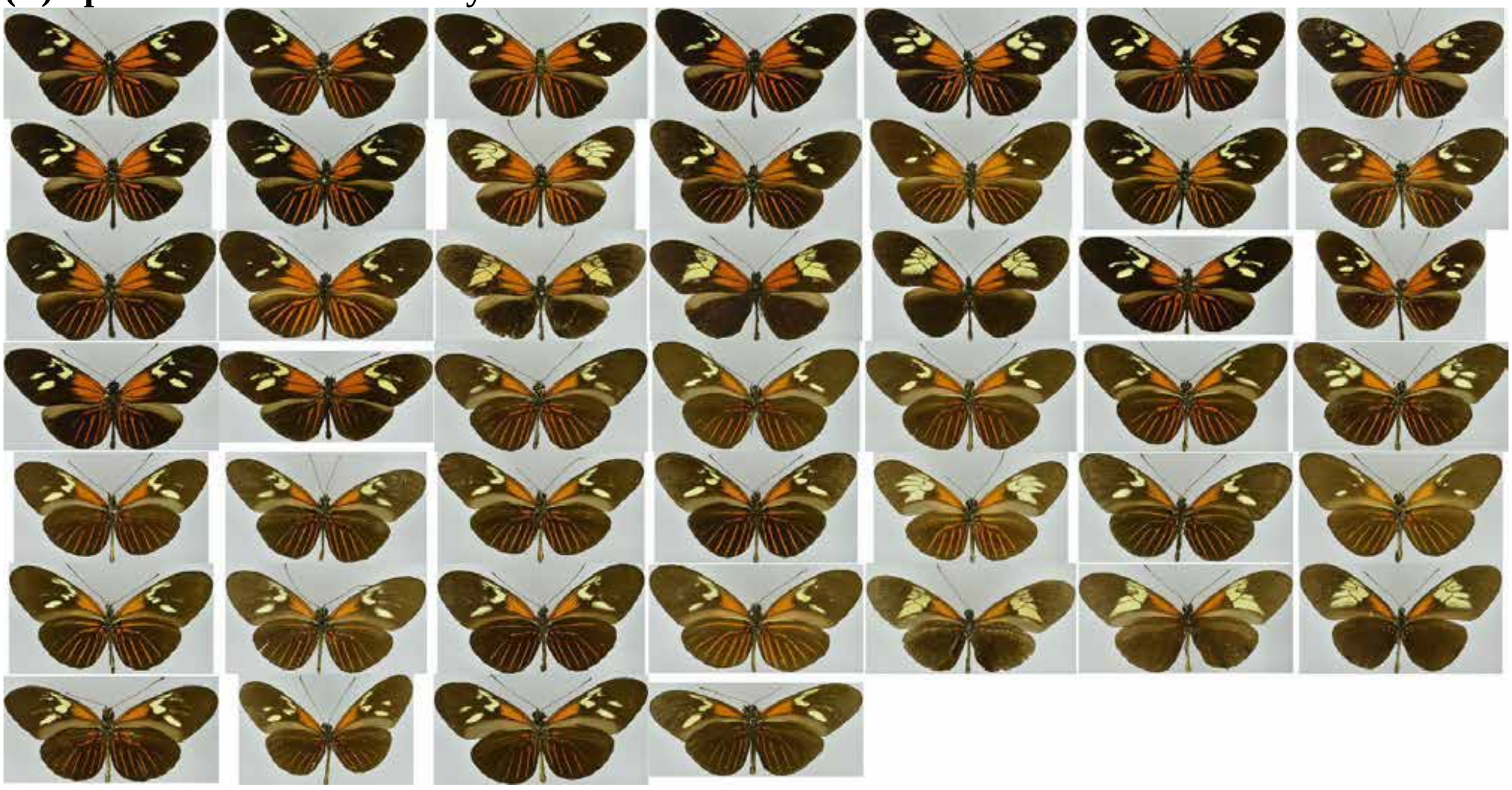

**3** ***Heliconius melpomene amandus***

**(B) Specimens identified as hybrids**

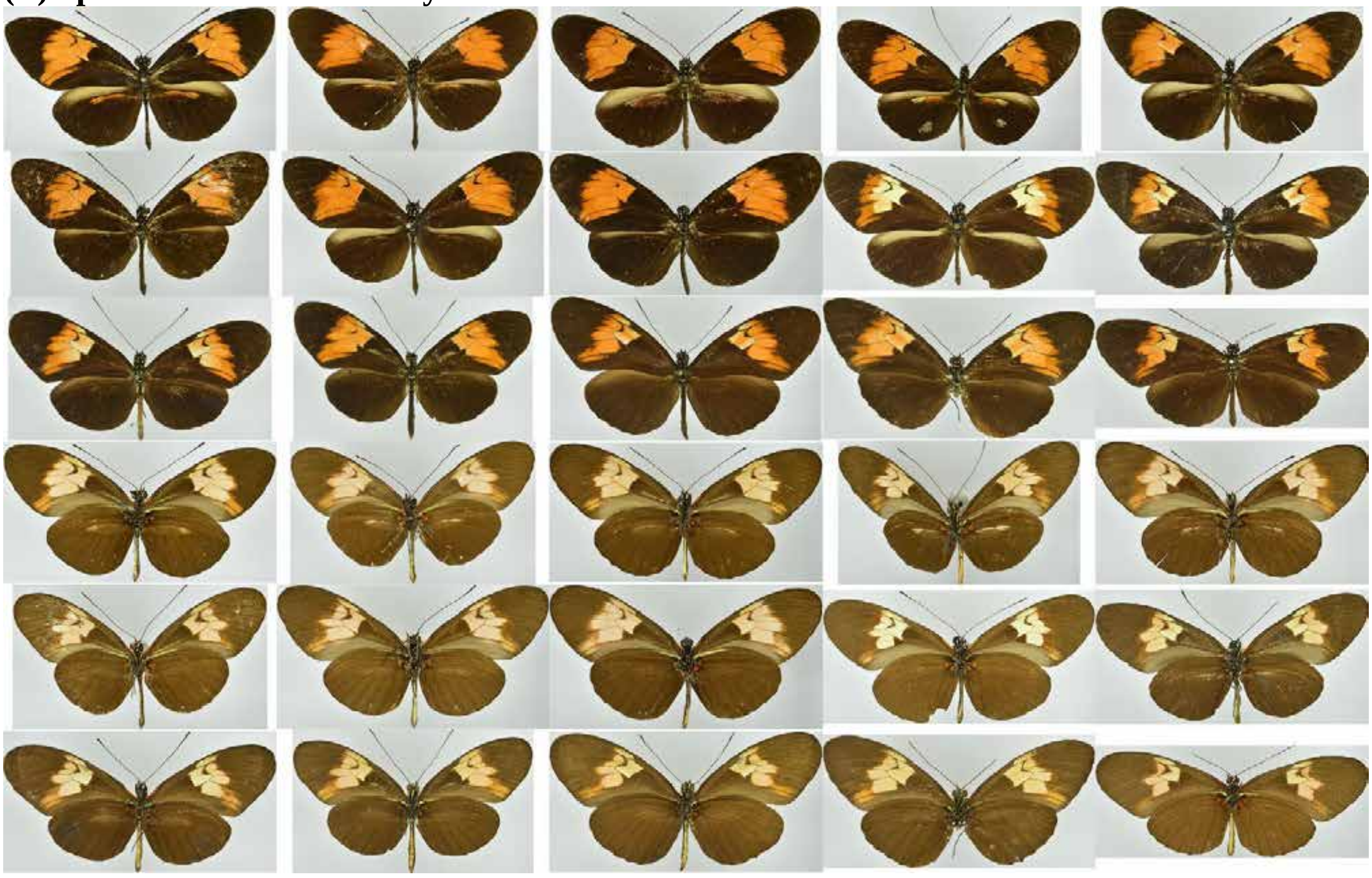

**4 *Heliconius melpomene amaryllis***

**(A) Specimens identified as valid subspecies or accepted synonym**

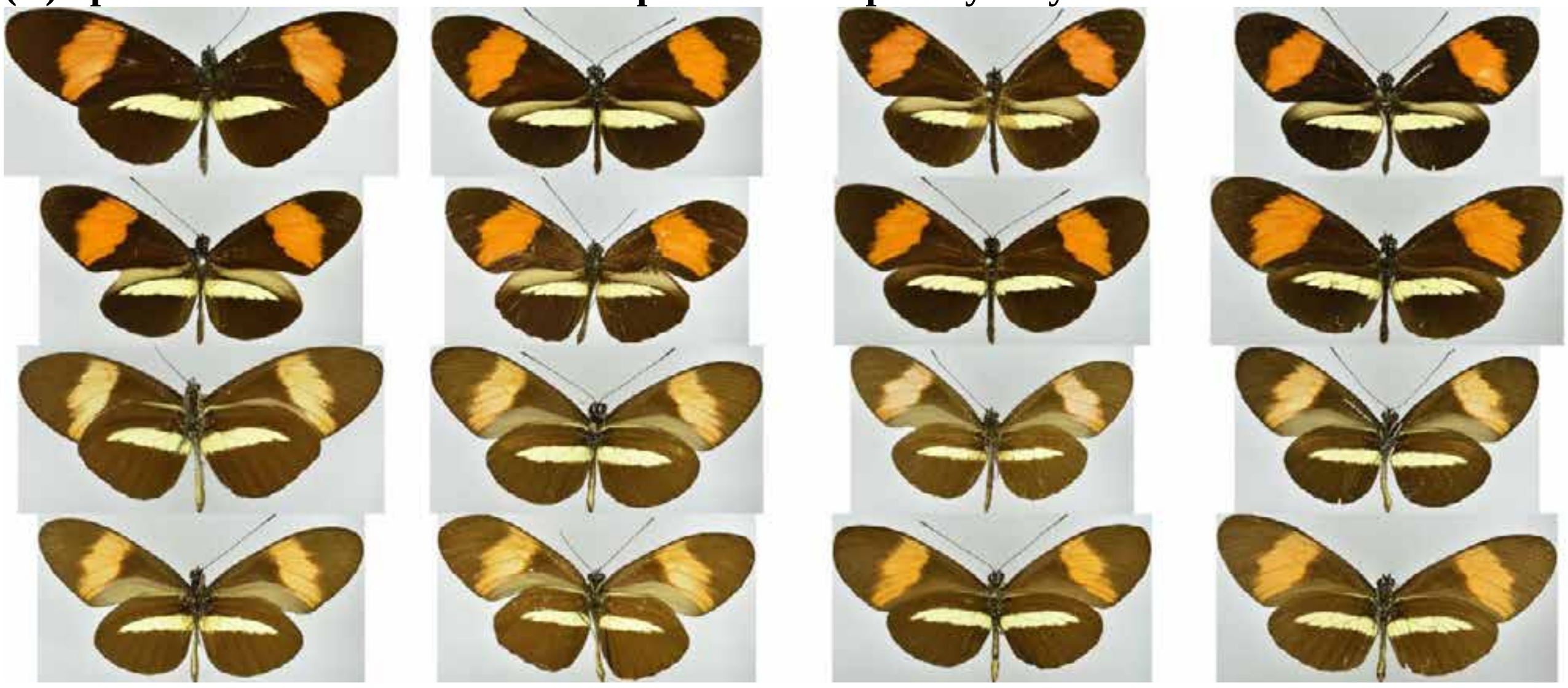

**(B) Specimens identified as hybrids**

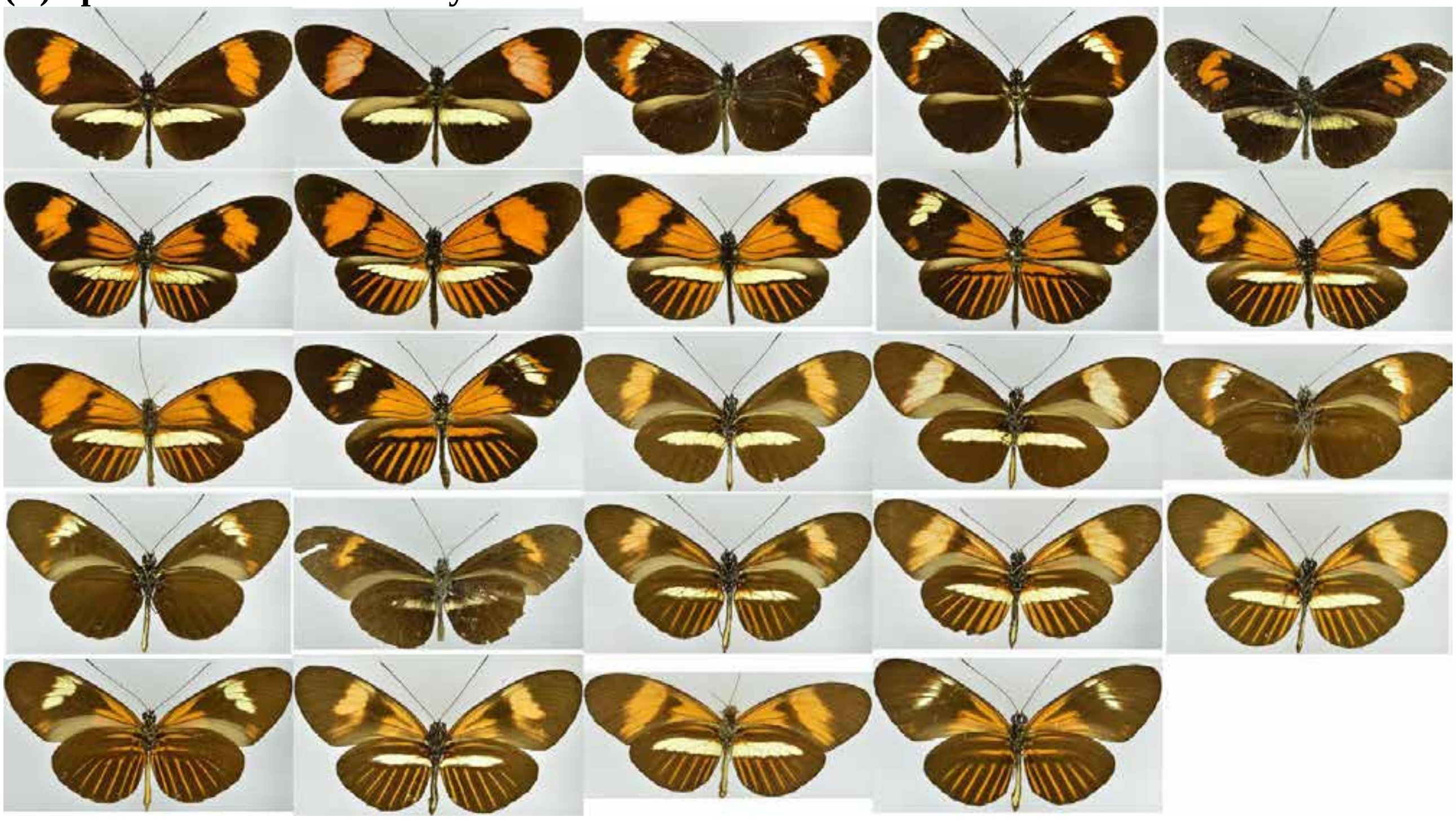

**5 *Heliconius erato amphitrite***
**(B) Specimens identified as hybrids**

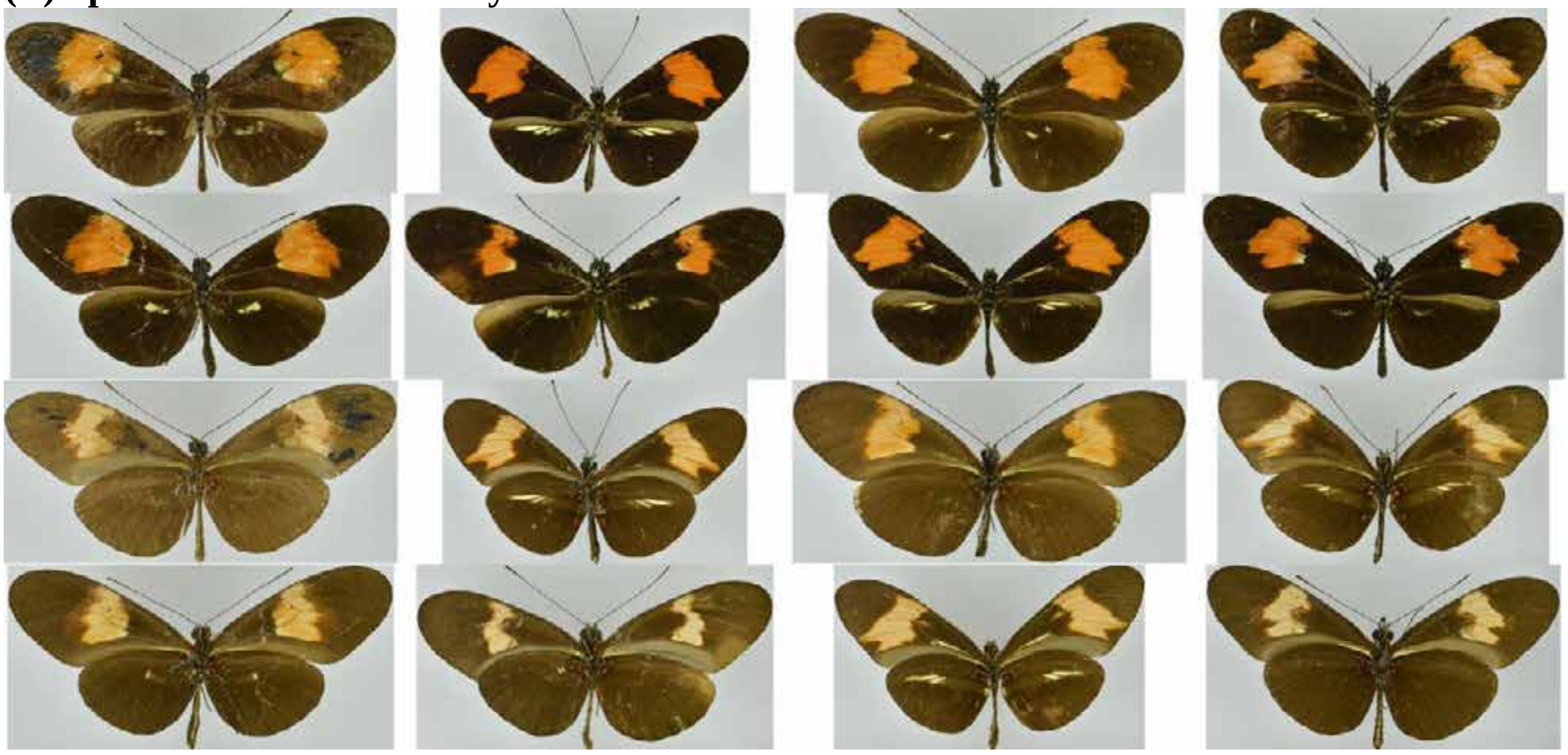

**6 *Heliconius melpomene burchelli***

**(A) Specimens identified as valid subspecies or accepted synonym**

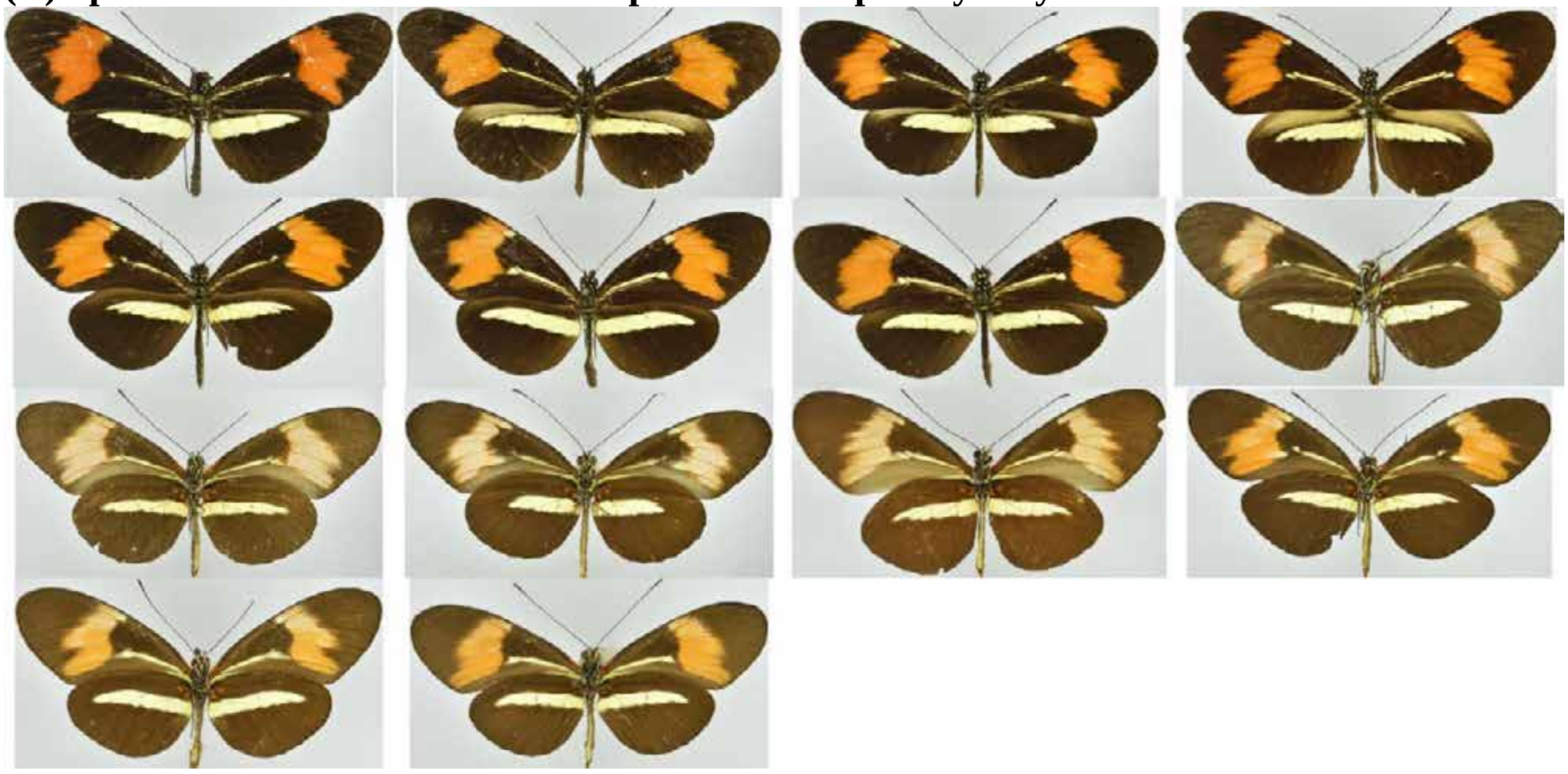

**(B) Specimens identified as hybrids**

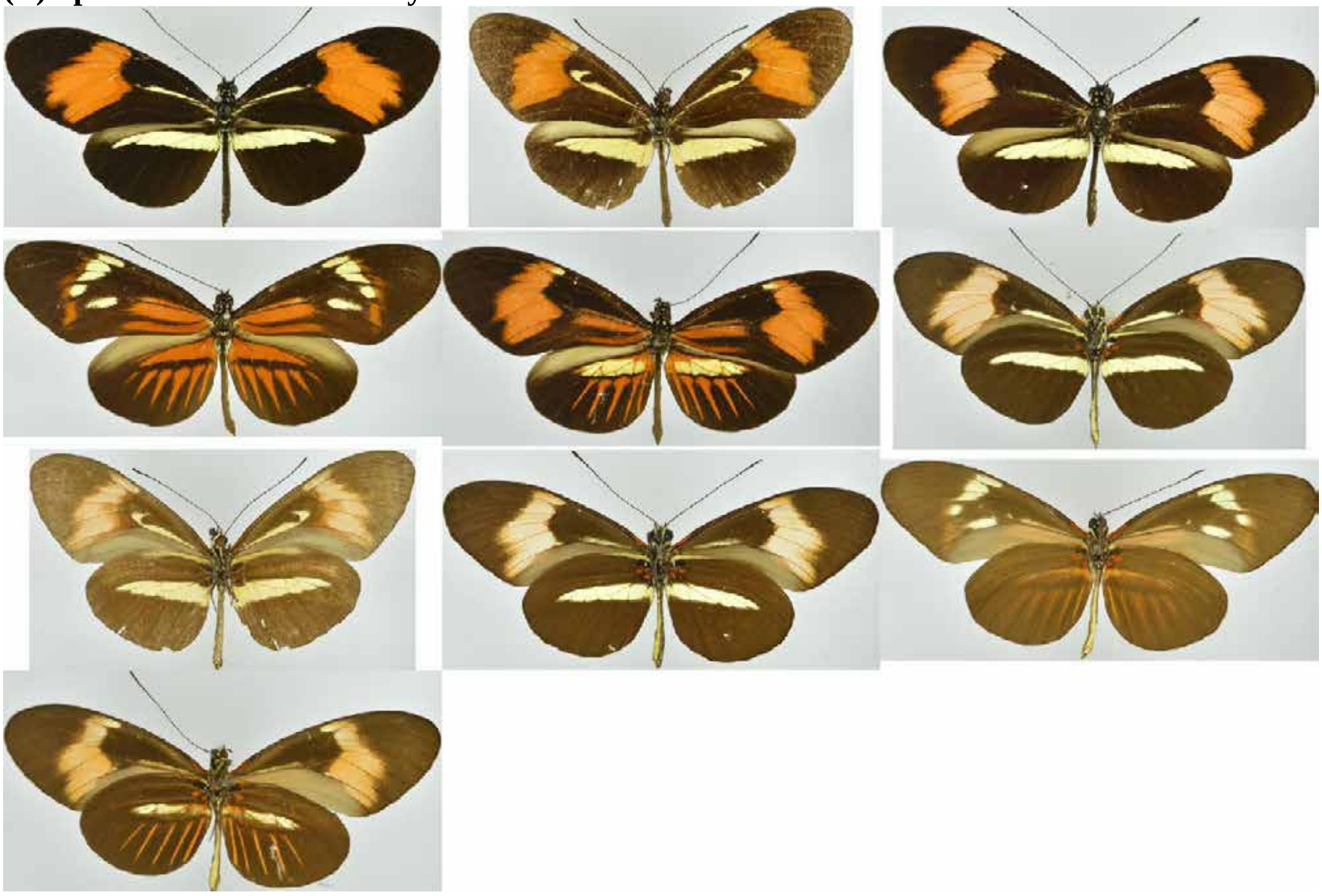

**7 *Heliconius erato colombina***

**(A) Specimens identified as valid subspecies or accepted synonym**

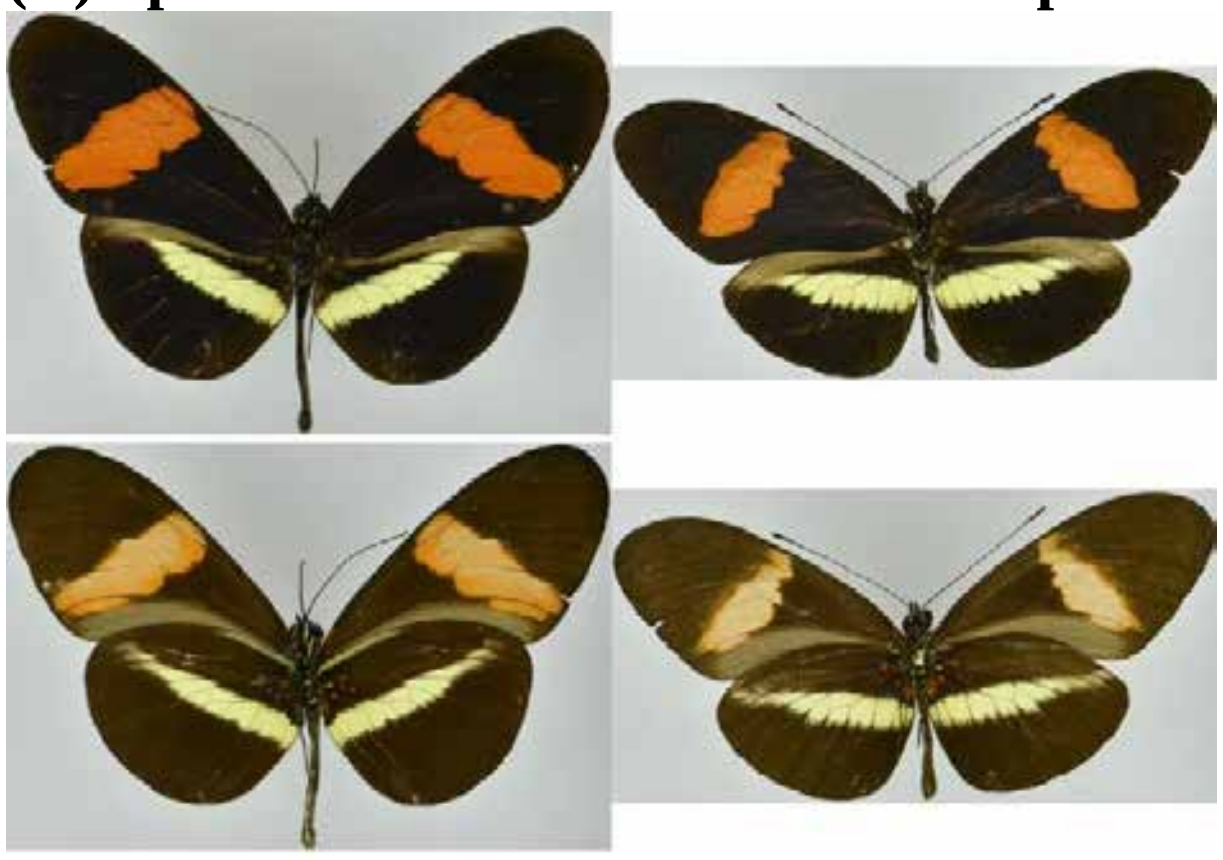

**8 *Heliconius erato cyrbia***

**(A) Specimens identified as valid subspecies or accepted synonym**

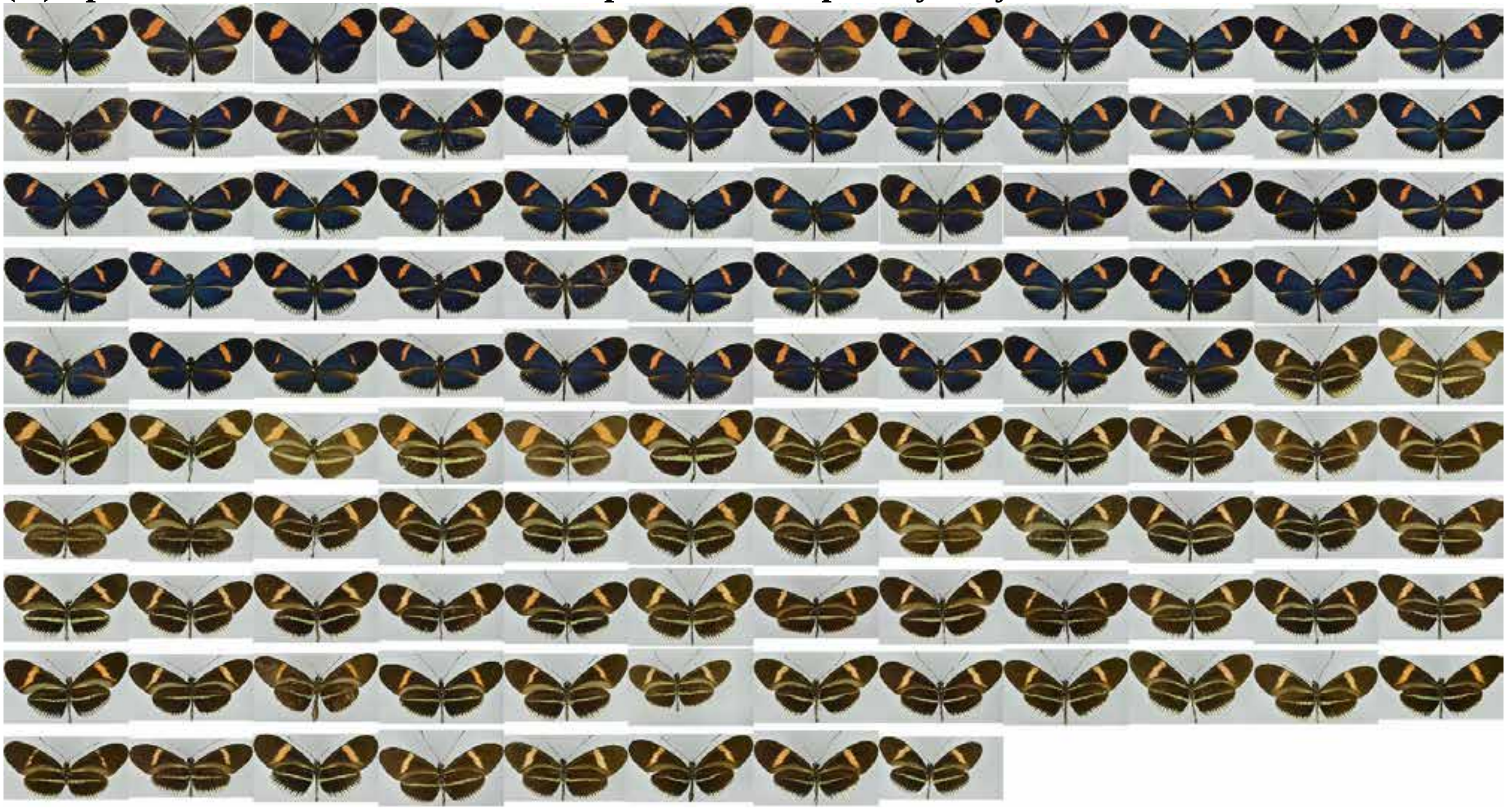

**(B) Specimens identified as hybrids**

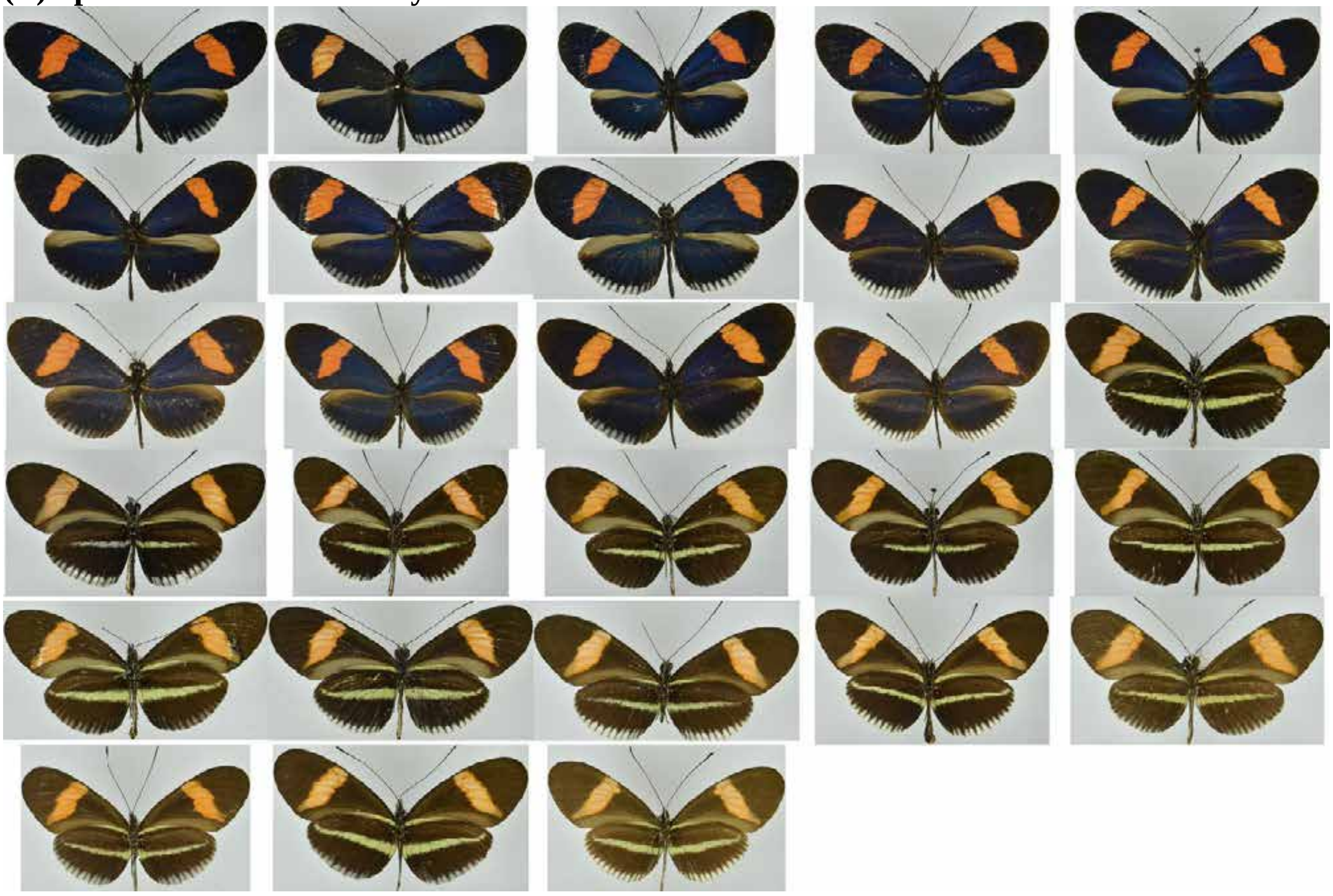

**9 *Heliconius melpomene cythera***

**(A) Specimens identified as valid subspecies or accepted synonym**

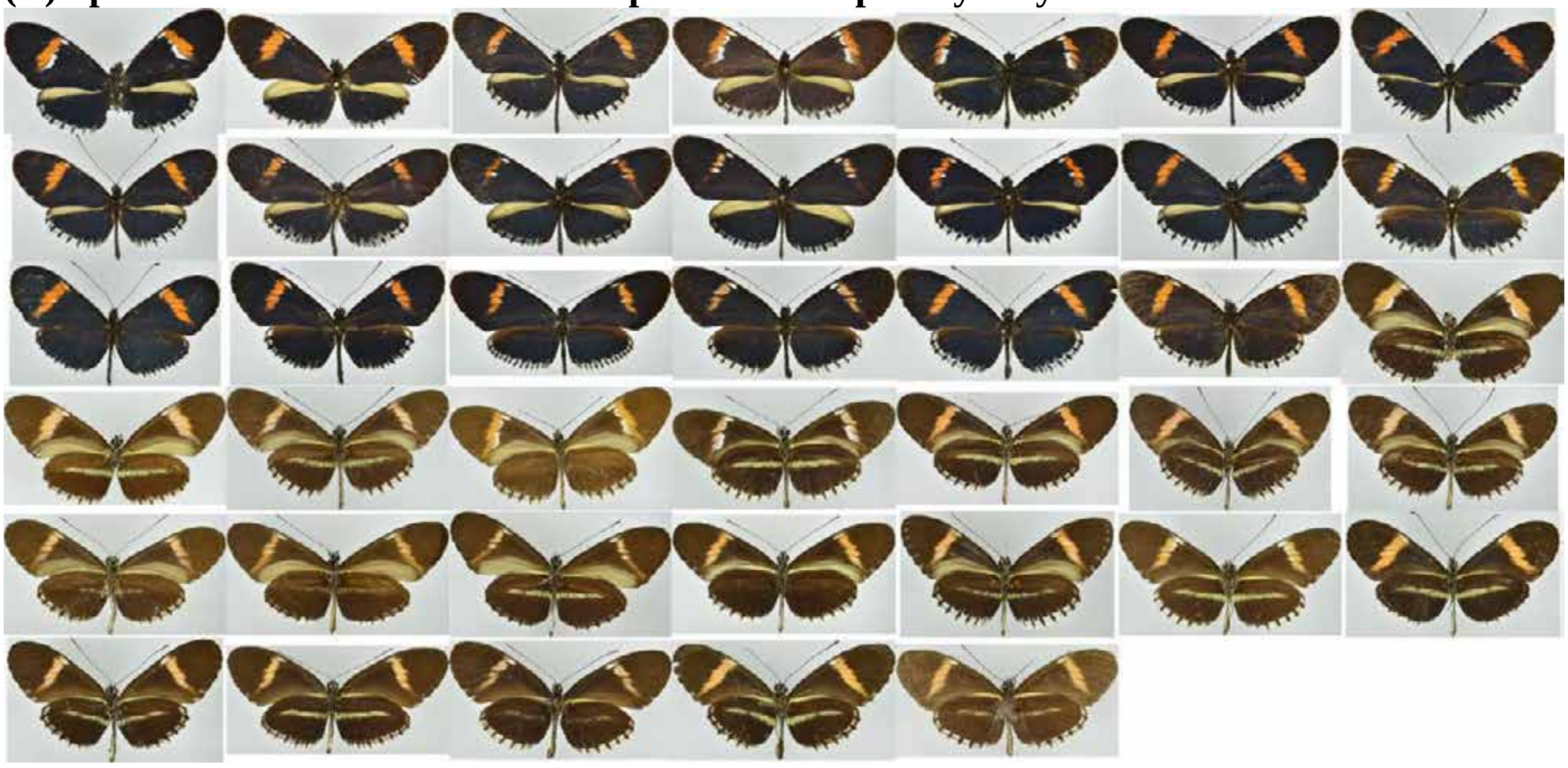

**(B) Specimens identified as hybrids**

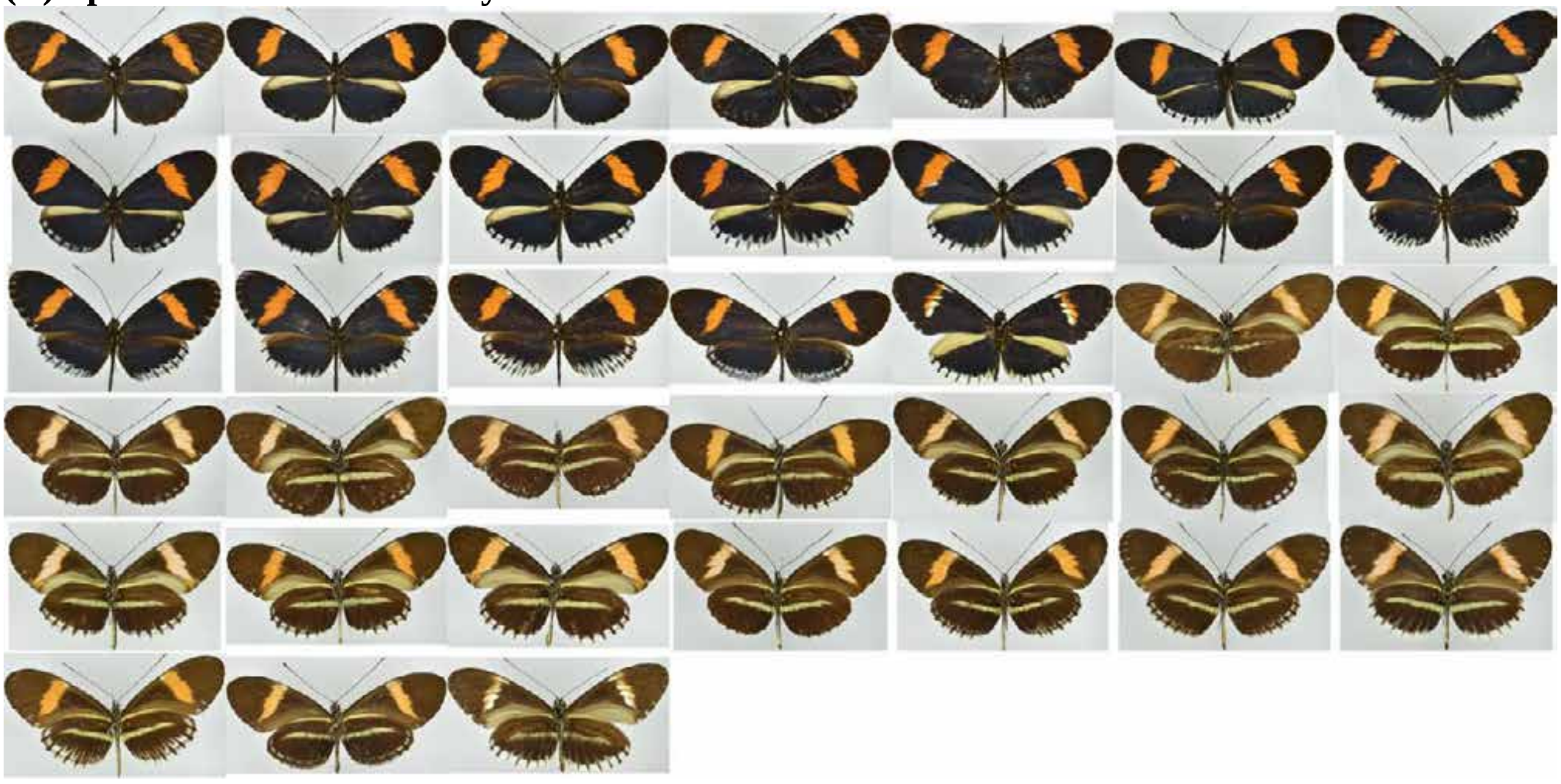

**10 *Heliconius erato demophoon***

**(A) Specimens identified as valid subspecies or accepted synonym**

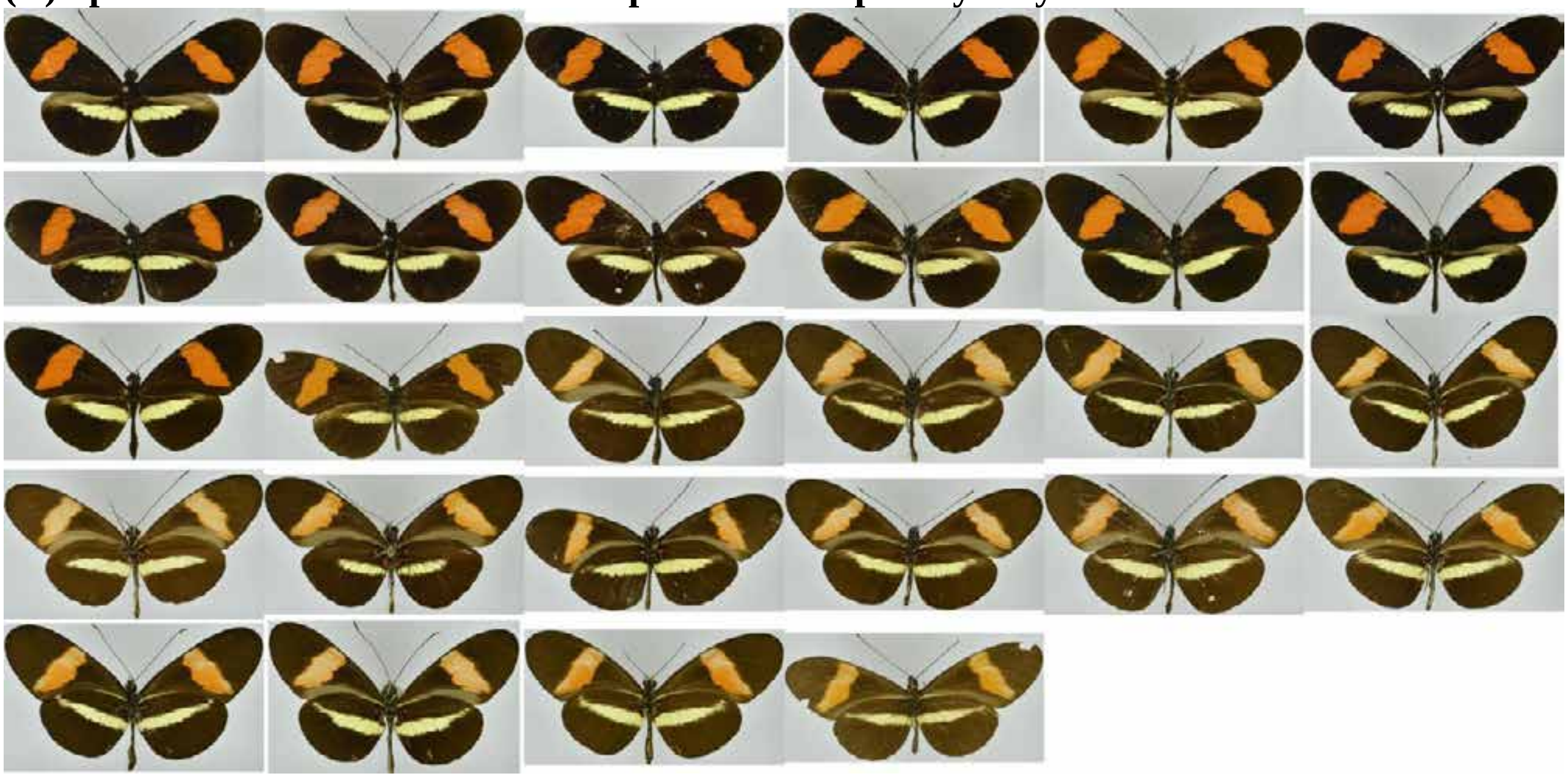

**11 *Heliconius erato dignus***

**(A) Specimens identified as valid subspecies or accepted synonym**

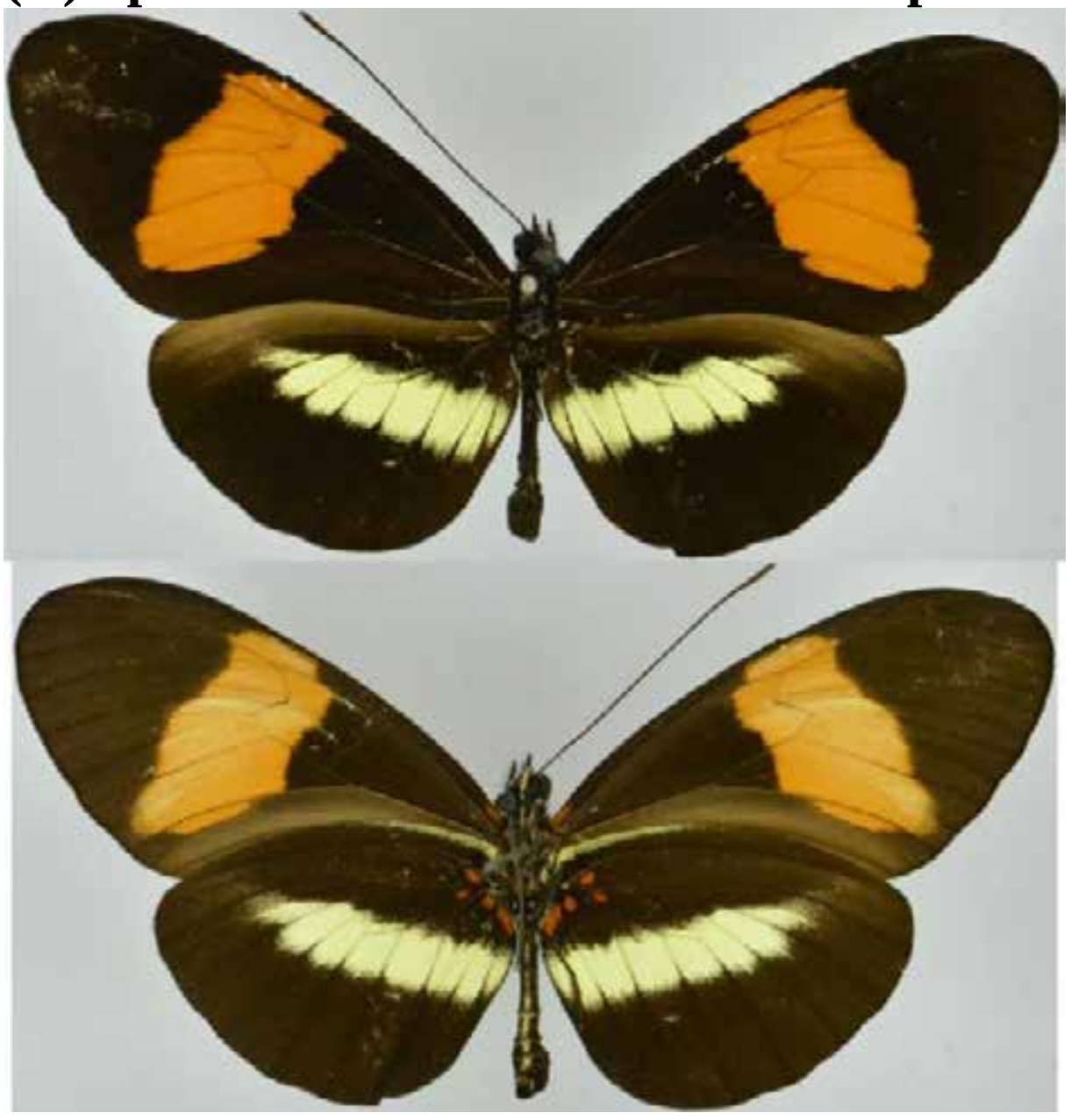

**(B) Specimens identified as hybrids**

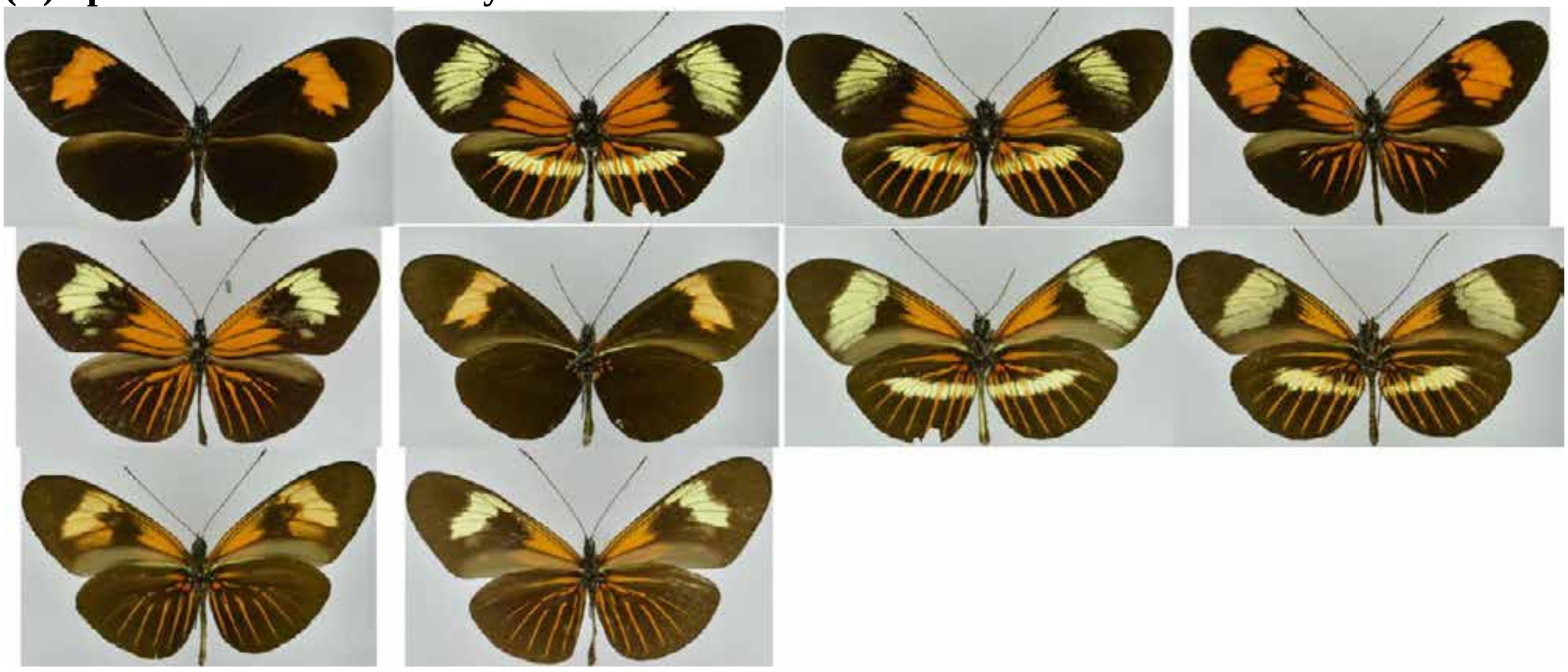

12 *Heliconius melpomene ecuadorensis*

**(A) Specimens identified as valid subspecies or accepted synonym**

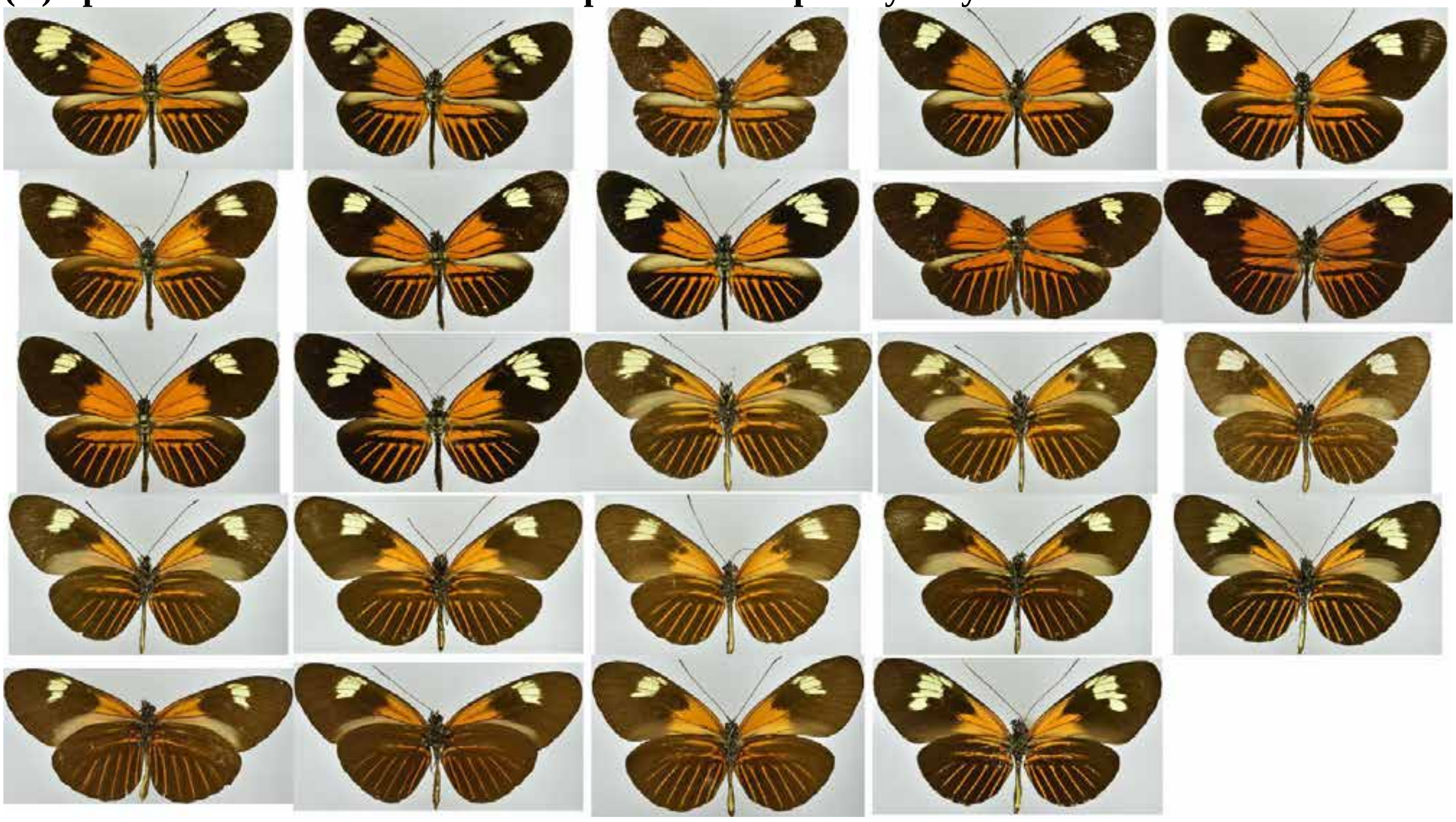

**(B) Specimens identified as hybrids**

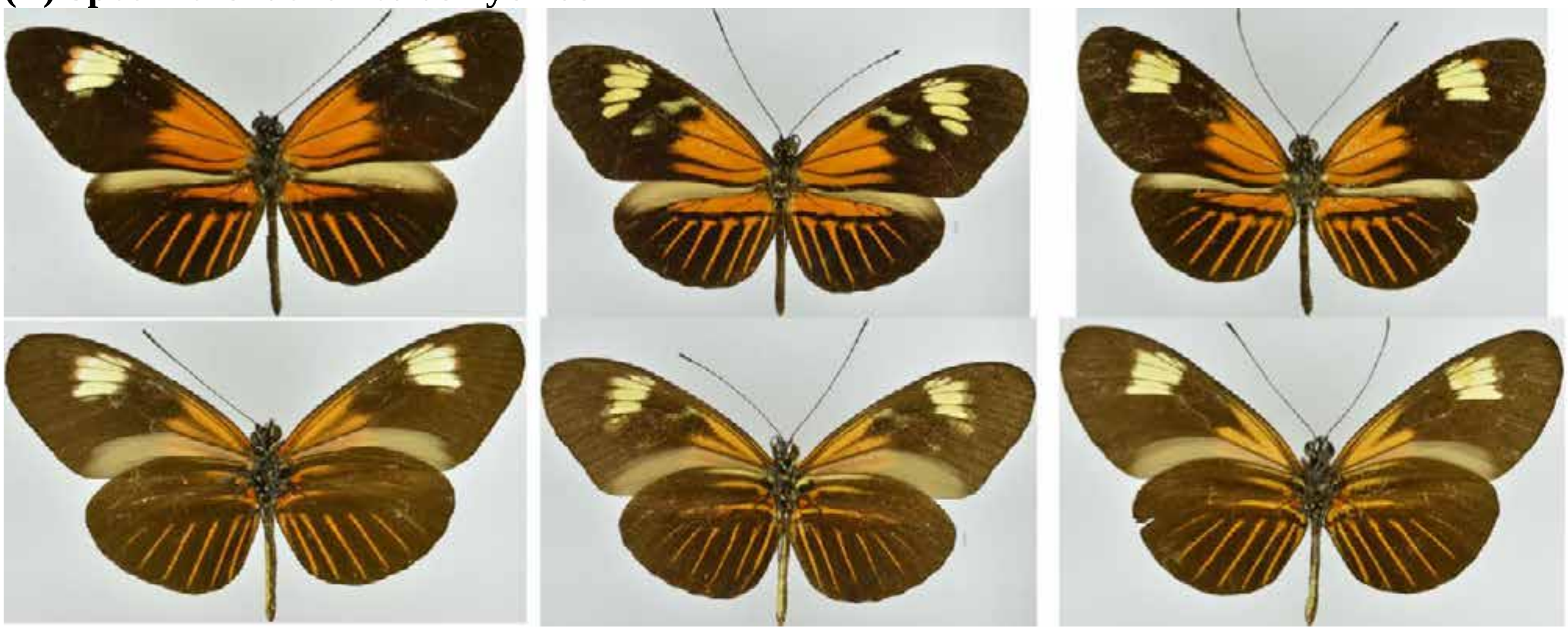

**13** ***Heliconius erato emma***

**(A) Specimens identified as valid subspecies or accepted synonym**

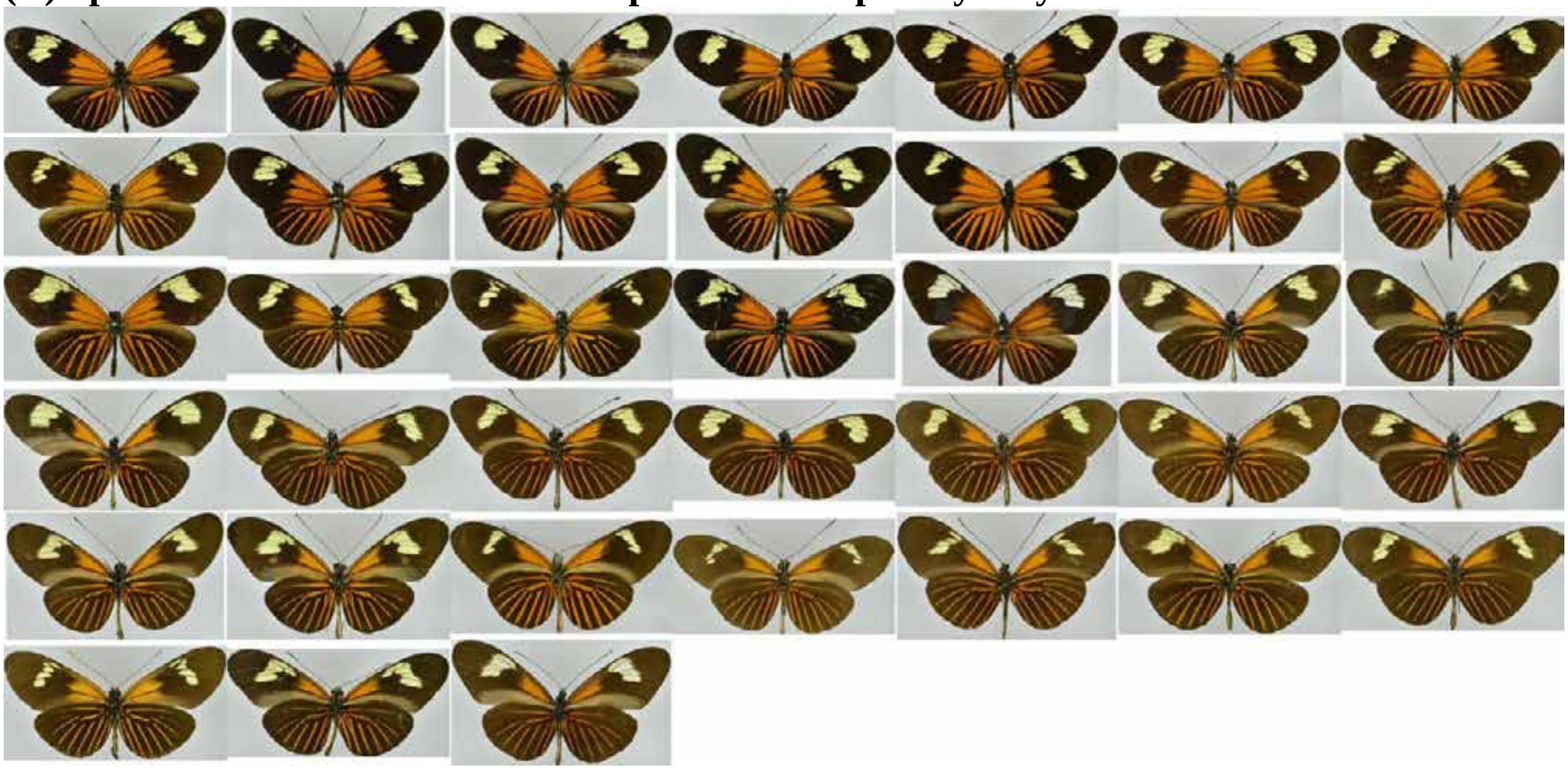

**(B) Specimens identified as hybrids**

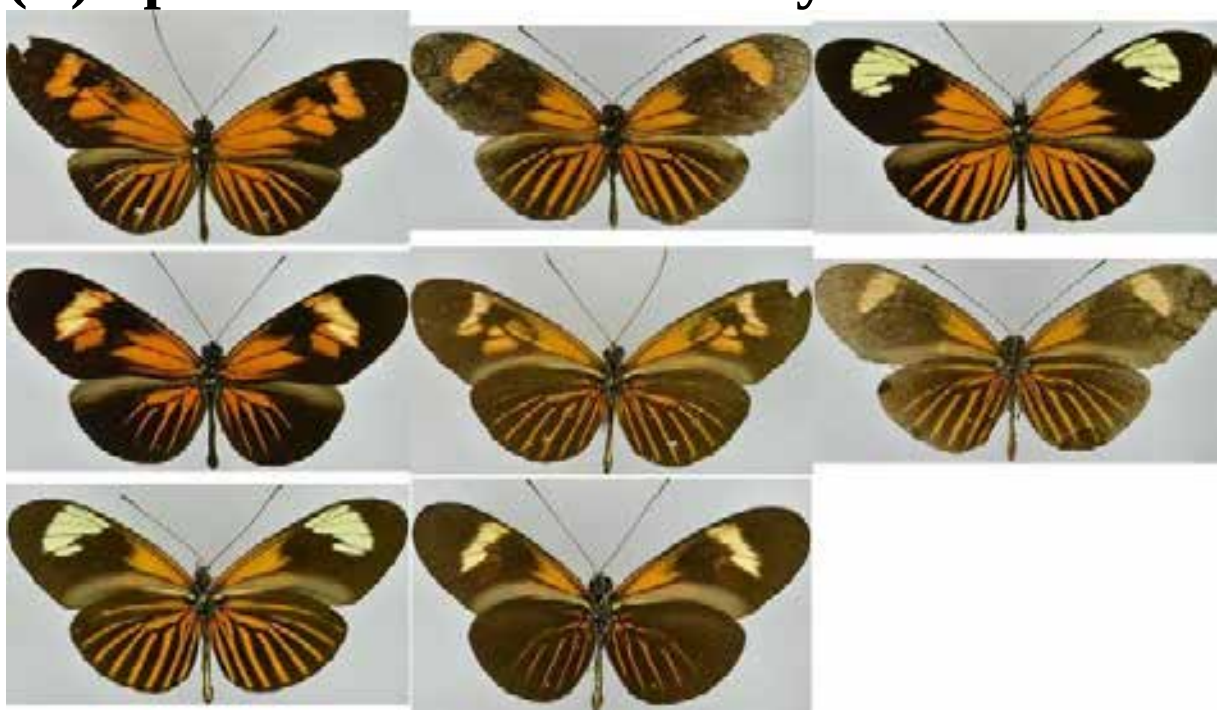

14 *Heliconius erato erato*

(A) Specimens identified as valid subspecies or accepted synonym

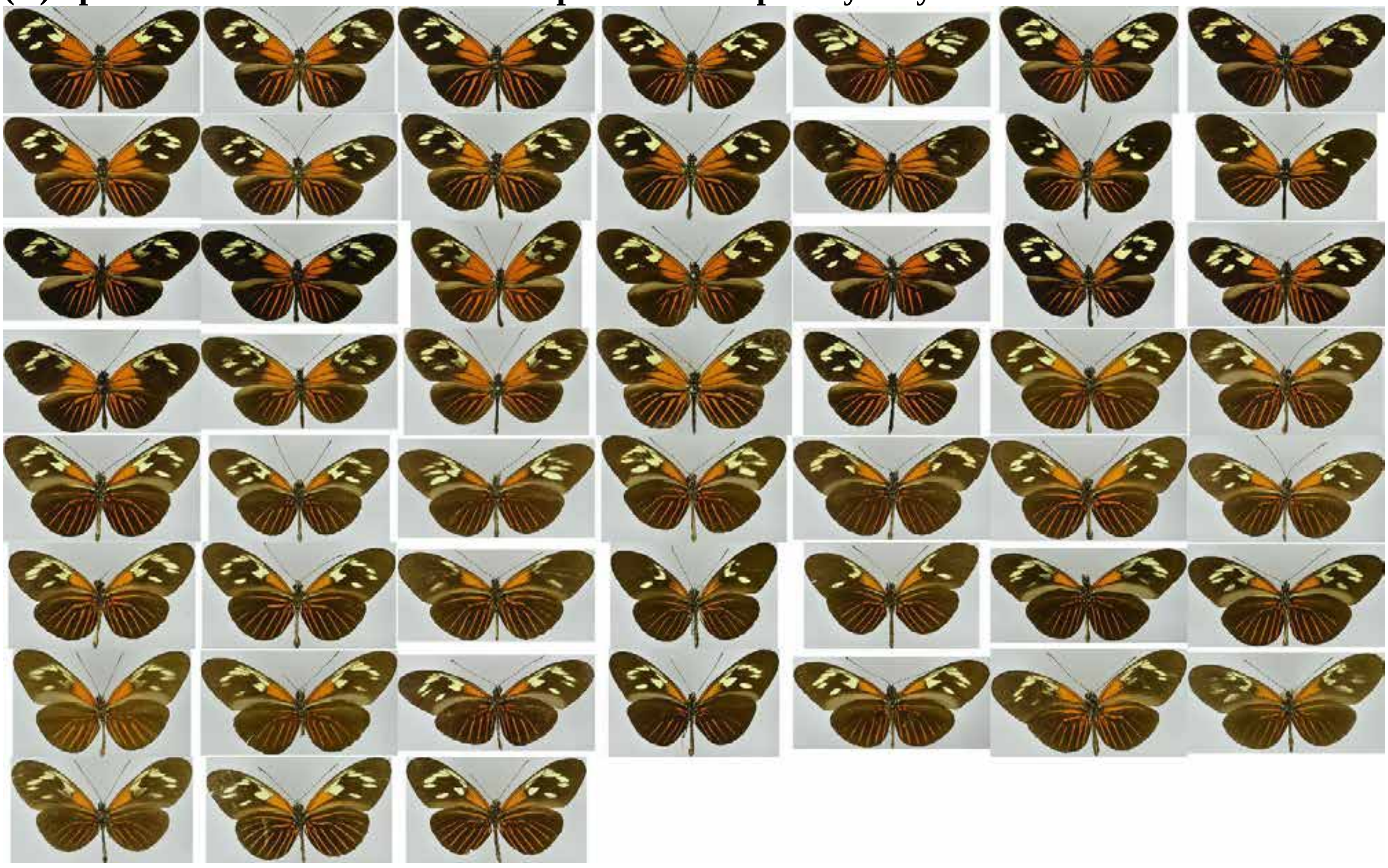

(B) Specimens identified as hybrids

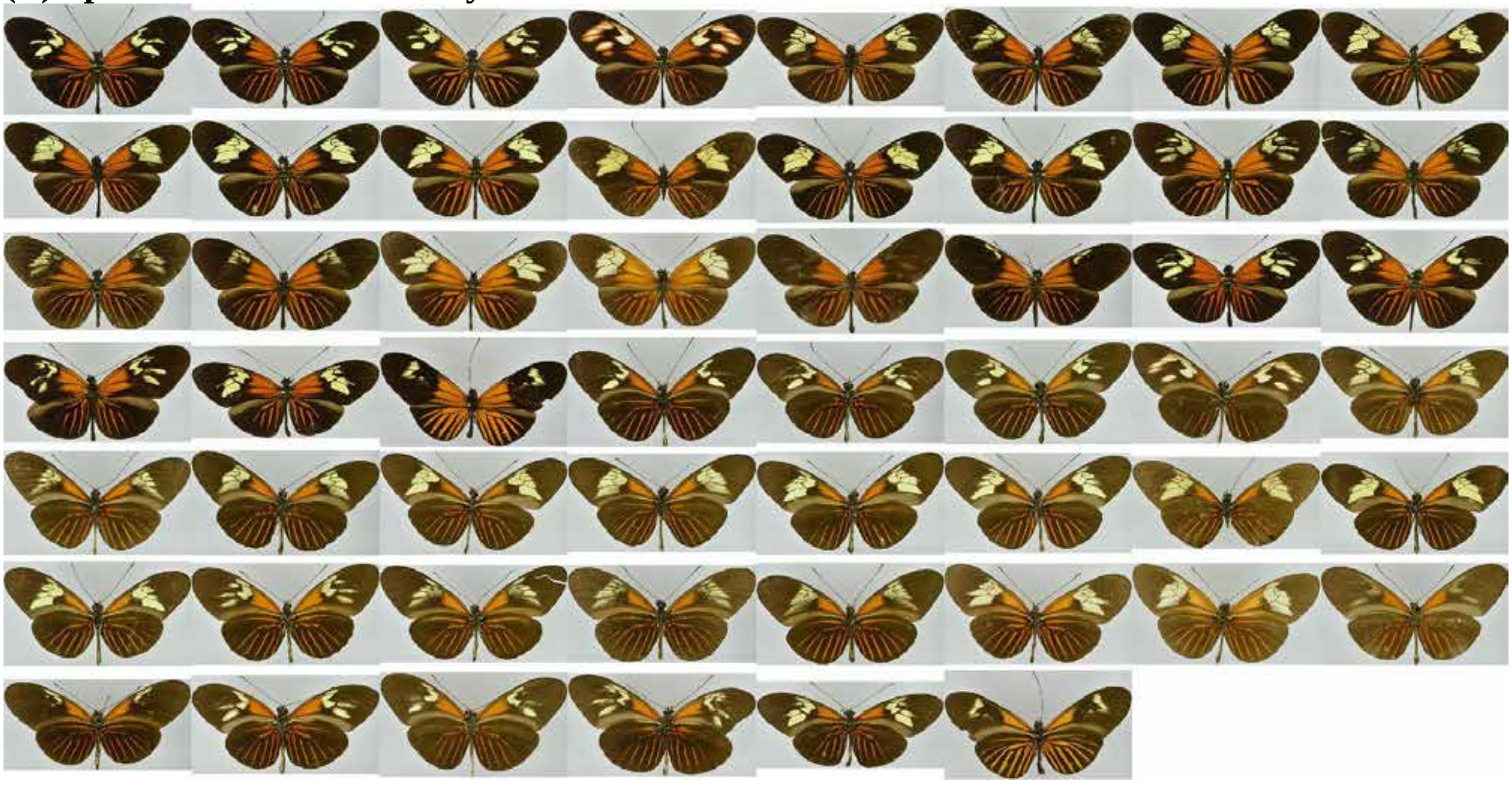

15 *Heliconius erato etylus*

(A) Specimens identified as valid subspecies or accepted synonym

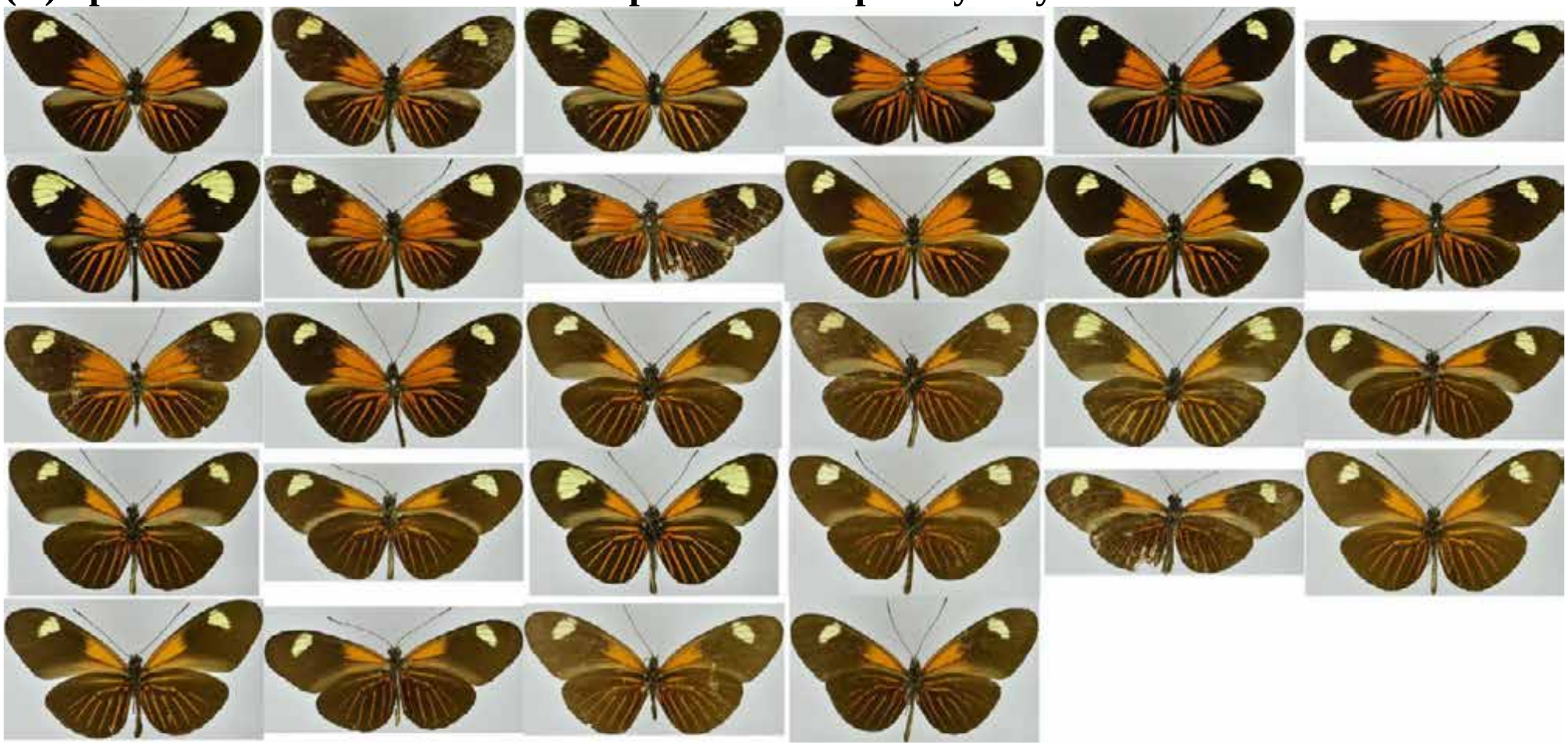

(B) Specimens identified as hybrids

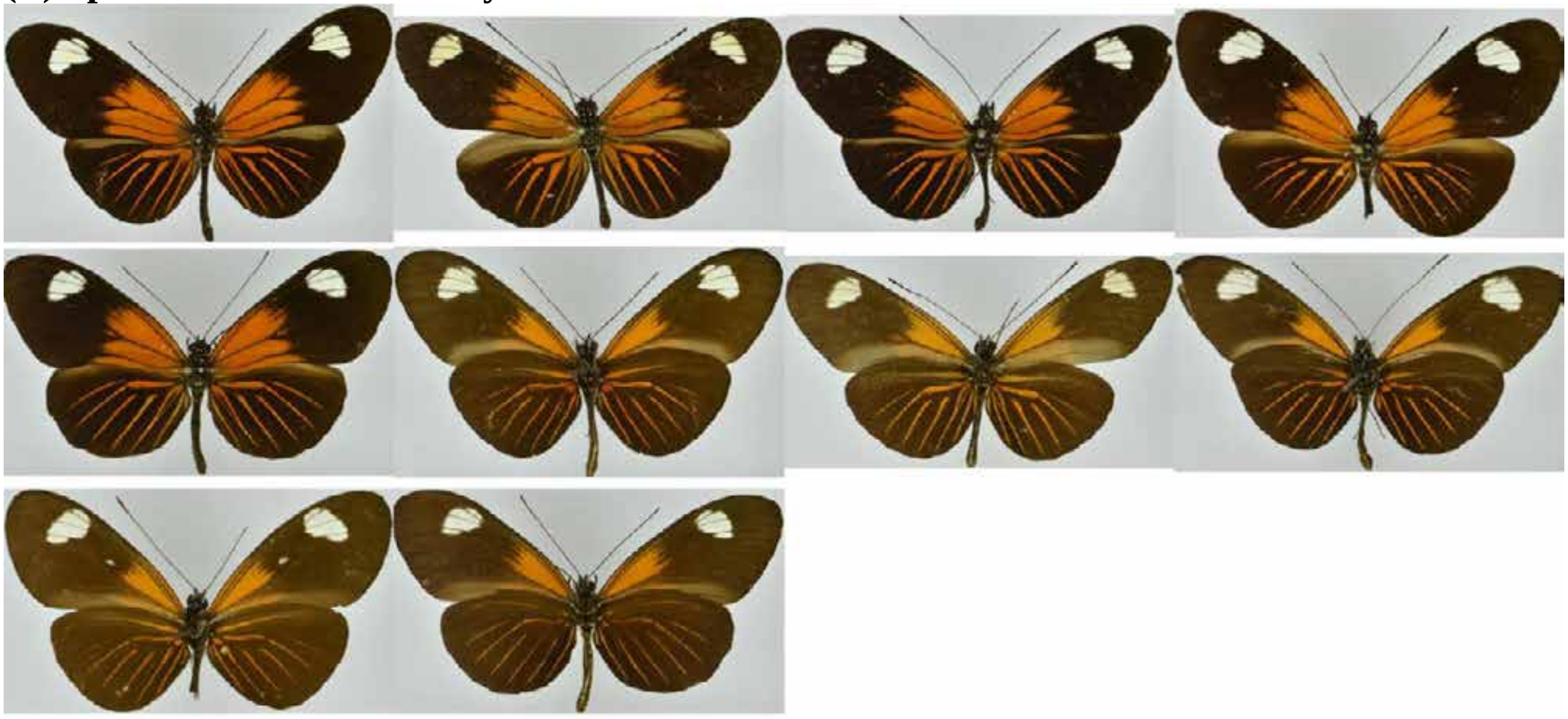

**16 *Heliconius erato favorinus***

**(A) Specimens identified as valid subspecies or accepted synonym**

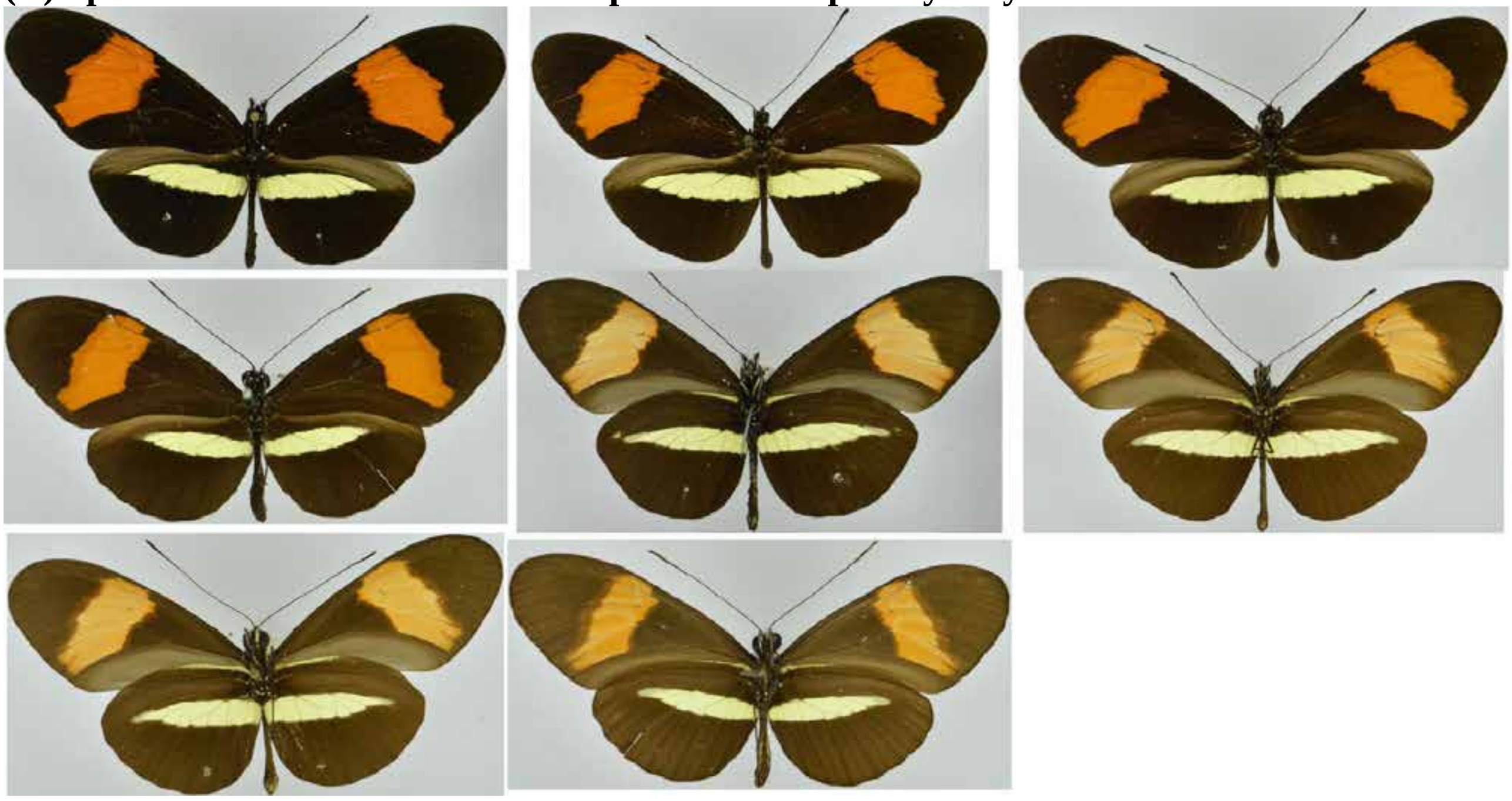

**(B) Specimens identified as hybrids**

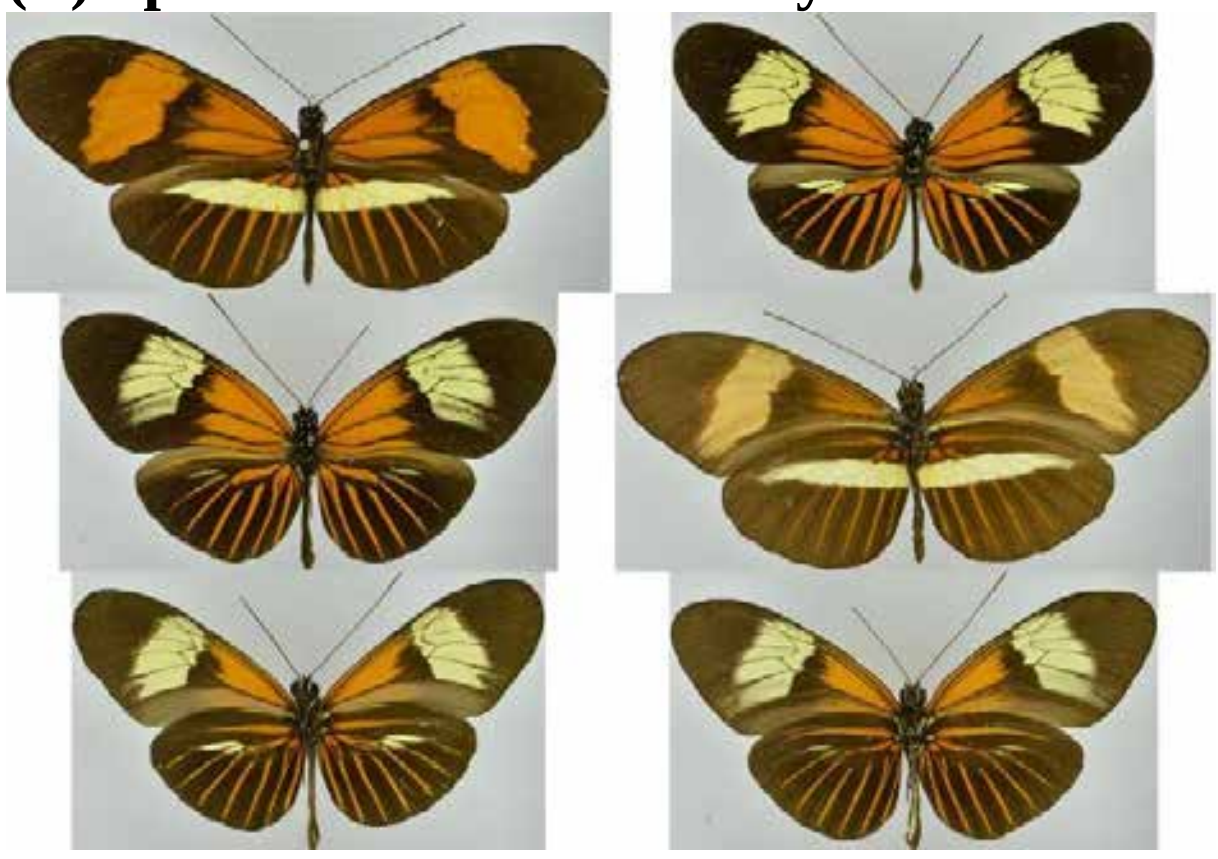

**17 *Heliconius erato favorinus x lativitta***
**(B) Specimens identified as hybrids**

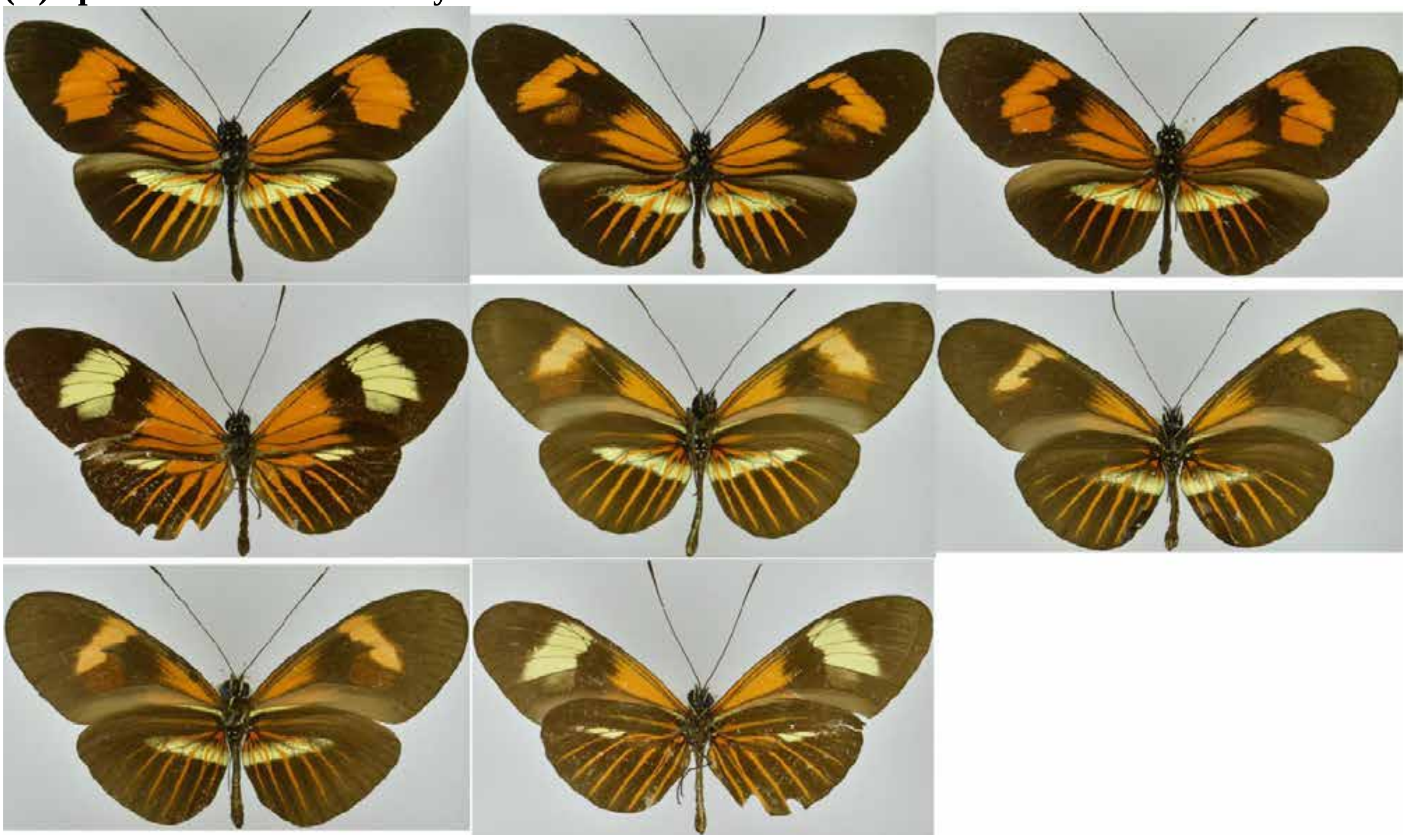

18 ***Heliconius melpomene flagrans***

**(A) Specimens identified as valid subspecies or accepted synonym**

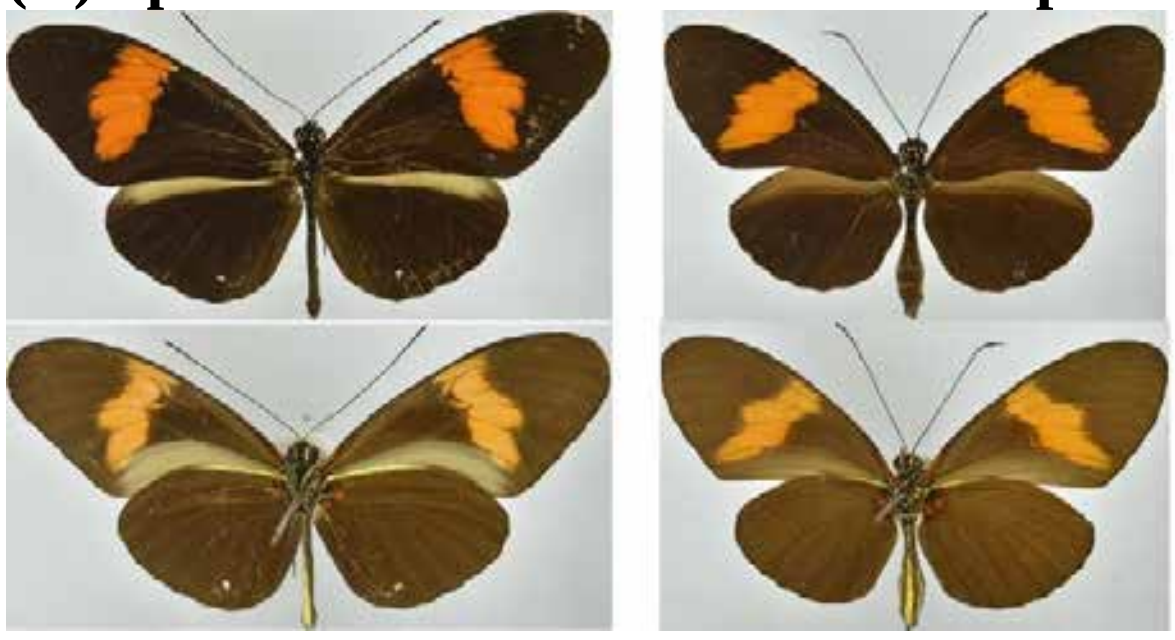

19 *Heliconius erato guarica*

(A) Specimens identified as valid subspecies or accepted synonym

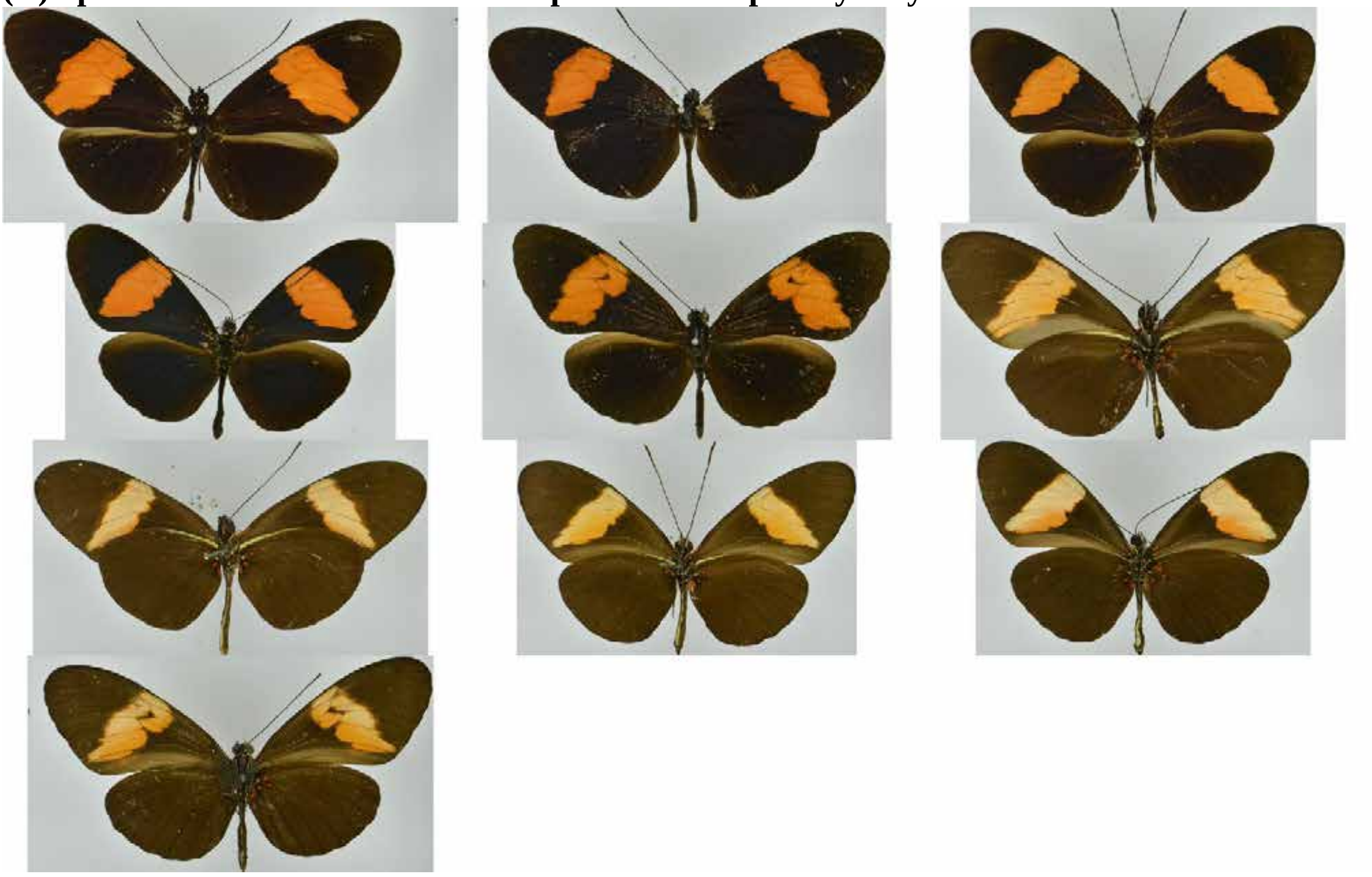

(B) Specimens identified as hybrids

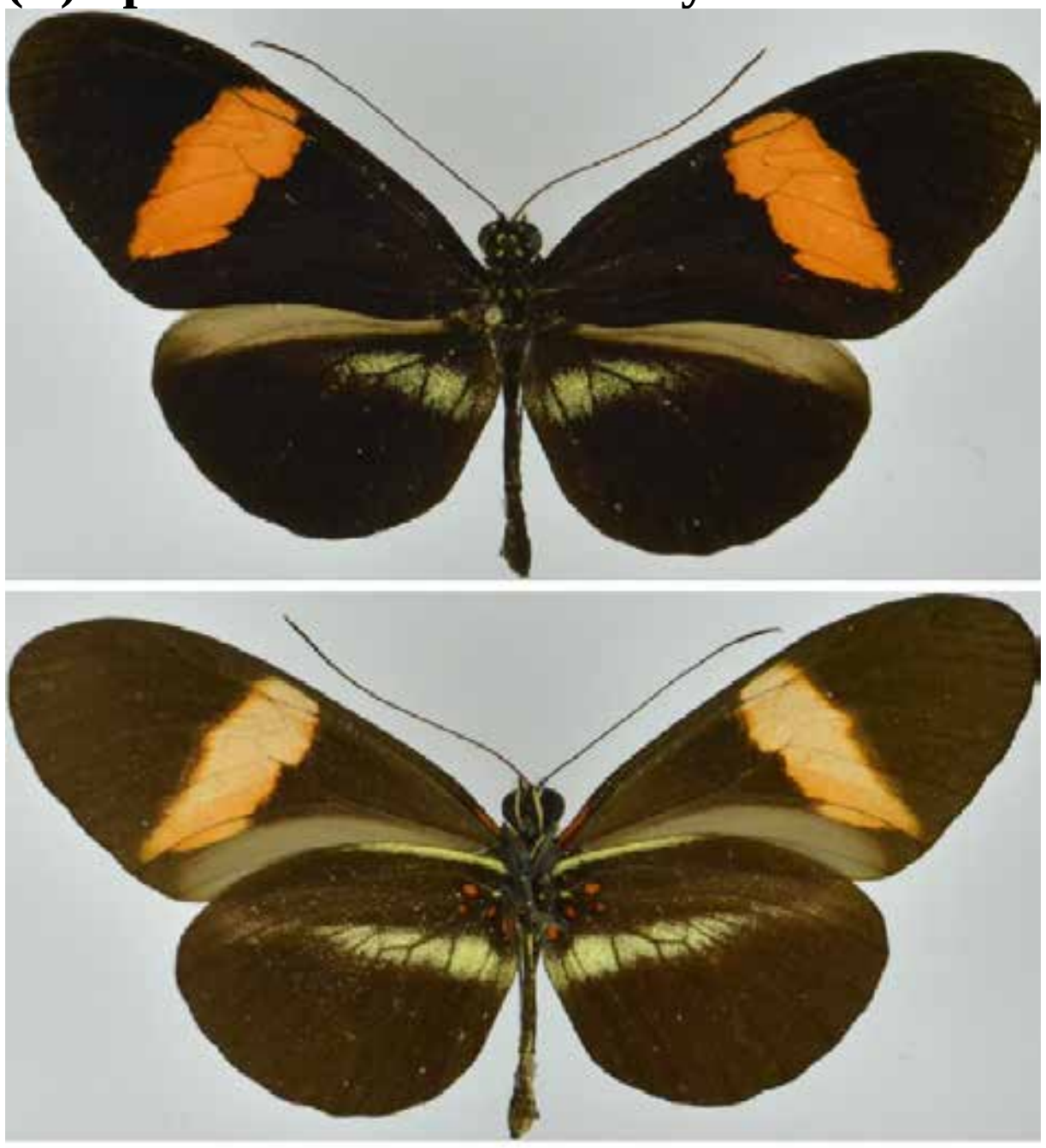

**20 *Heliconius erato hydara***

**(A) Specimens identified as valid subspecies or accepted synonym**

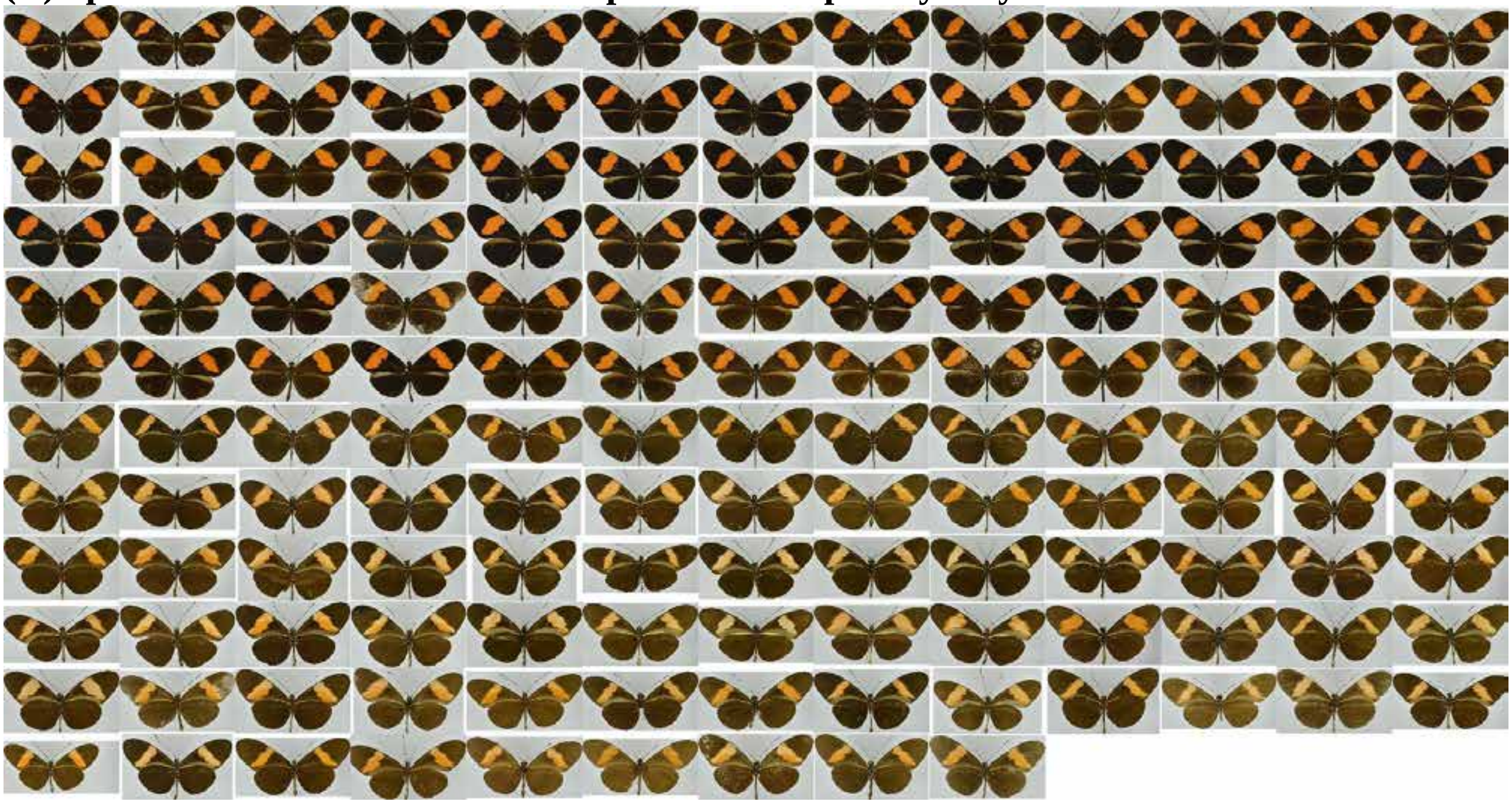

**(B) Specimens identified as hybrids**

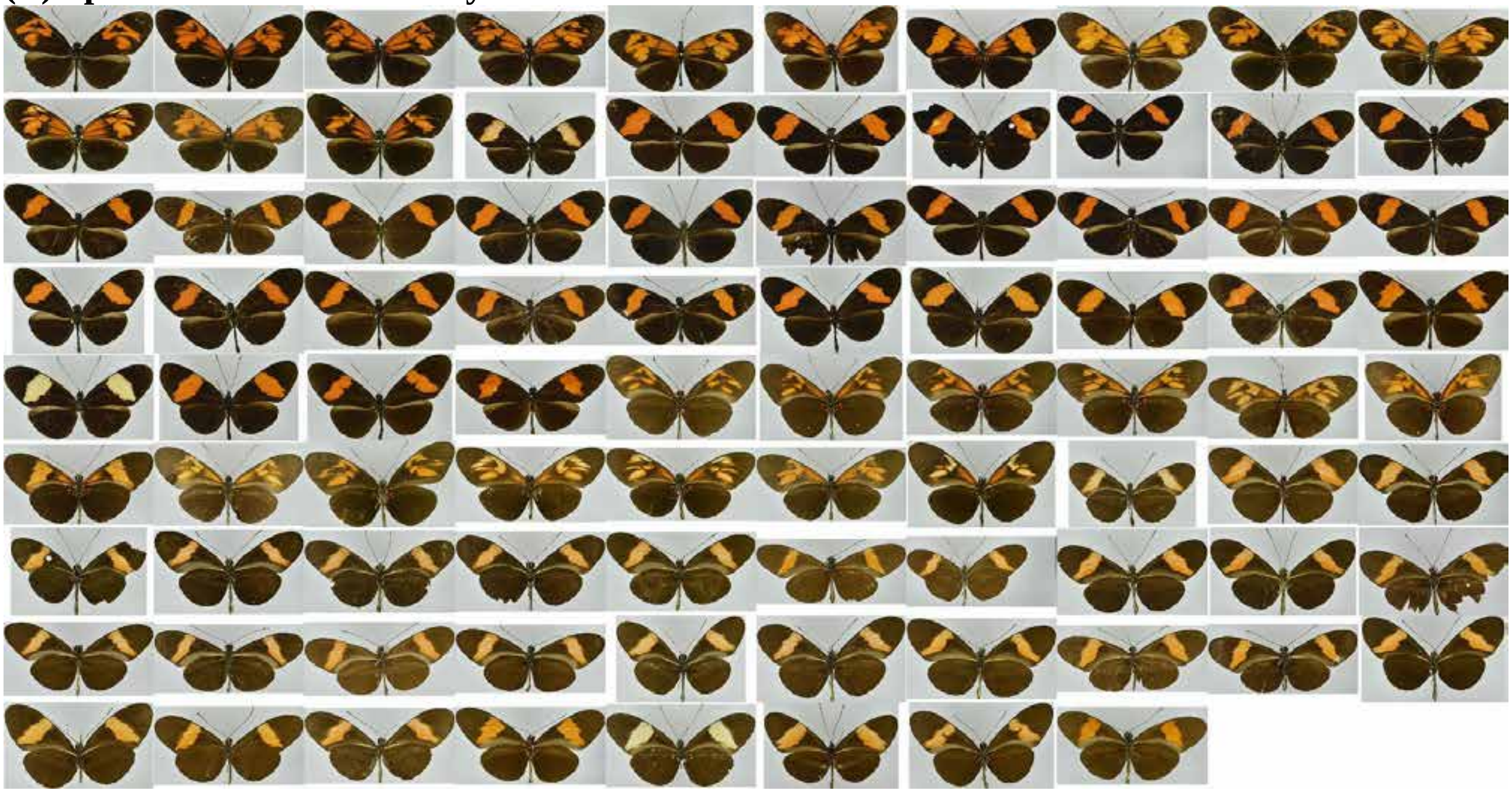

**21 *Heliconius erato lativitta***

**(A) Specimens identified as valid subspecies or accepted synonym**

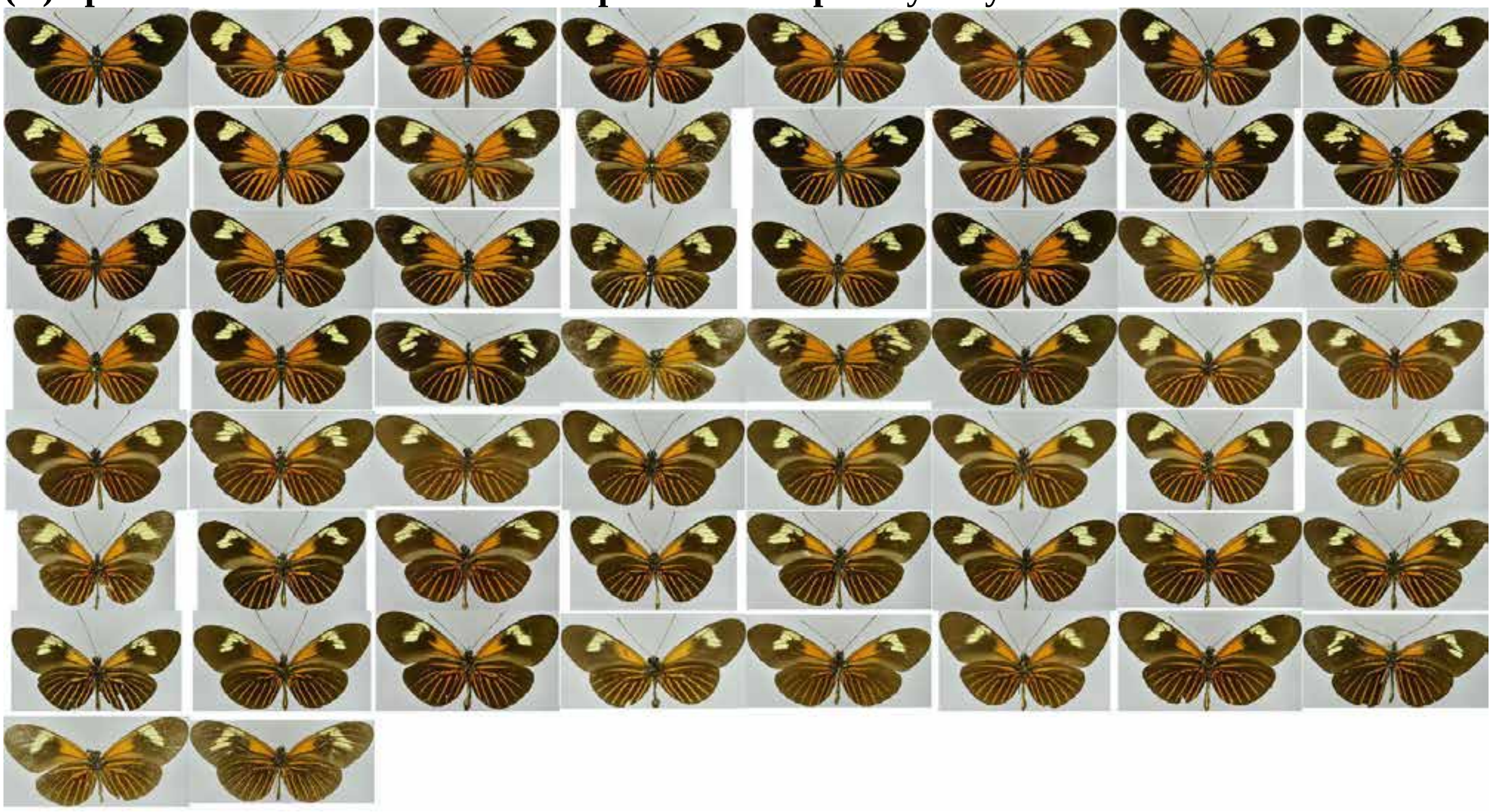

**(B) Specimens identified as hybrids**

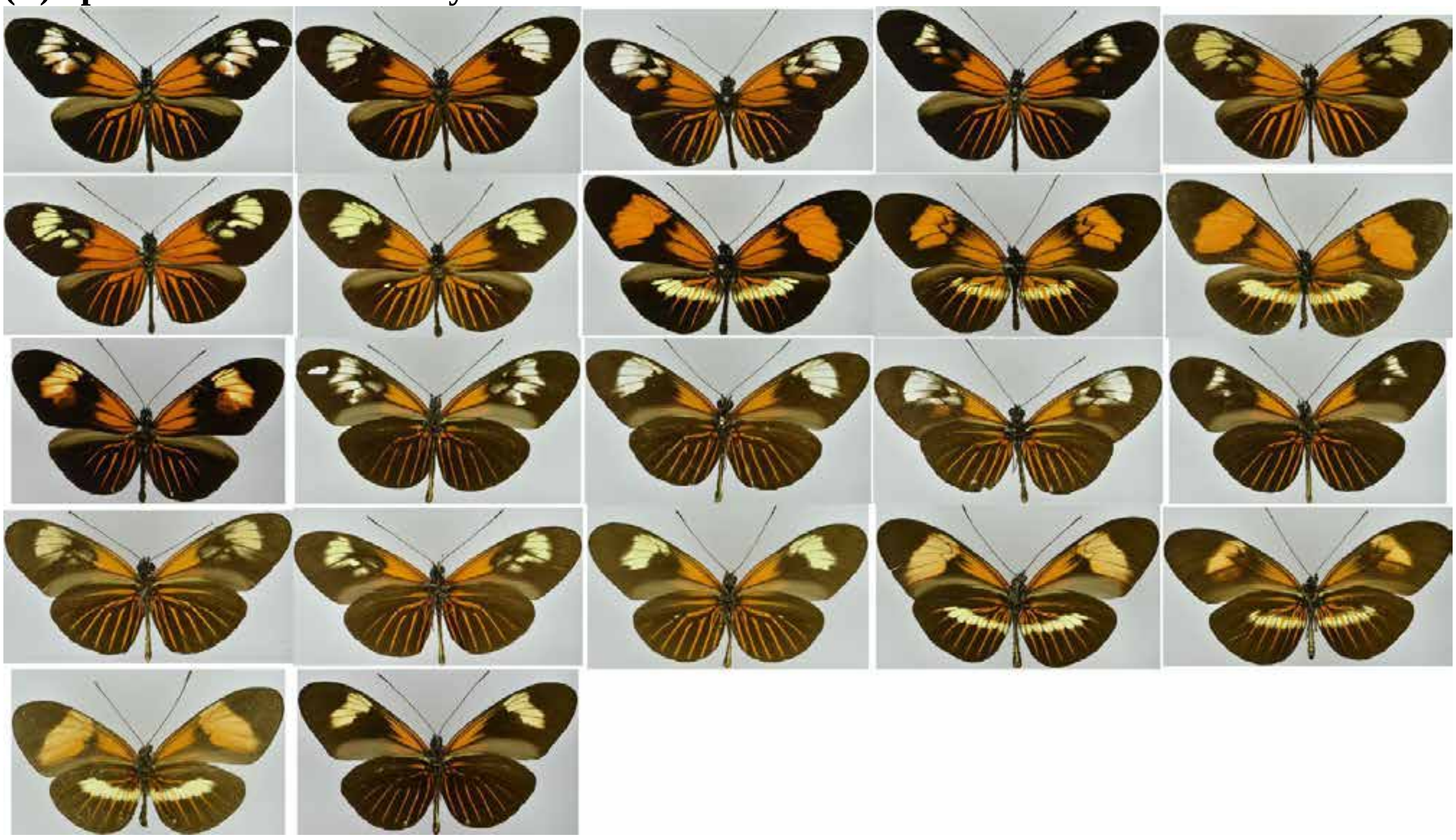

**22 *Heliconius erato luscombei***

**(A) Specimens identified as valid subspecies or accepted synonym**

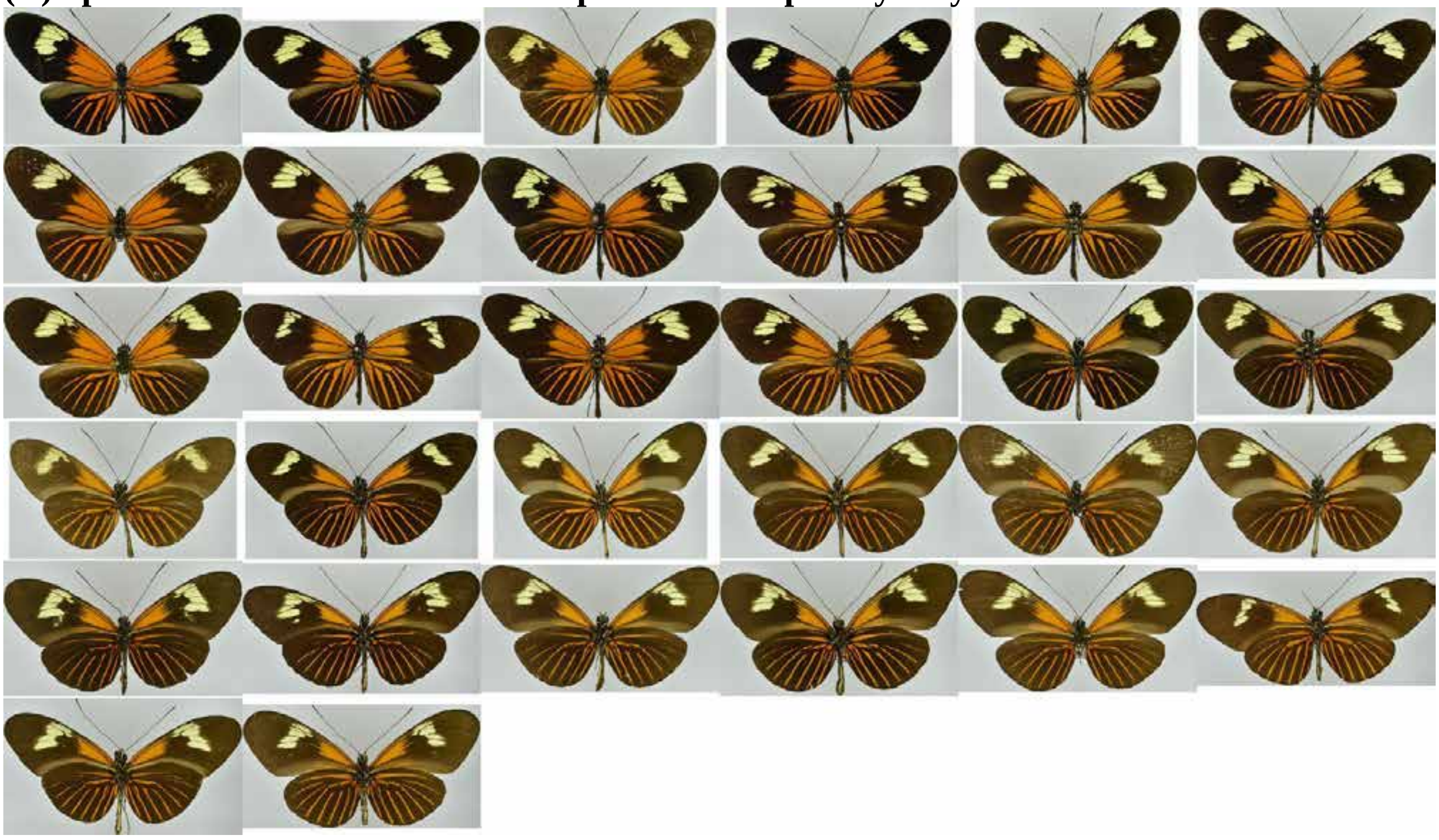

**23 *Heliconius melpomene malleti***

**(A) Specimens identified as valid subspecies or accepted synonym**

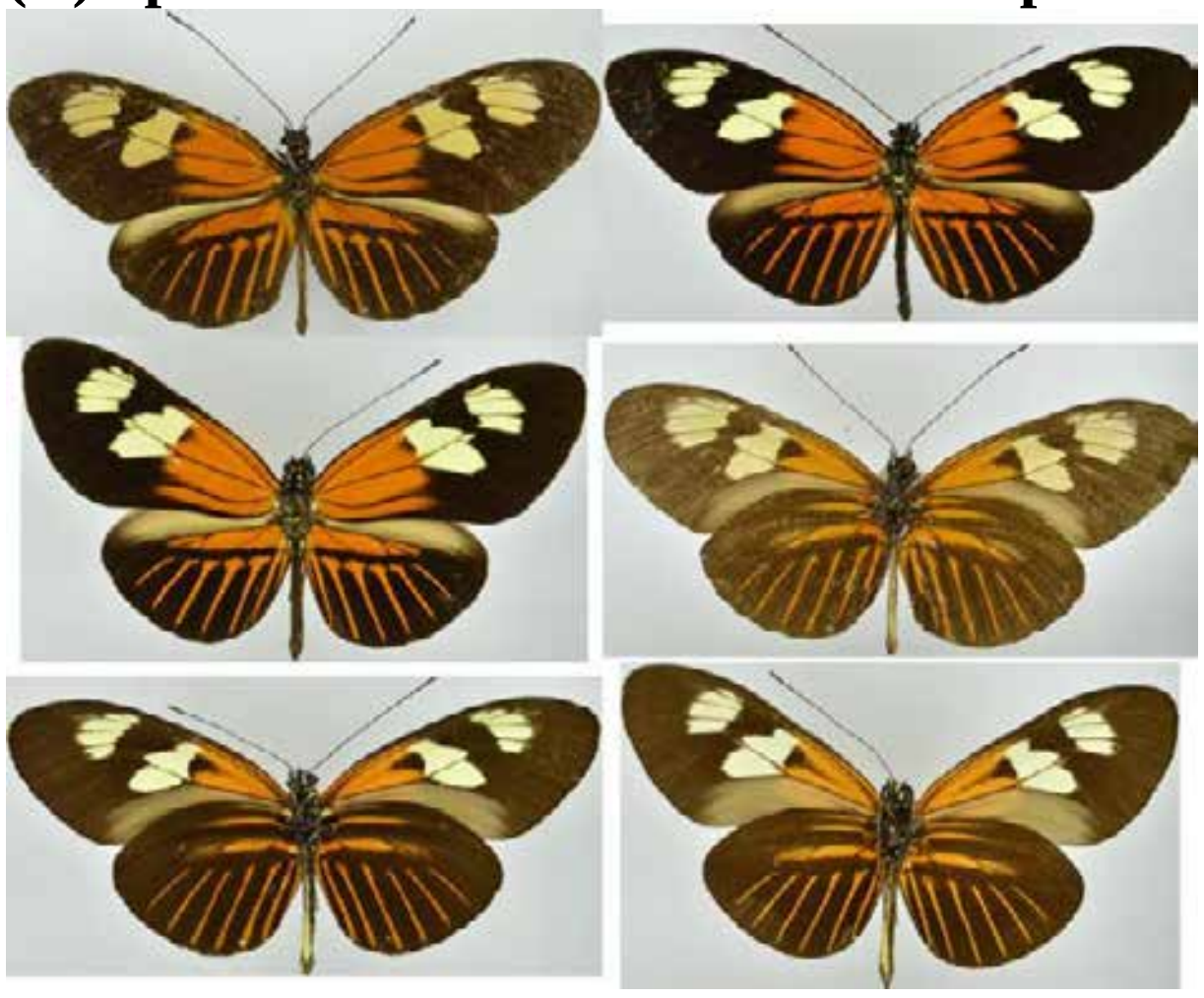

**(B) Specimens identified as hybrids**

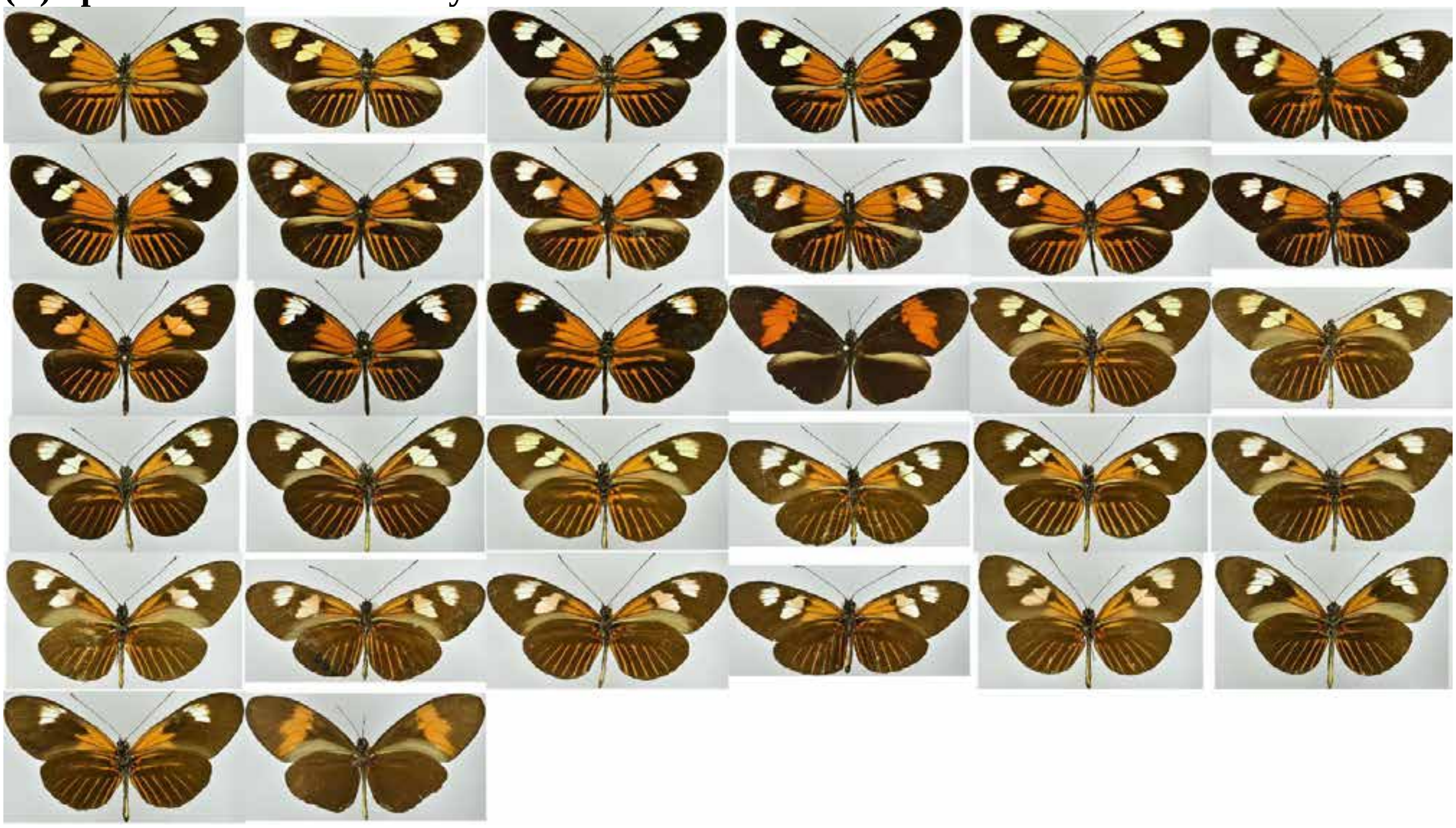

24 *Heliconius melpomene melpomene*

(A) Specimens identified as valid subspecies or accepted synonym

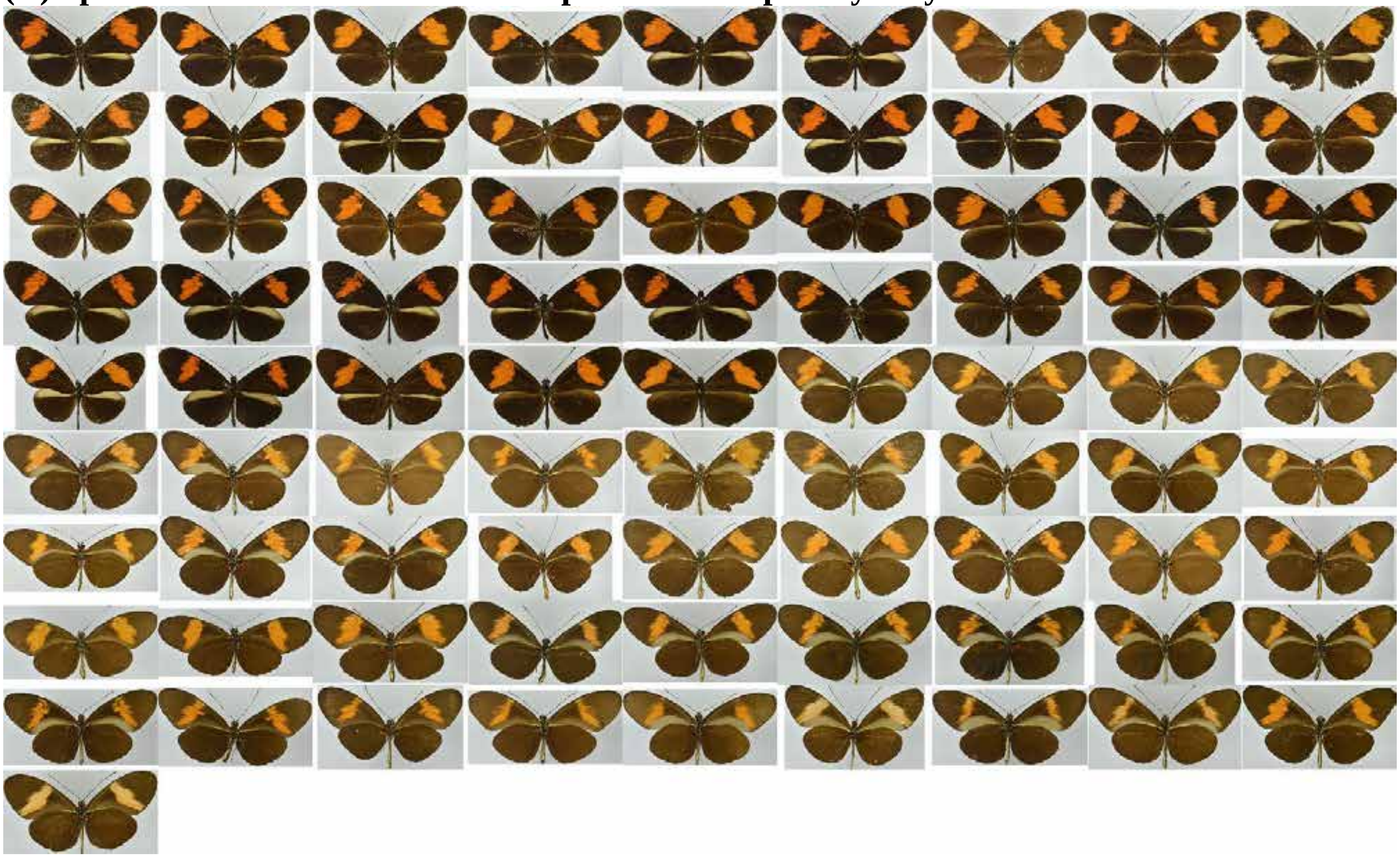

(B) Specimens identified as hybrids

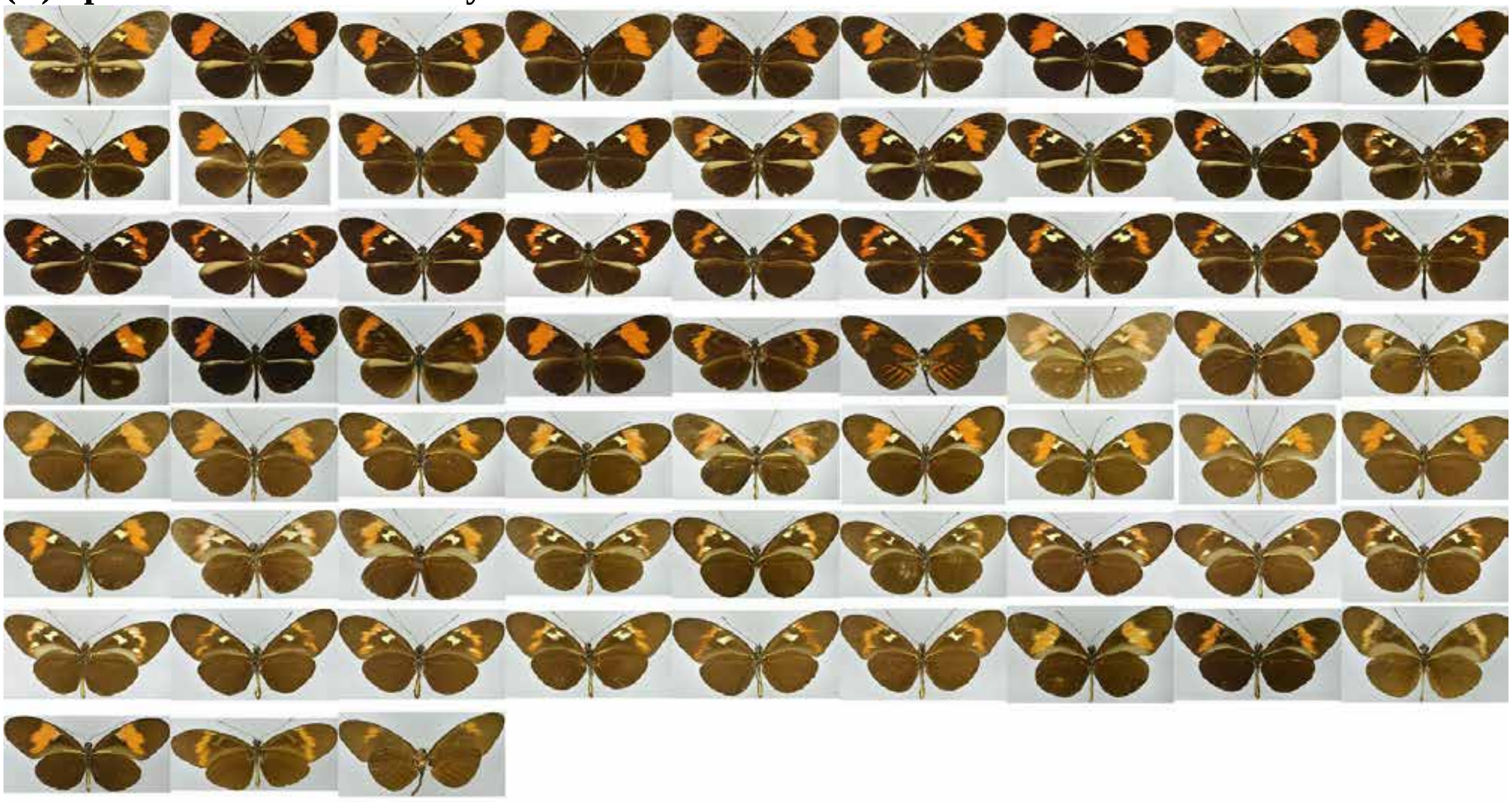

**25 *Heliconius erato microclea***

**(A) Specimens identified as valid subspecies or accepted synonym**

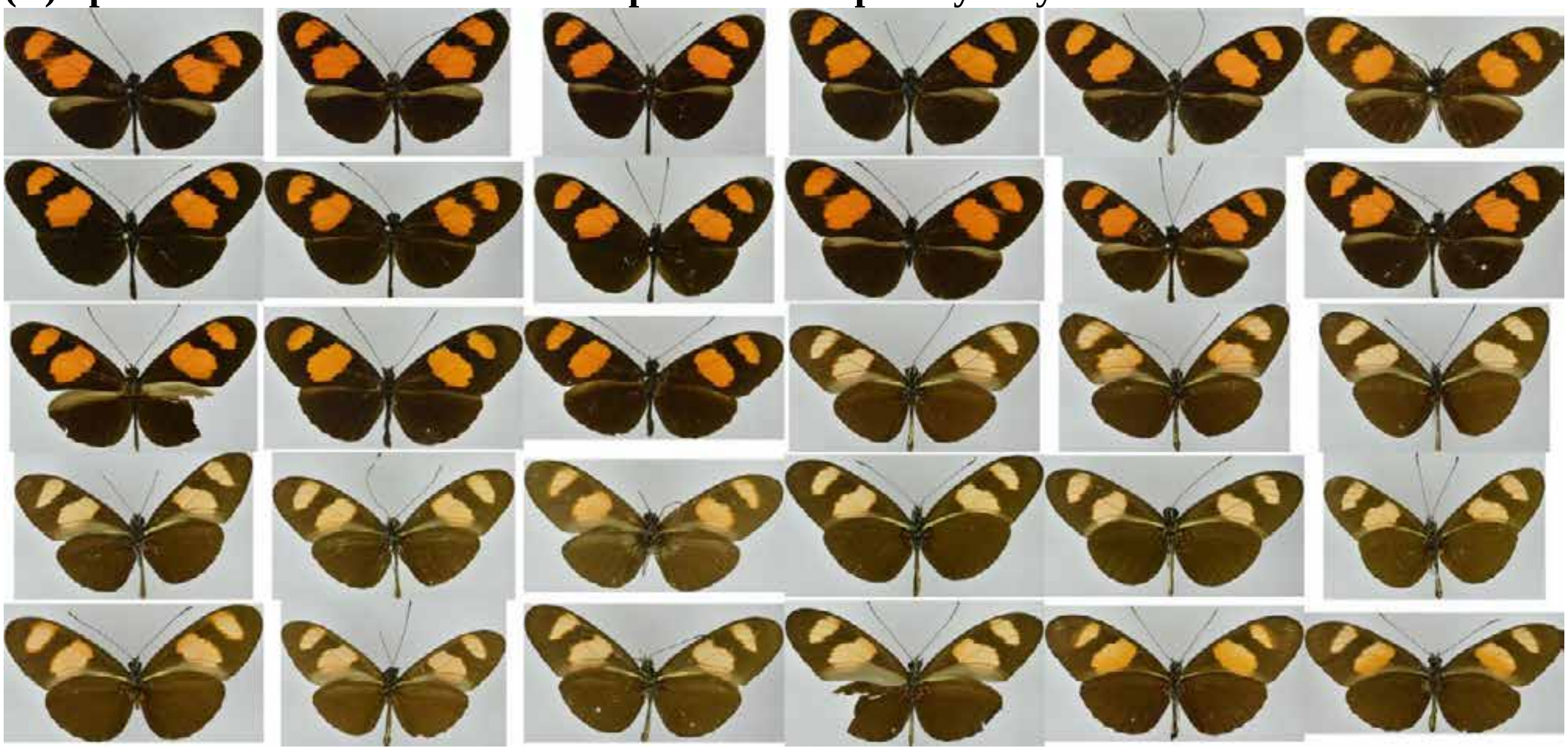

**26 *Heliconius melpomene nanna***

**(A) Specimens identified as valid subspecies or accepted synonym**

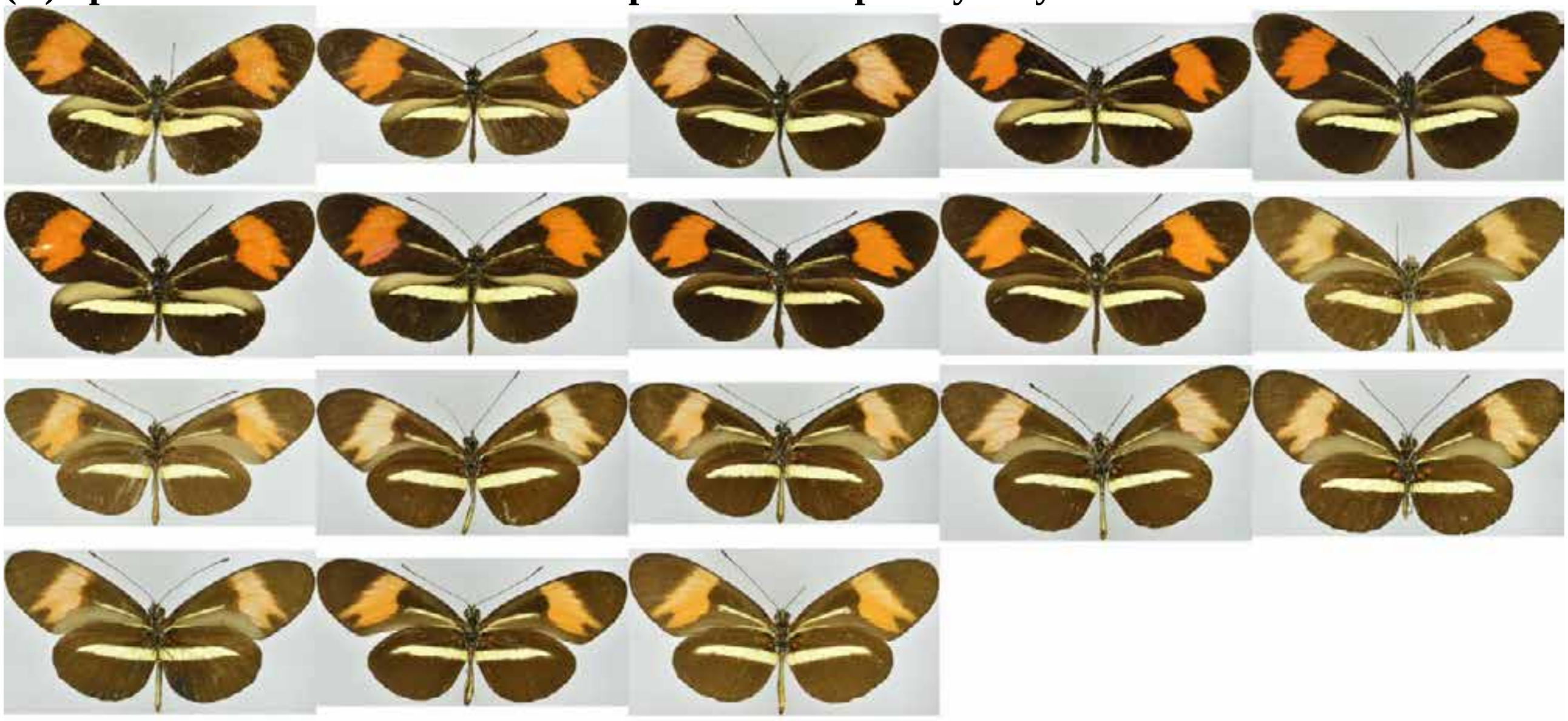

**(B) Specimens identified as hybrids**

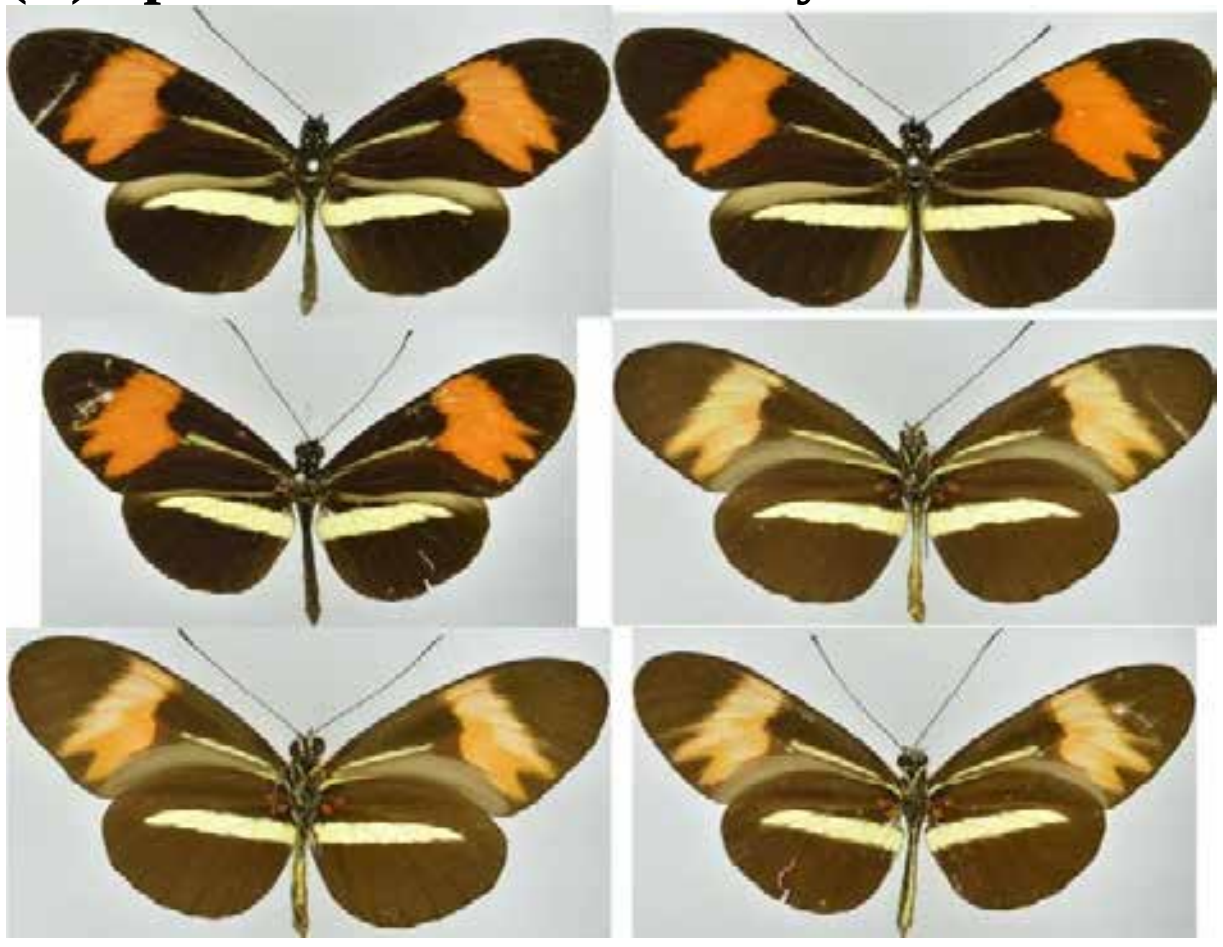

27 *Heliconius erato notabilis*

**(A) Specimens identified as valid subspecies or accepted synonym**

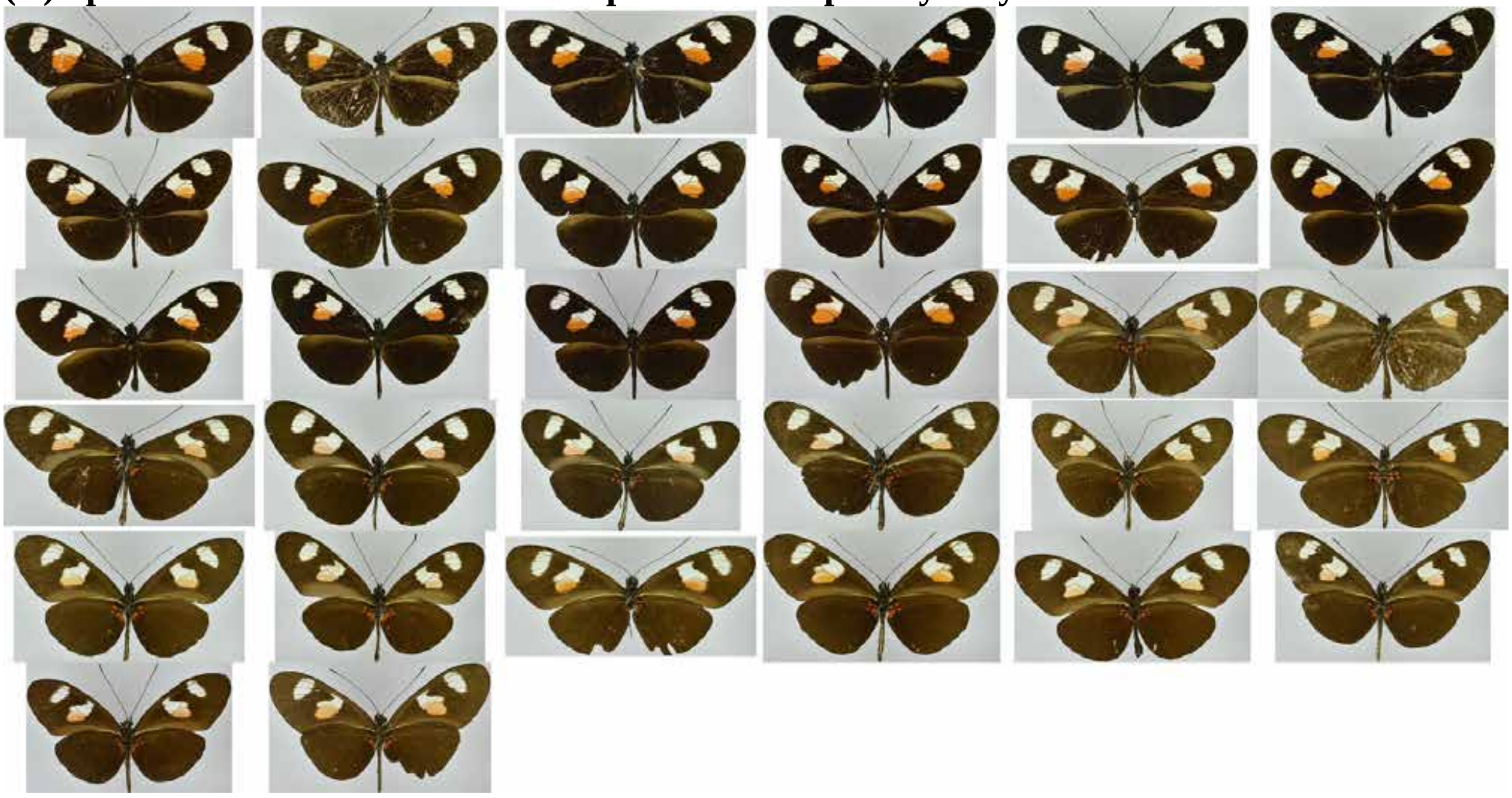

**(B) Specimens identified as hybrids**

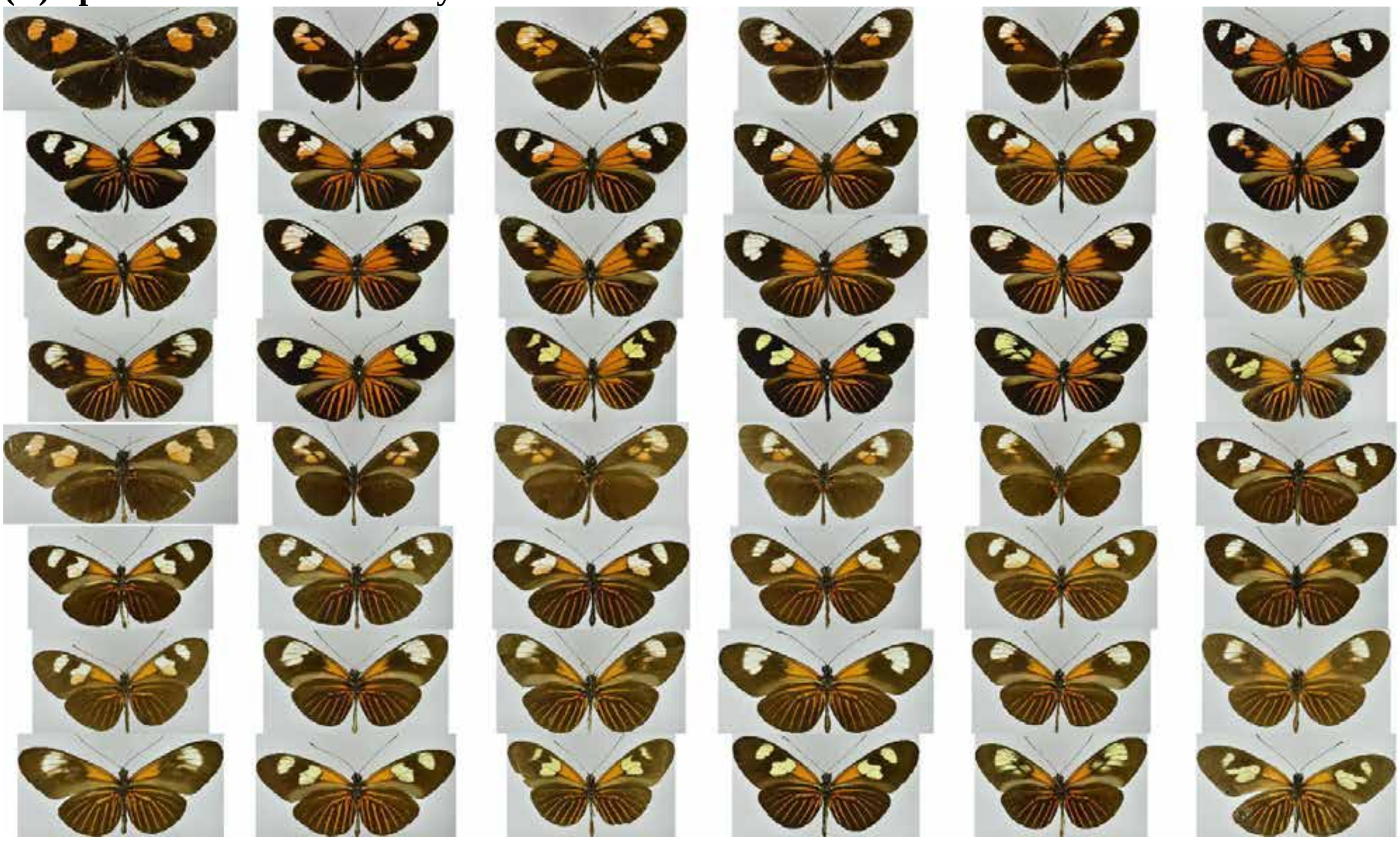

**28 *Heliconius melpomene penelope***

**(B) Specimens identified as hybrids**

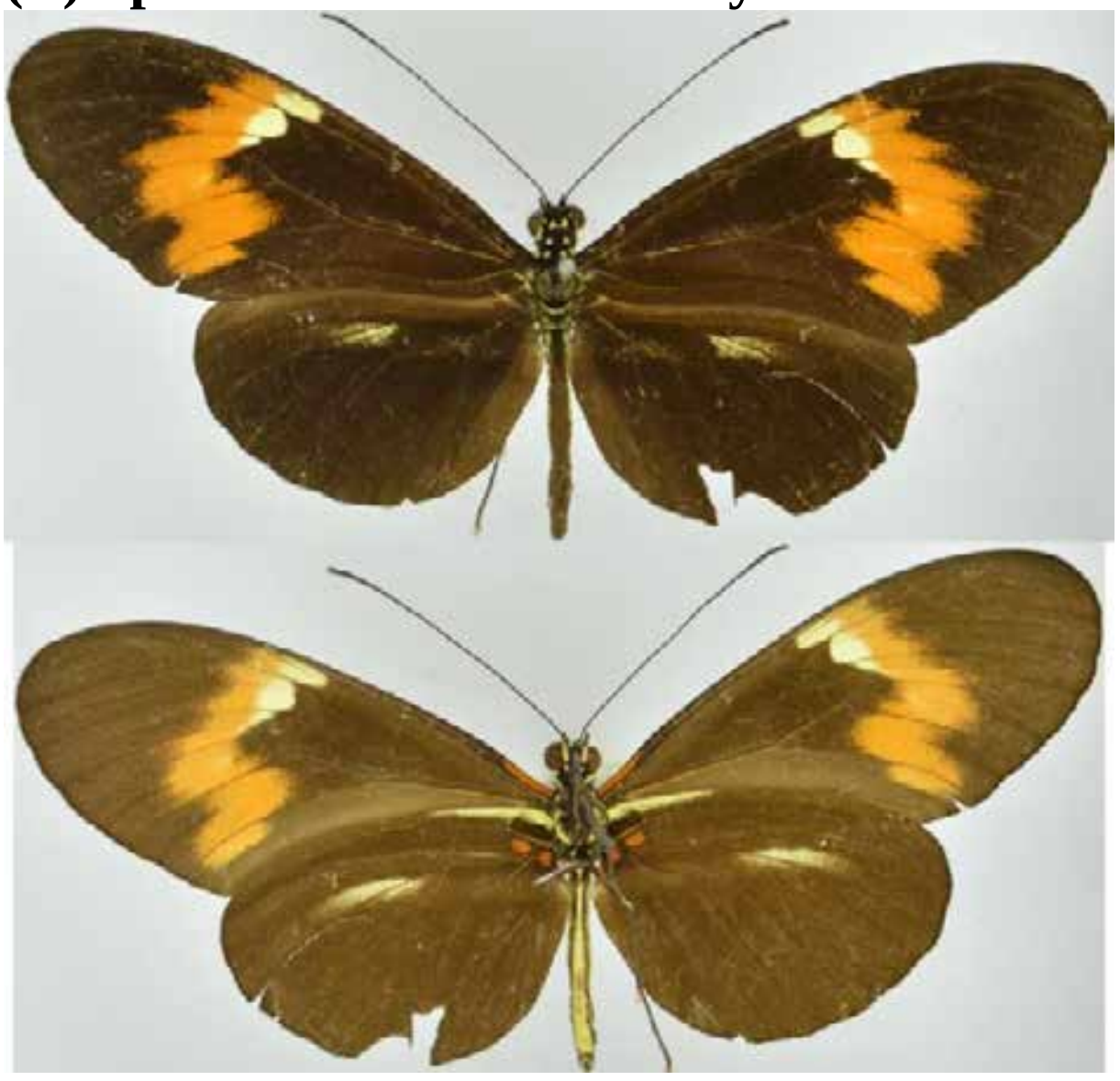

**29 *Heliconius erato petiverana***

**(A) Specimens identified as valid subspecies or accepted synonym**

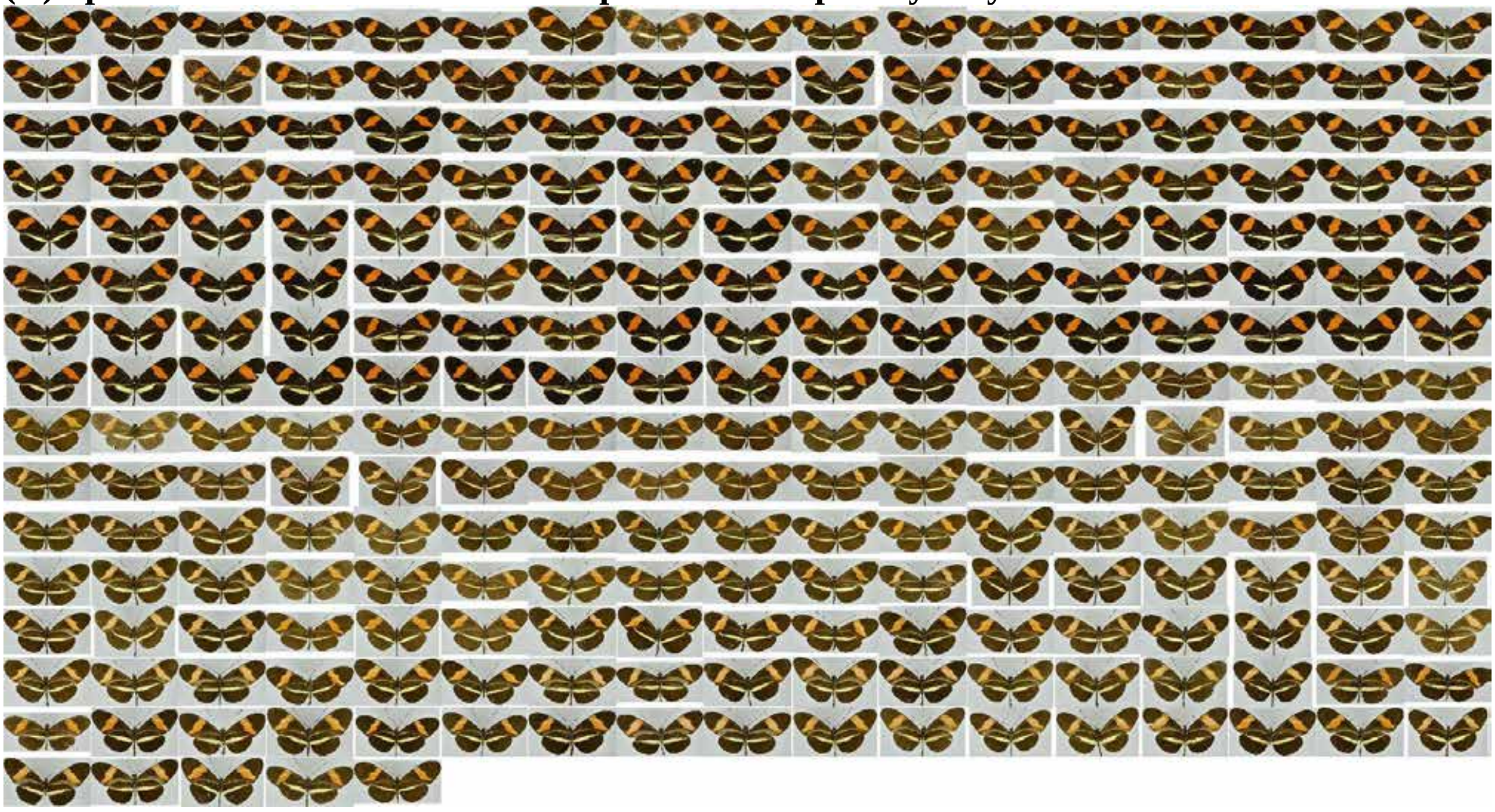

**30 *Heliconius erato phyllis***

**(A) Specimens identified as valid subspecies or accepted synonym**

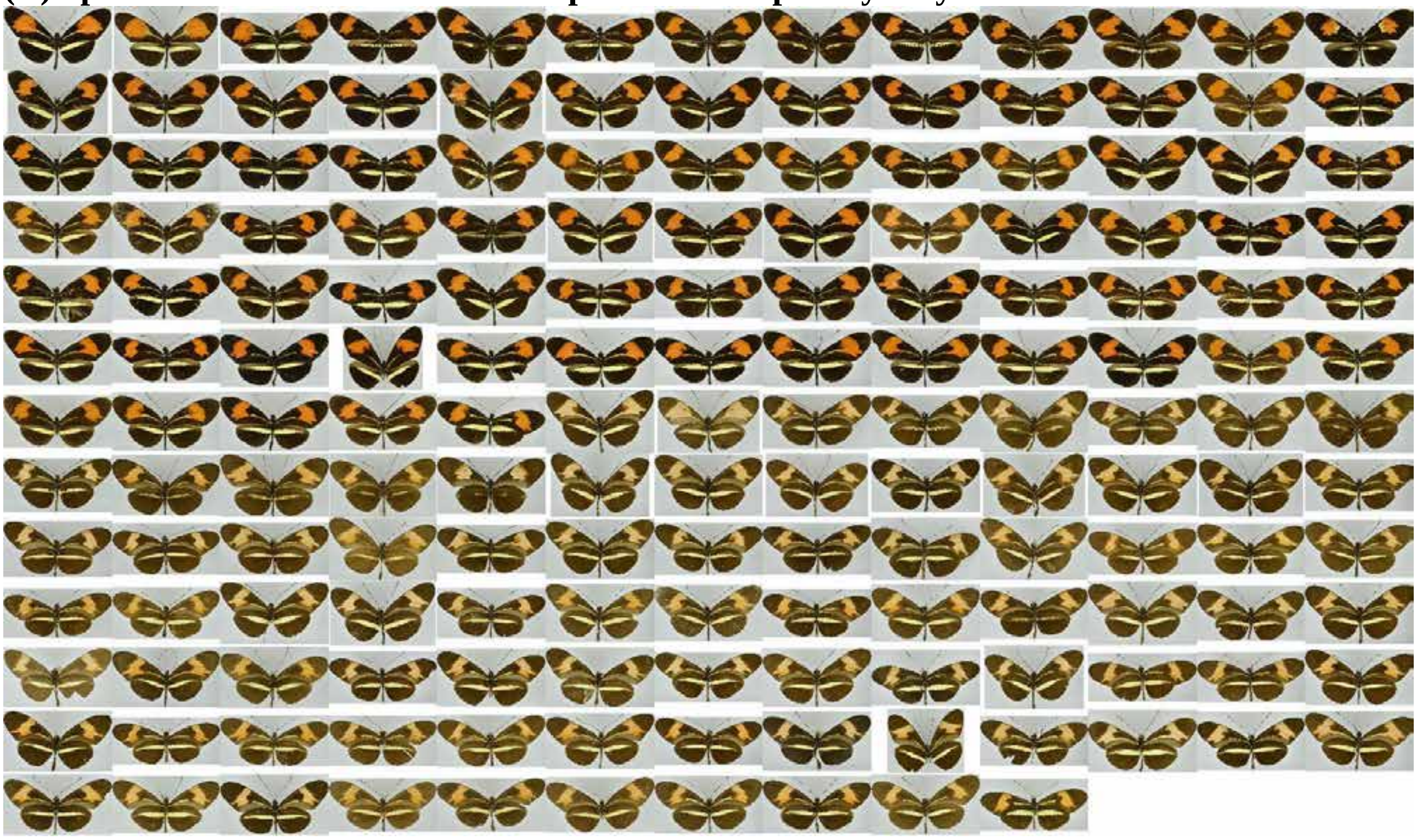

**(B) Specimens identified as hybrids**

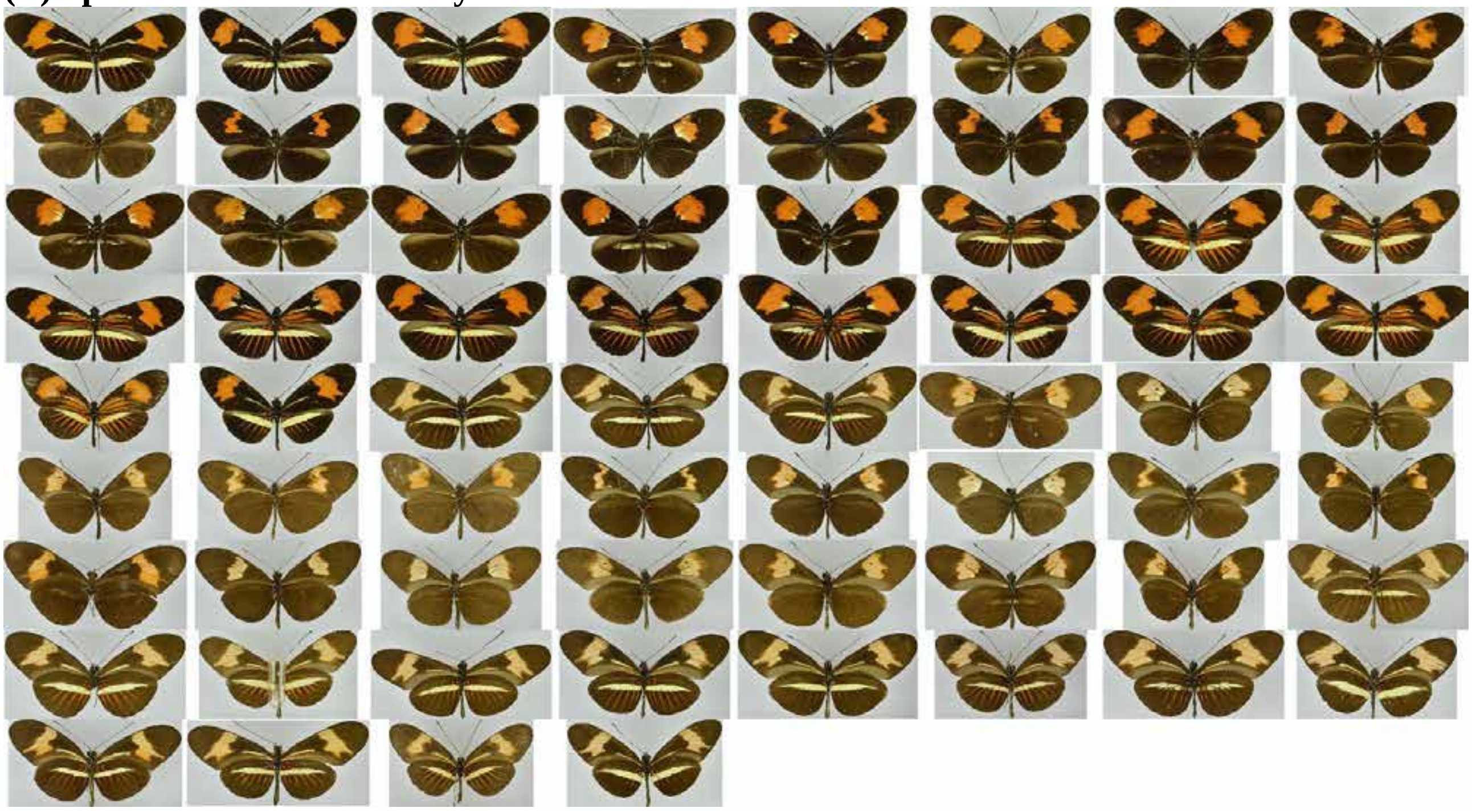

**31 *Heliconius melpomene plesseni***

**(A) Specimens identified as valid subspecies or accepted synonym**

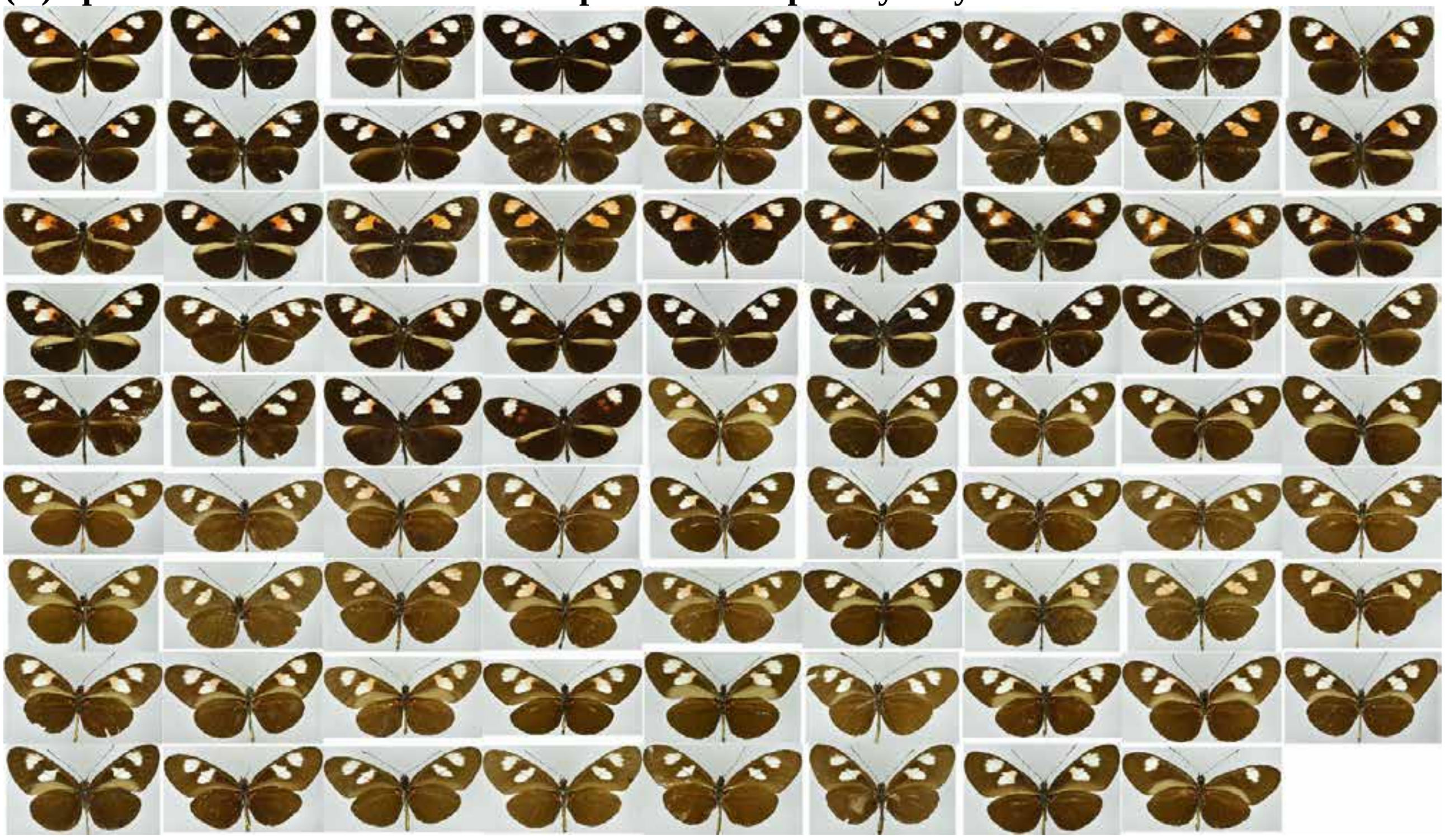

**(B) Specimens identified as hybrids**

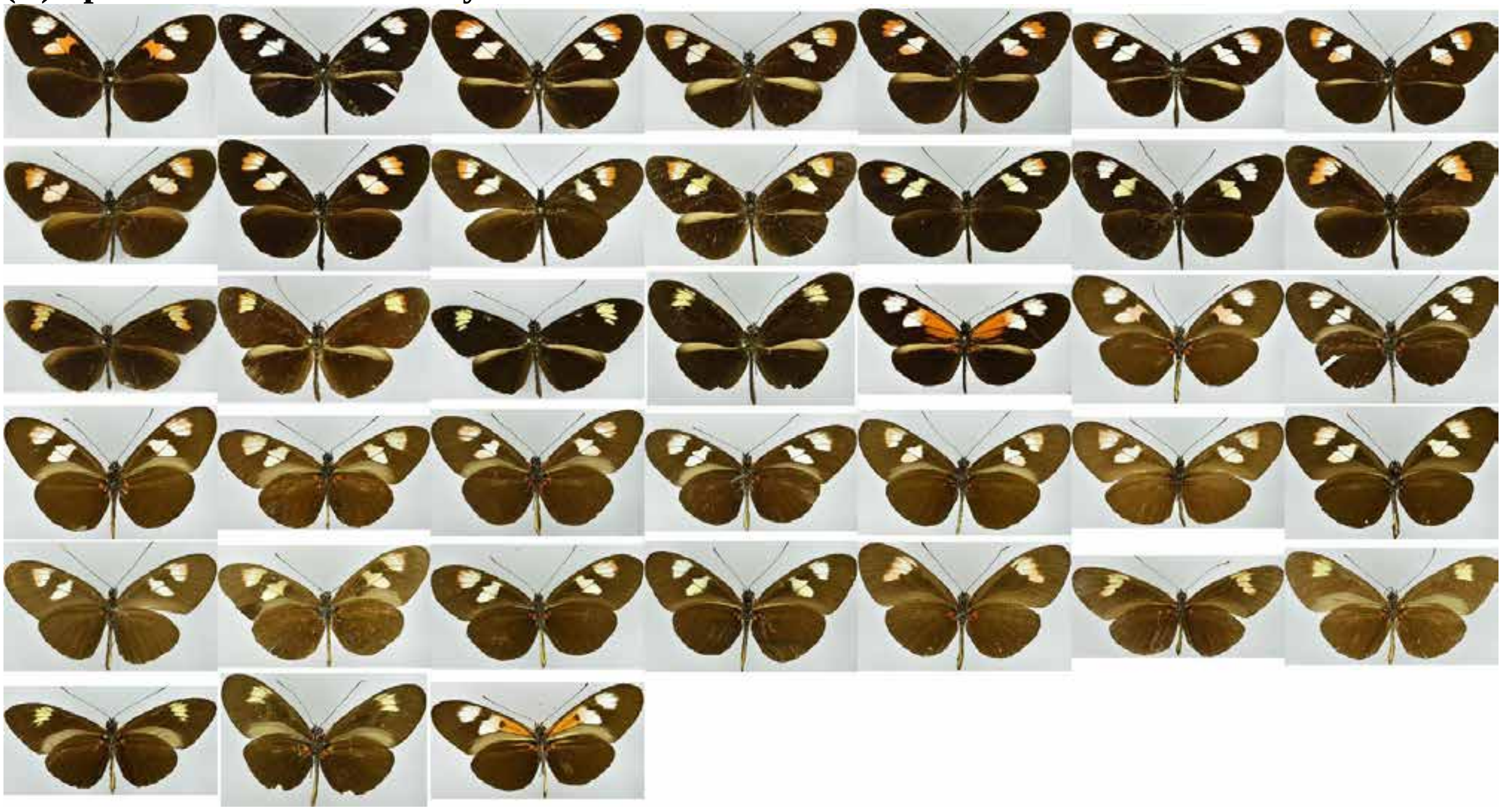

**32 *Heliconius melpomene rosina***

**(A) Specimens identified as valid subspecies or accepted synonym**

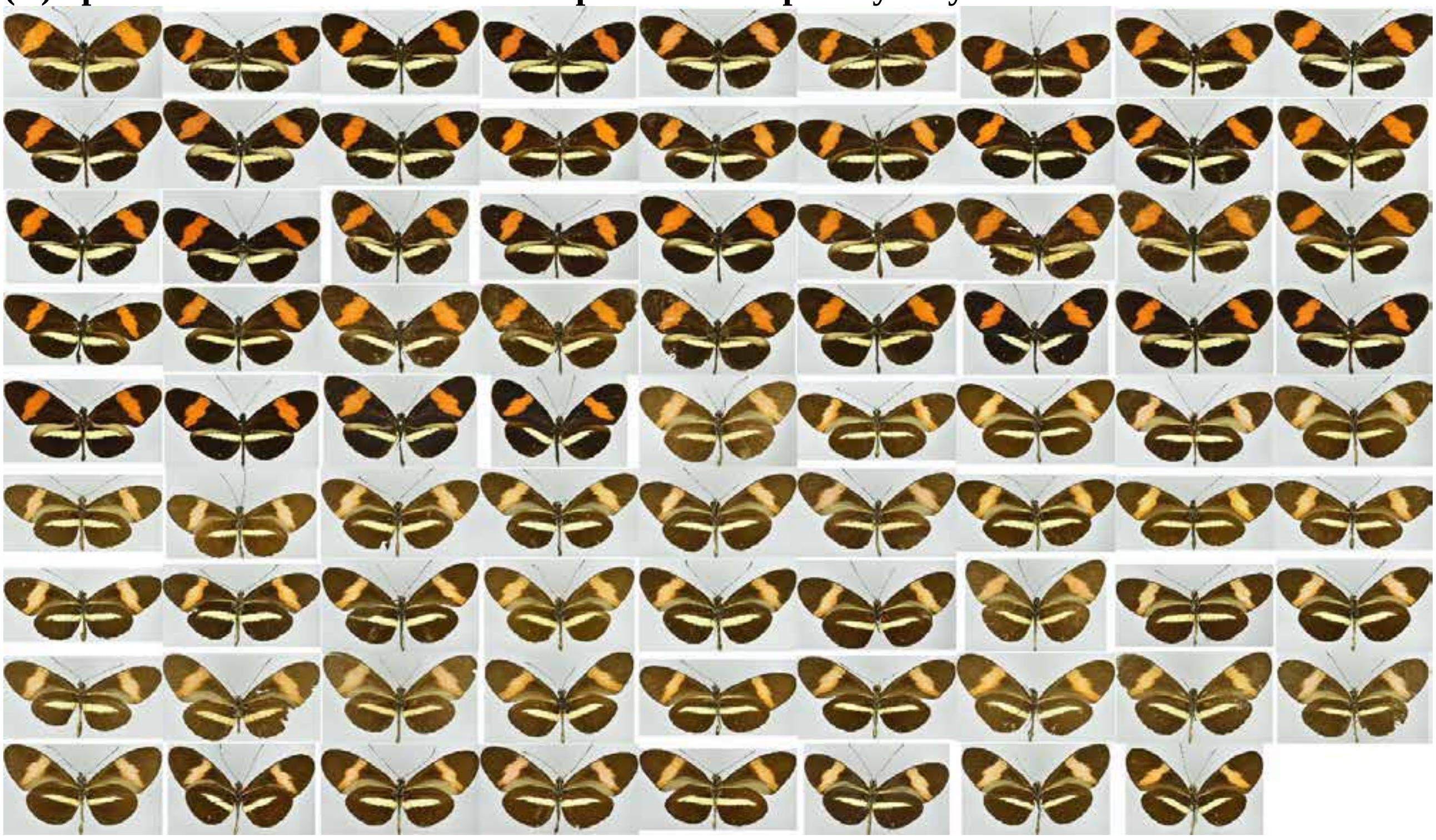

**(B) Specimens identified as hybrids**

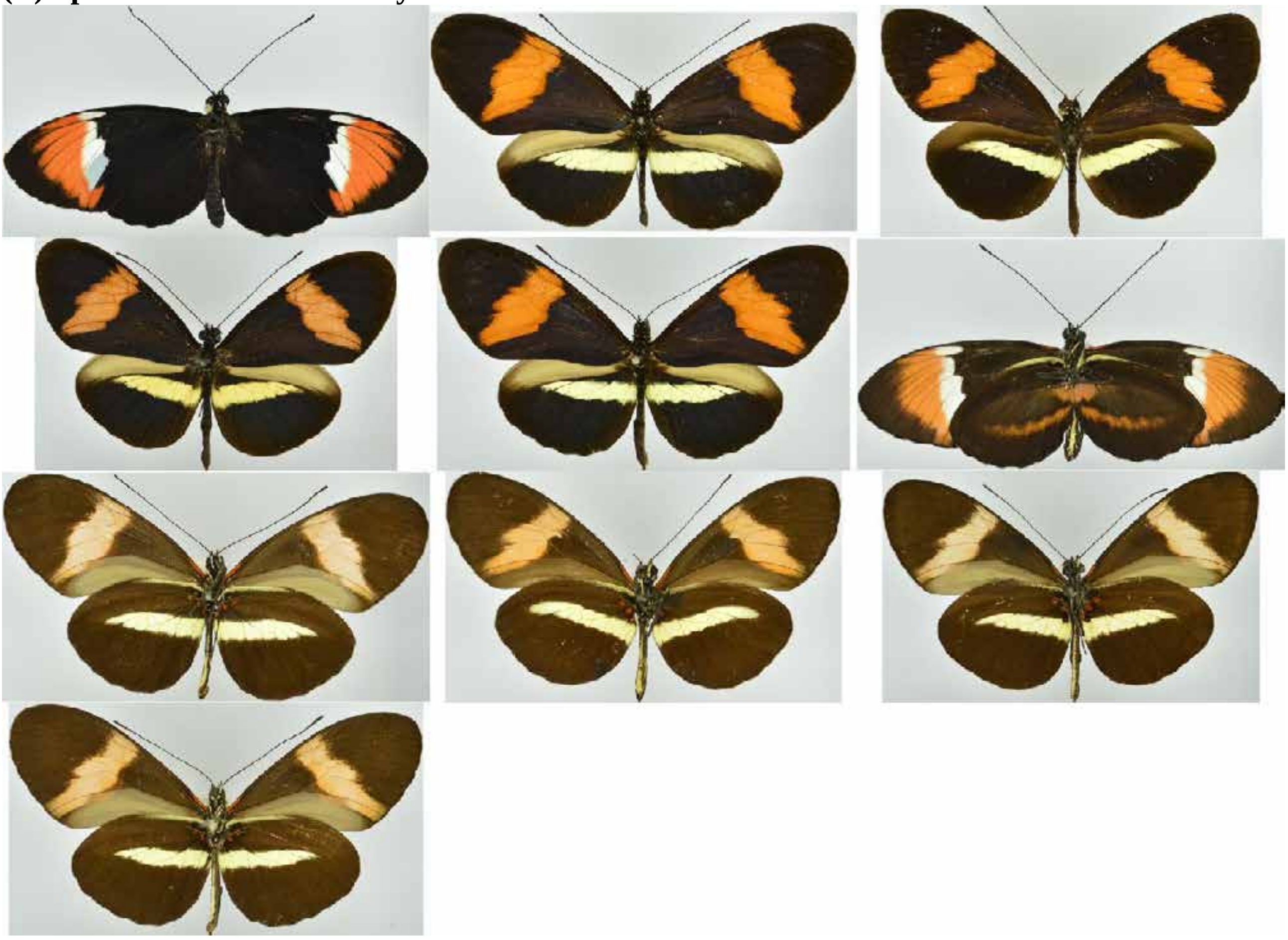

33 *Heliconius melpomene schunkei*

**(A) Specimens identified as valid subspecies or accepted synonym**

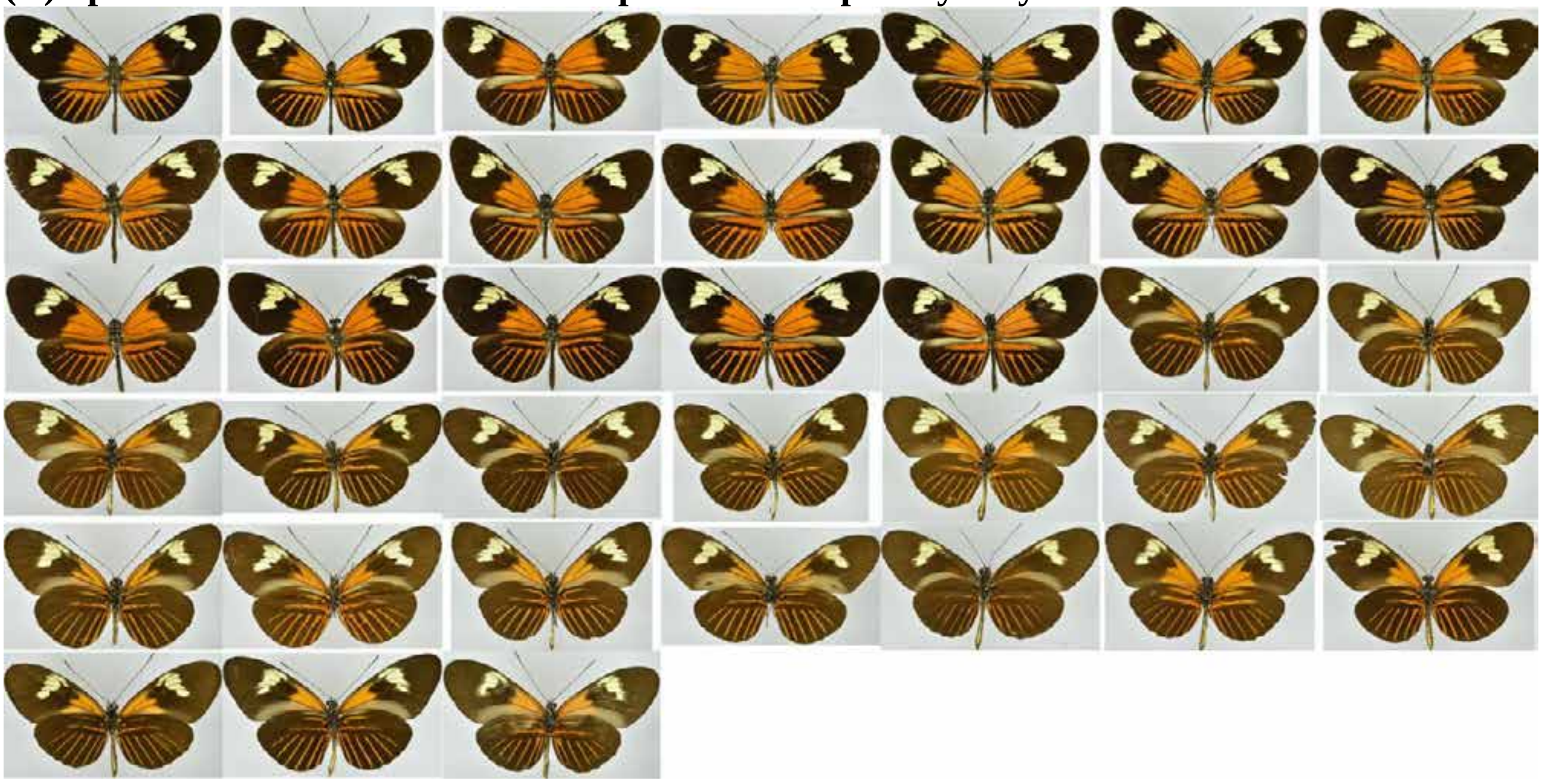

**(B) Specimens identified as hybrids**

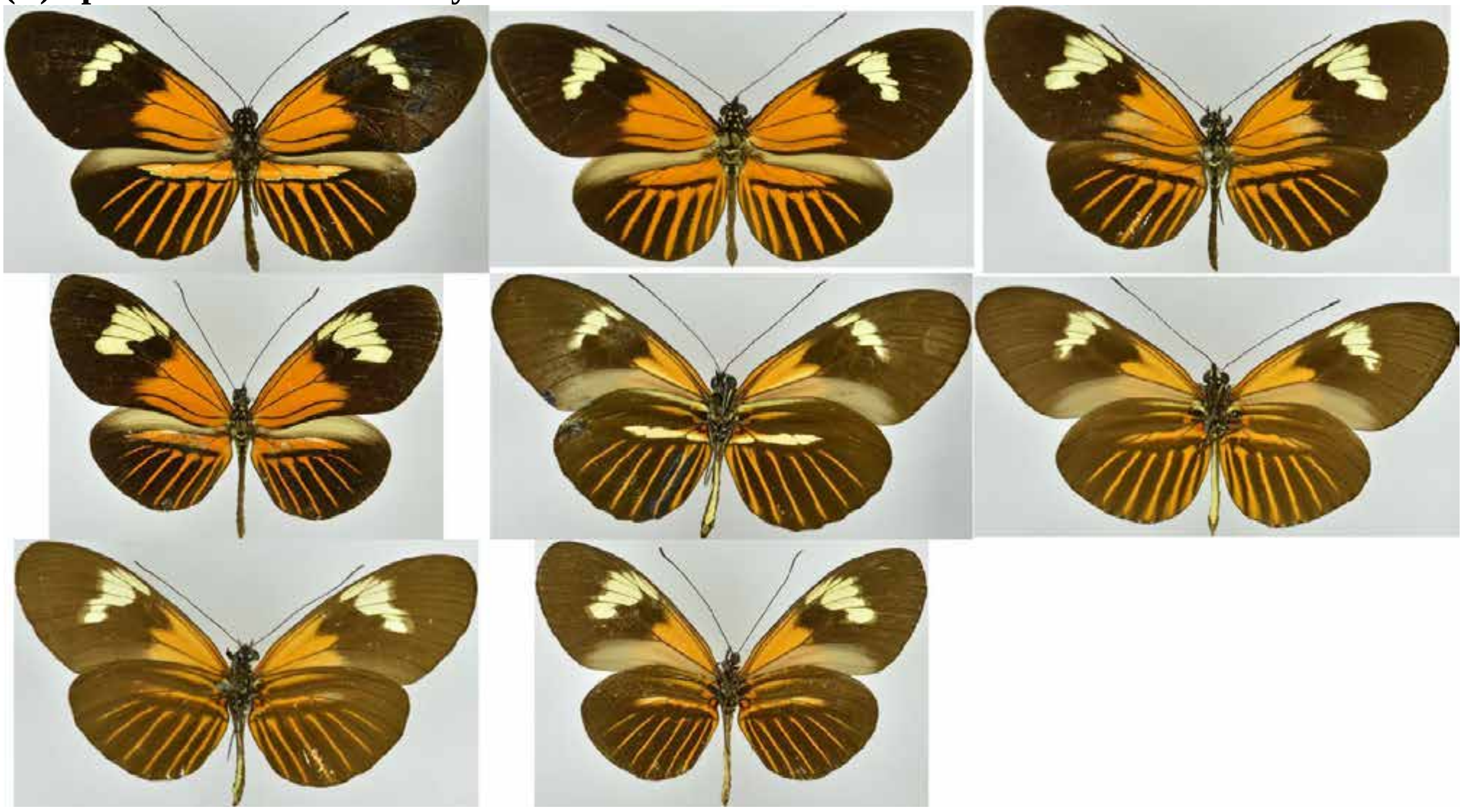

**34 *Heliconius melpomene thelxiopeia***

**(A) Specimens identified as valid subspecies or accepted synonym**

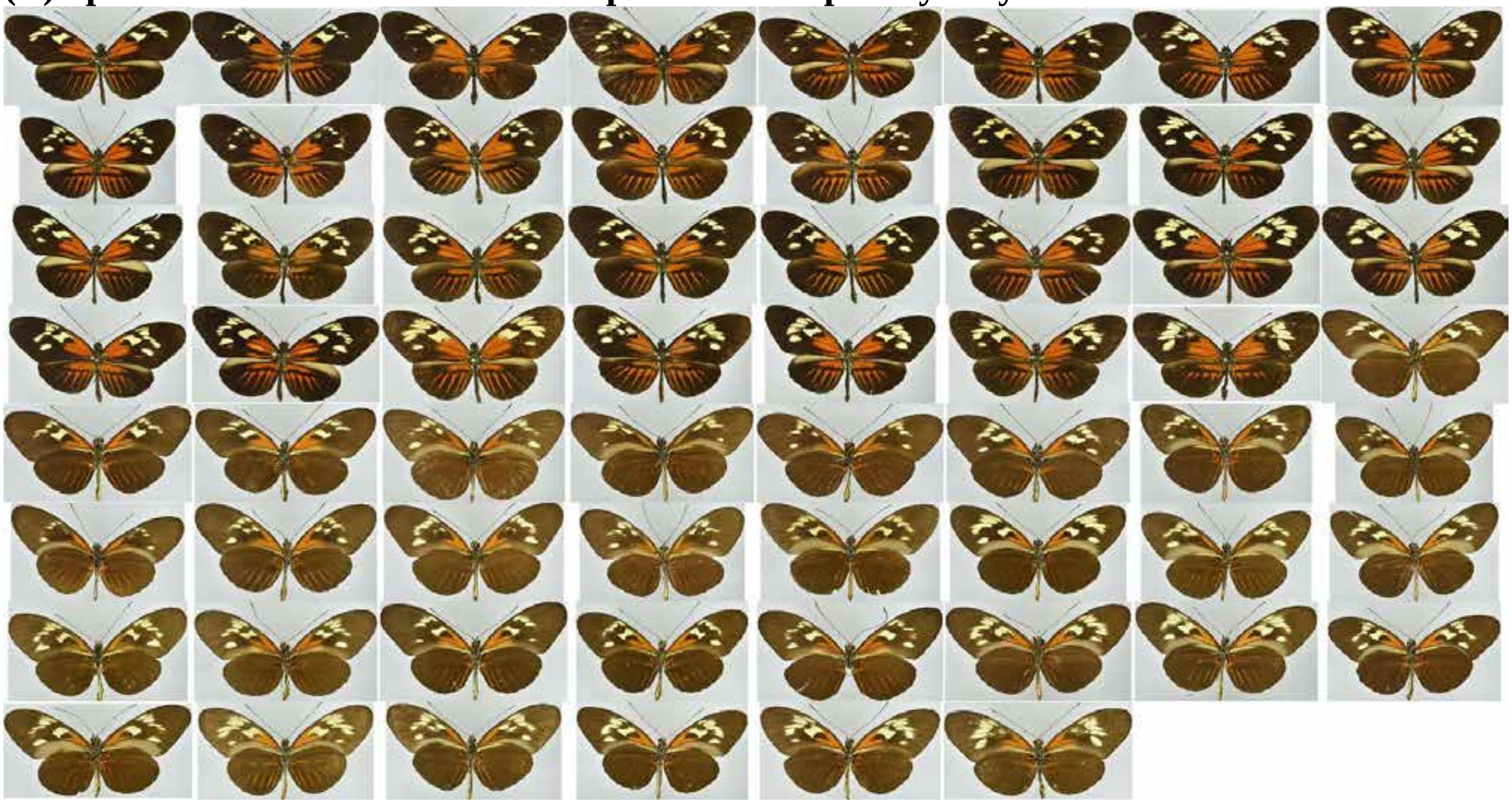

**(B) Specimens identified as hybrids**

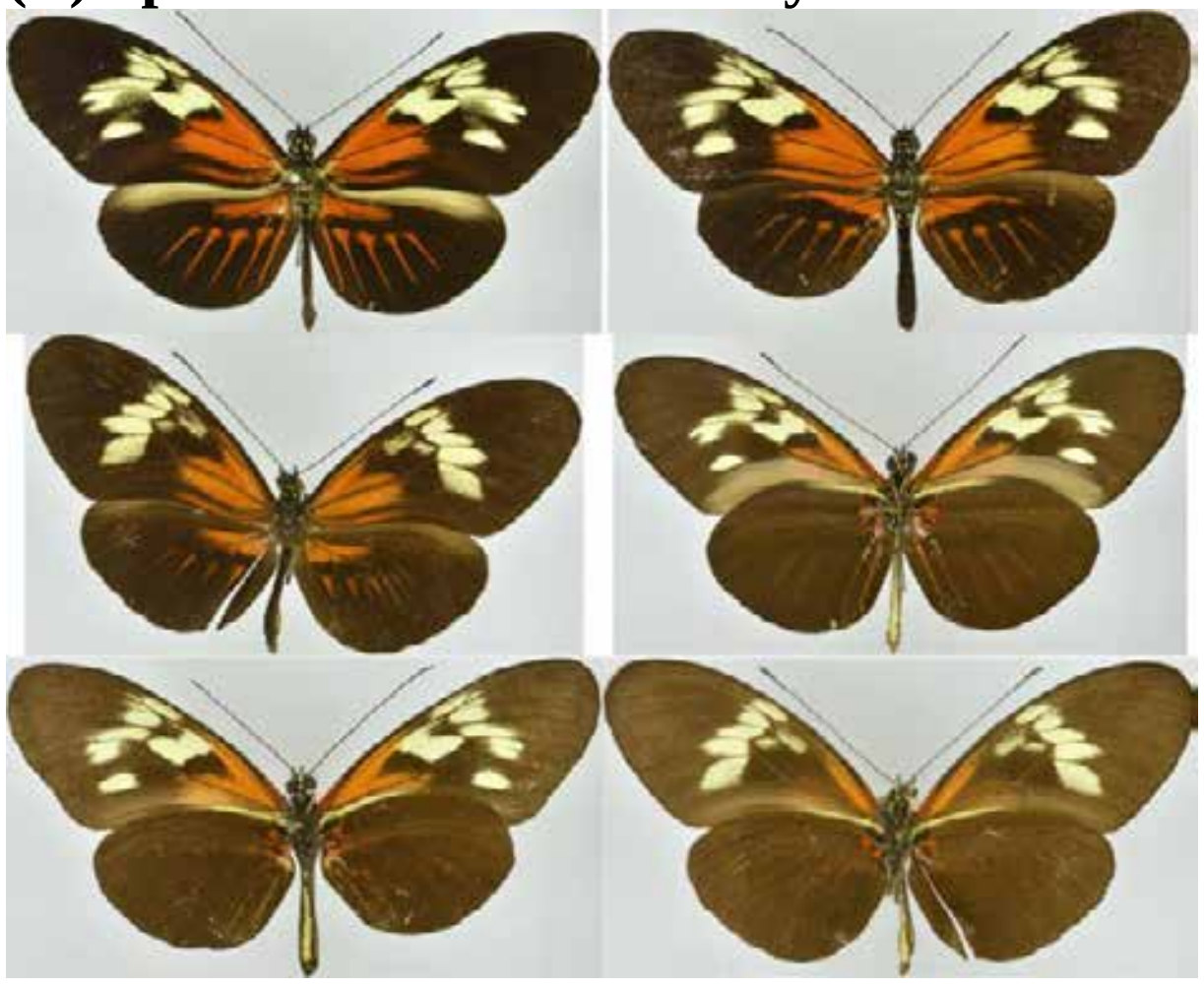

**35 *Heliconius erato venus***

**(A) Specimens identified as valid subspecies or accepted synonym**

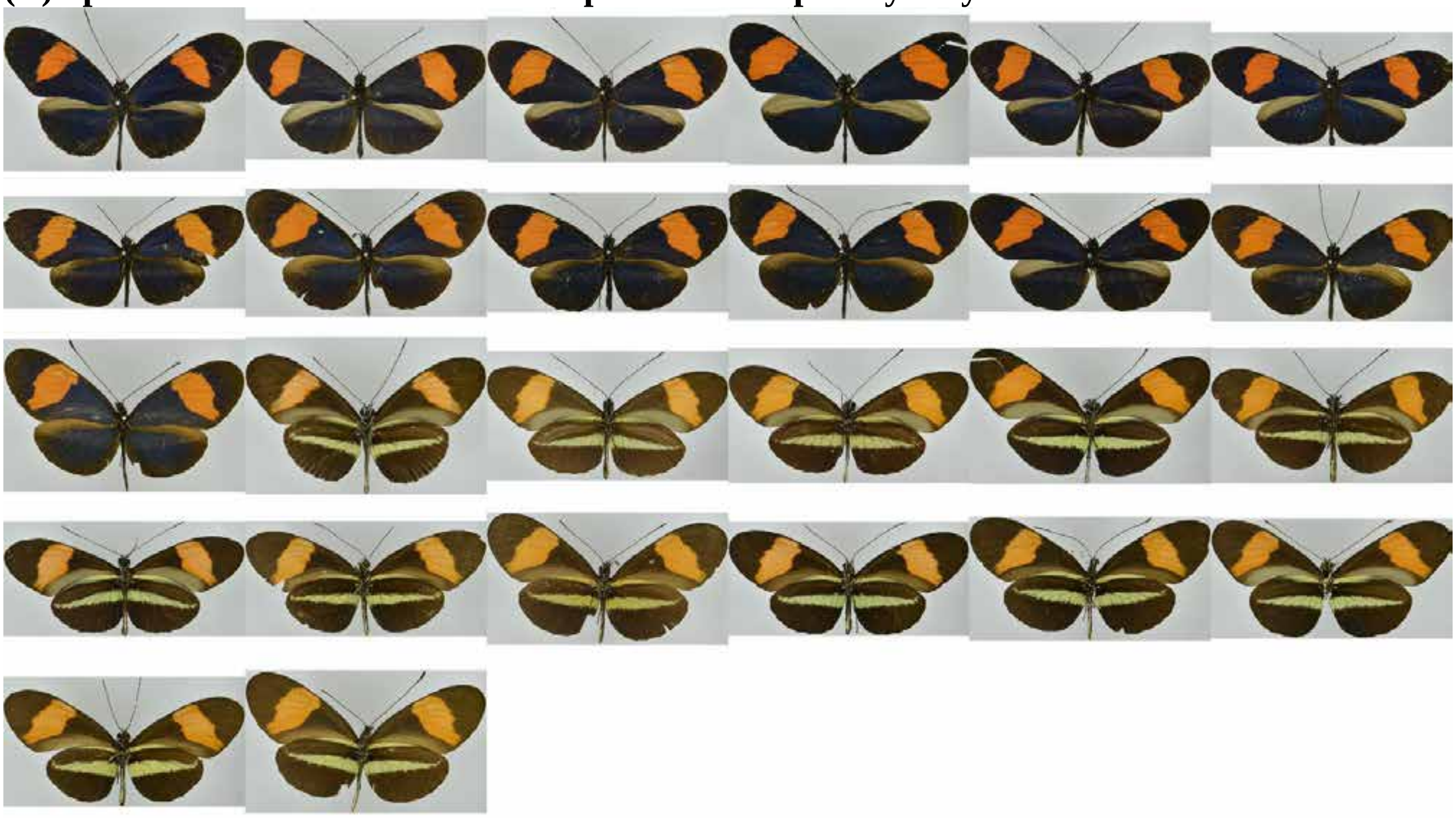

**36** ***Heliconius erato venustus***

**(A) Specimens identified as valid subspecies or accepted synonym**

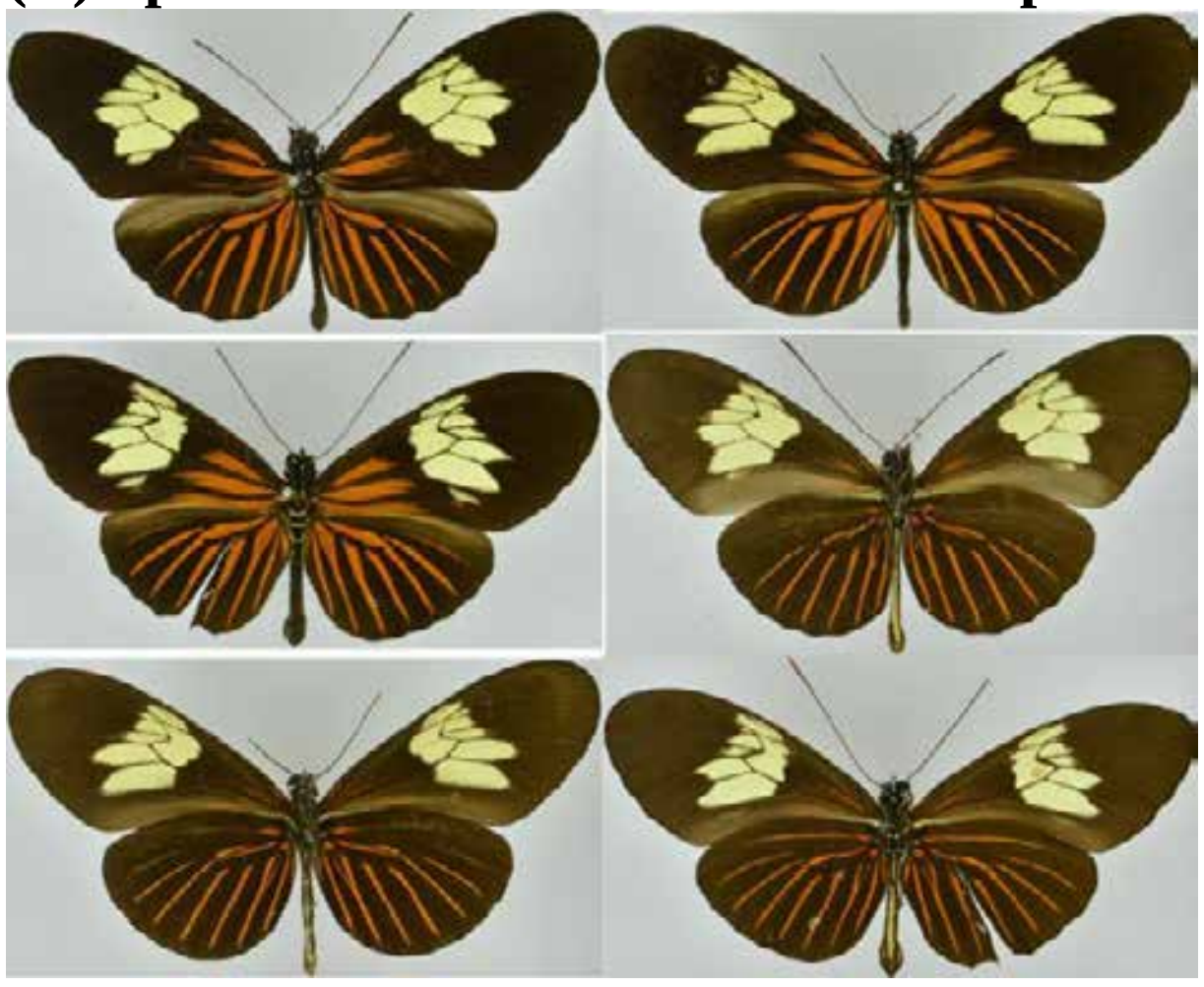

**(B) Specimens identified as hybrids**

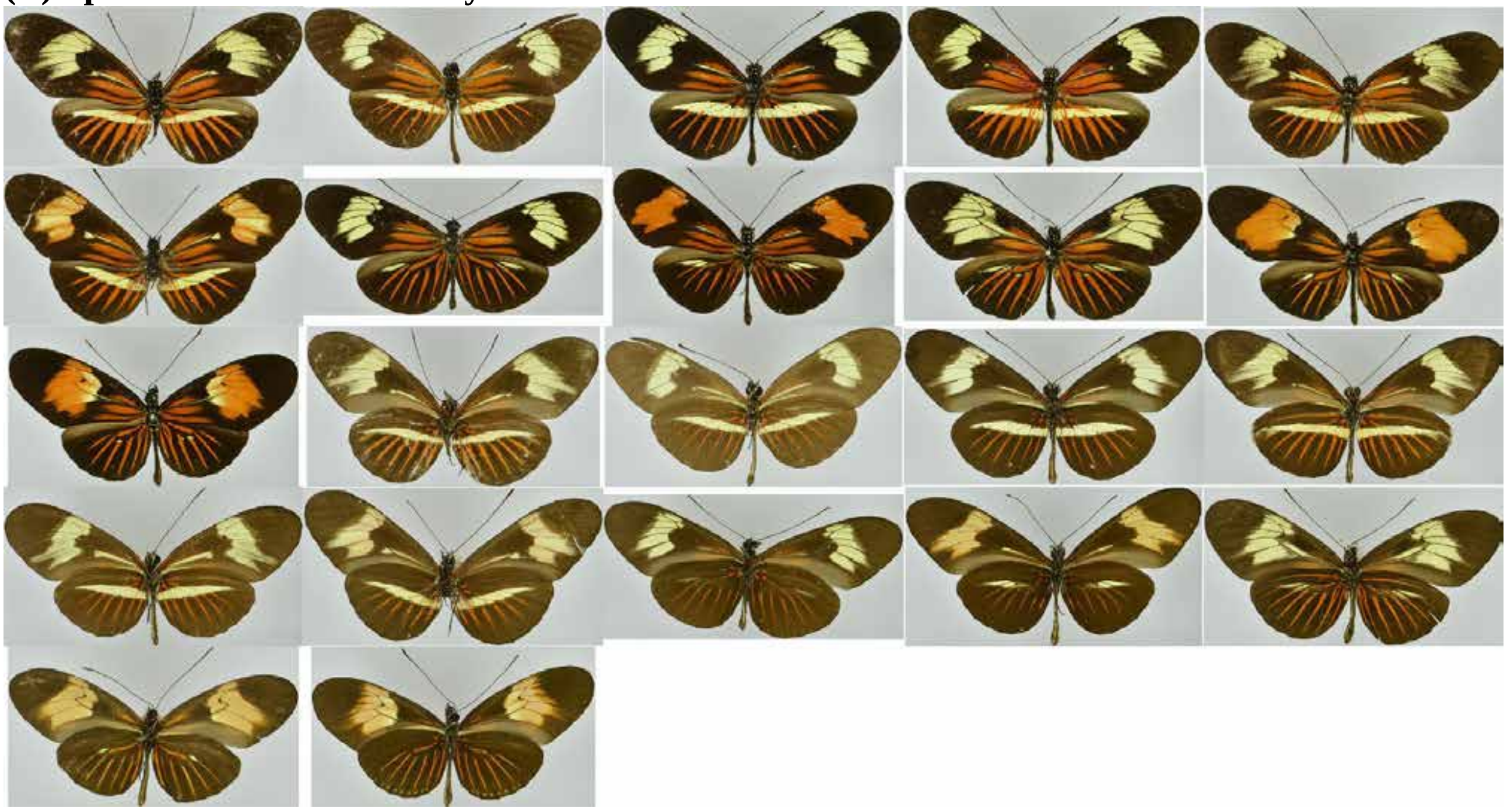

37 *Heliconius melpomene vulcanus*

(A) Specimens identified as valid subspecies or accepted synonym

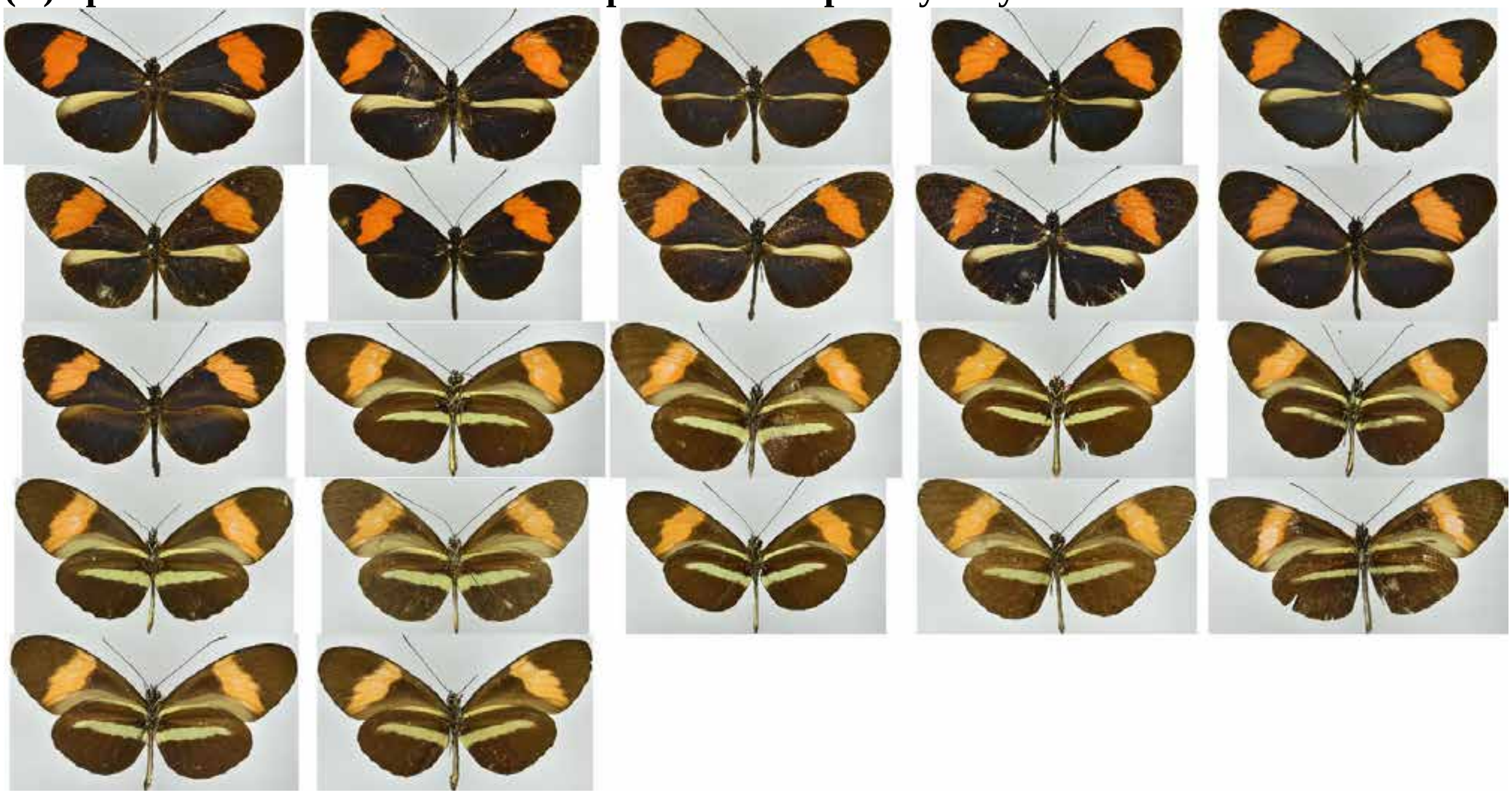

(B) Specimens identified as hybrids

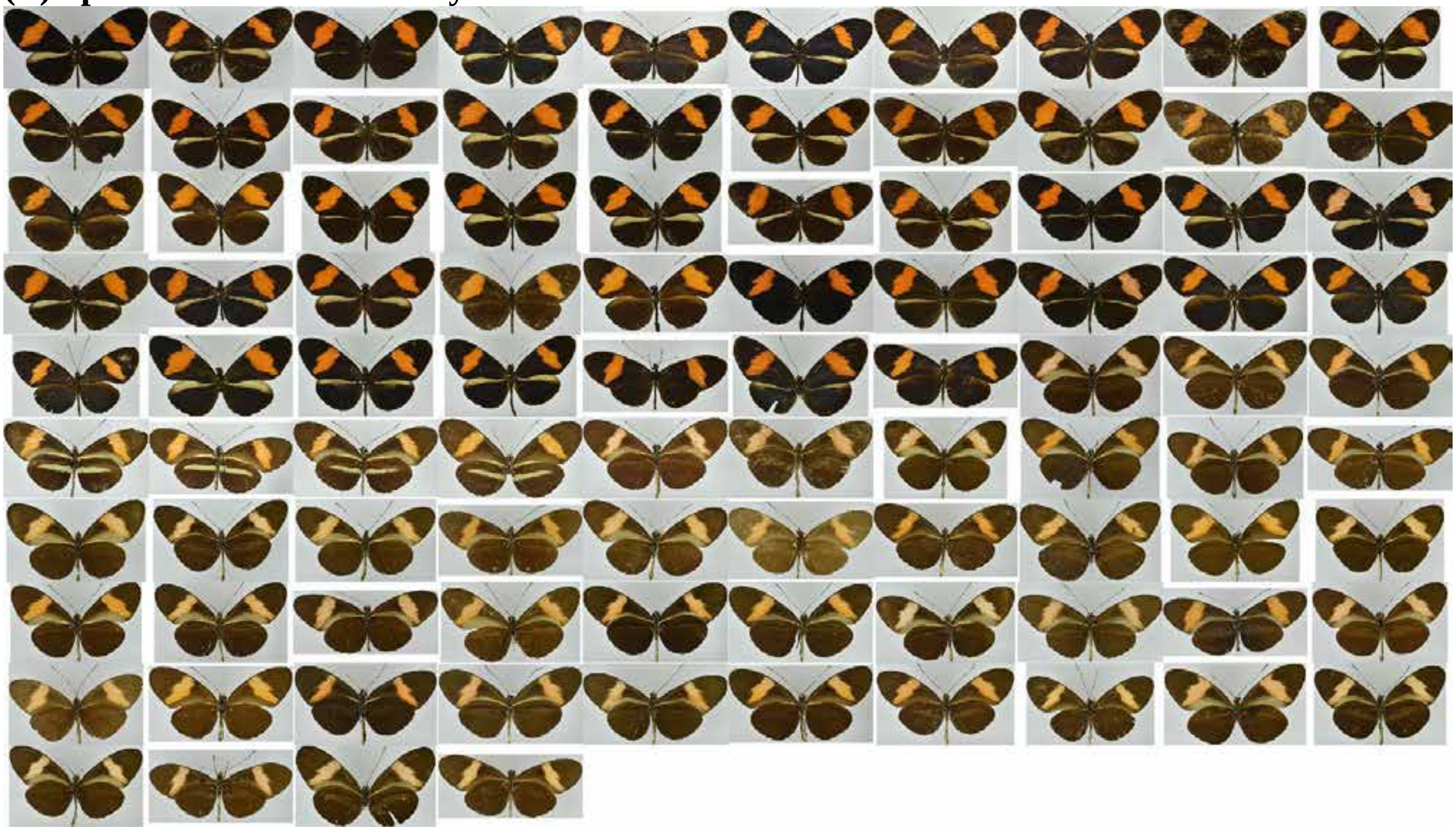

**38 *Heliconius melpomene xenoclea***

**(A) Specimens identified as valid subspecies or accepted synonym**

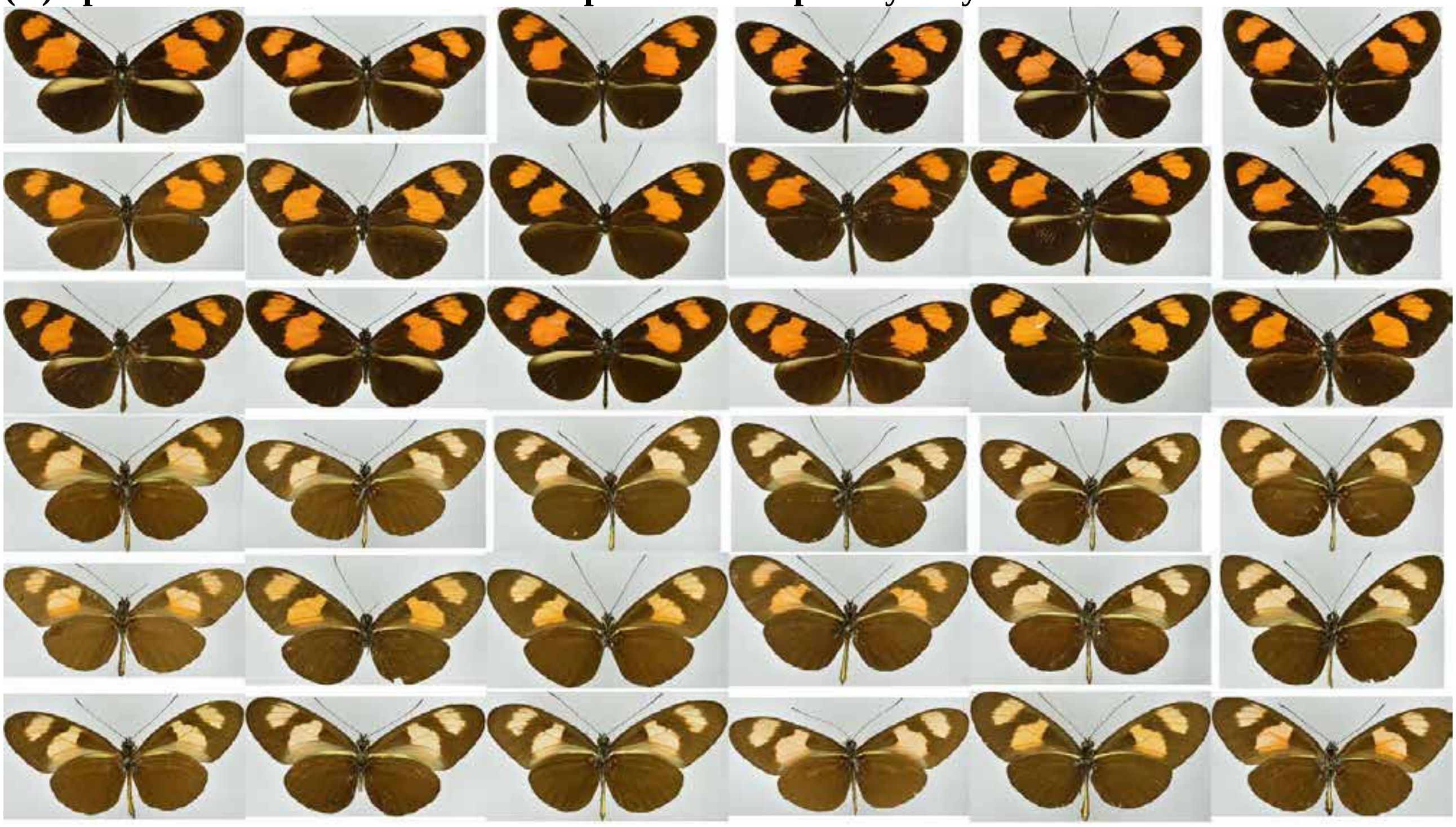

**(B) Specimens identified as hybrids**

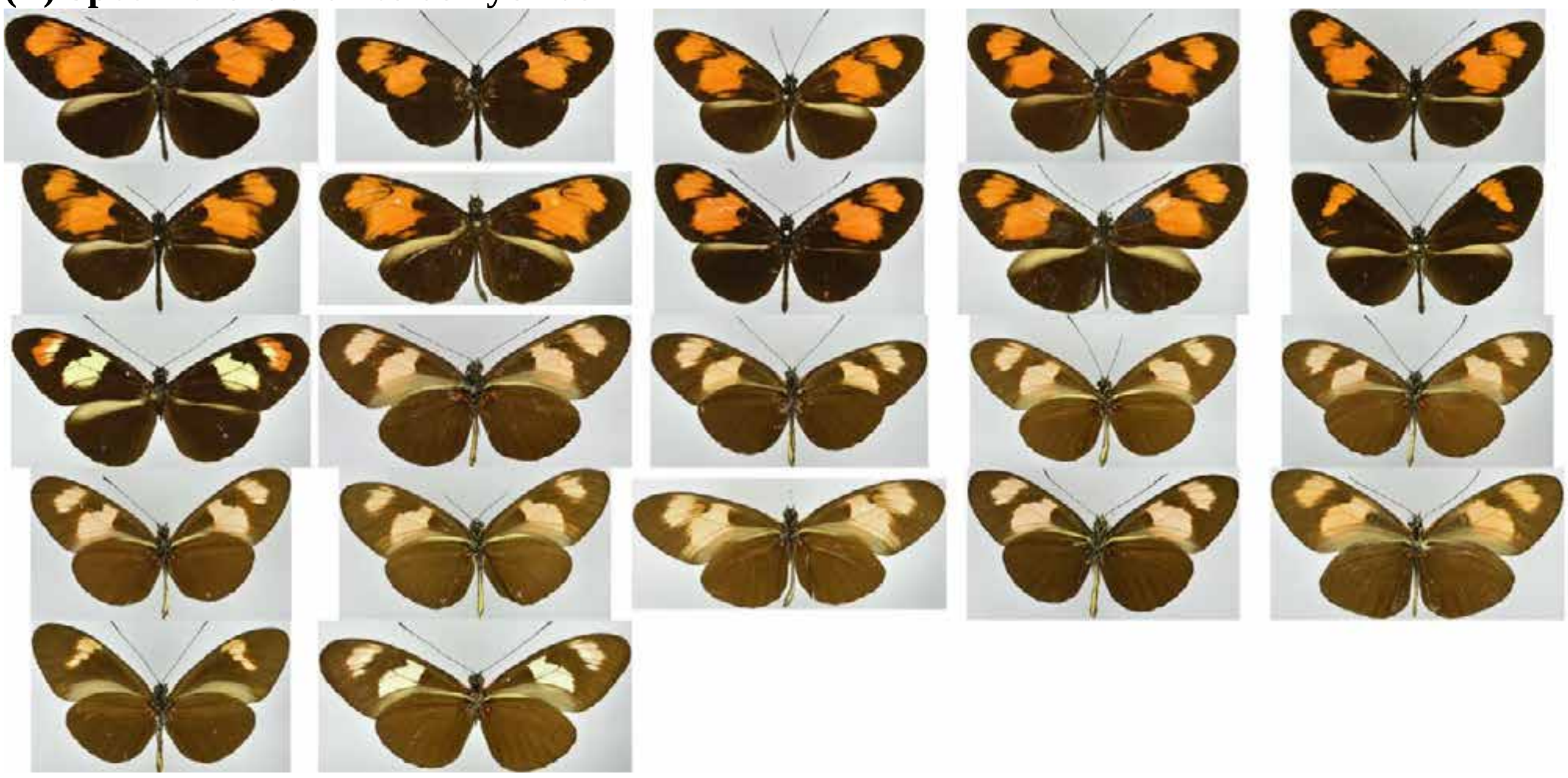

**Fig. S4. Collections of specimen photographs used in this study, grouped by subspecies.** Photographs show all dorsal views, followed by all ventral views (with the same specimen order, left to right, top to bottom). Subspecies identification followed NHM specimen labels and the taxonomy of Lamas (2014) (*30*). Individual image files used in deep learning, with specimen numbers corresponding Table S2, can be downloaded from the Dryad Data Repository: doi:10.5061/dryad.2hp1978. **(A)** Specimens included in the reduced dataset comprising valid subspecies and synonyms only. **(B)** Specimens excluded from the reduced dataset based on taxonomic labelling as hypothesised hybrids (e.g. based on their phenotype and/or capture locality) and a visual screen. Hybrid status was recorded (Table S2) based on additional taxonomic information from specimen labels, NHM records, the taxonomic checklist of Lamas, 2004 (*30*) and www.butterfliesofamerica.com. Photo Credits: Robyn Crowther and Sophie Ledger, Natural History Museum, London.

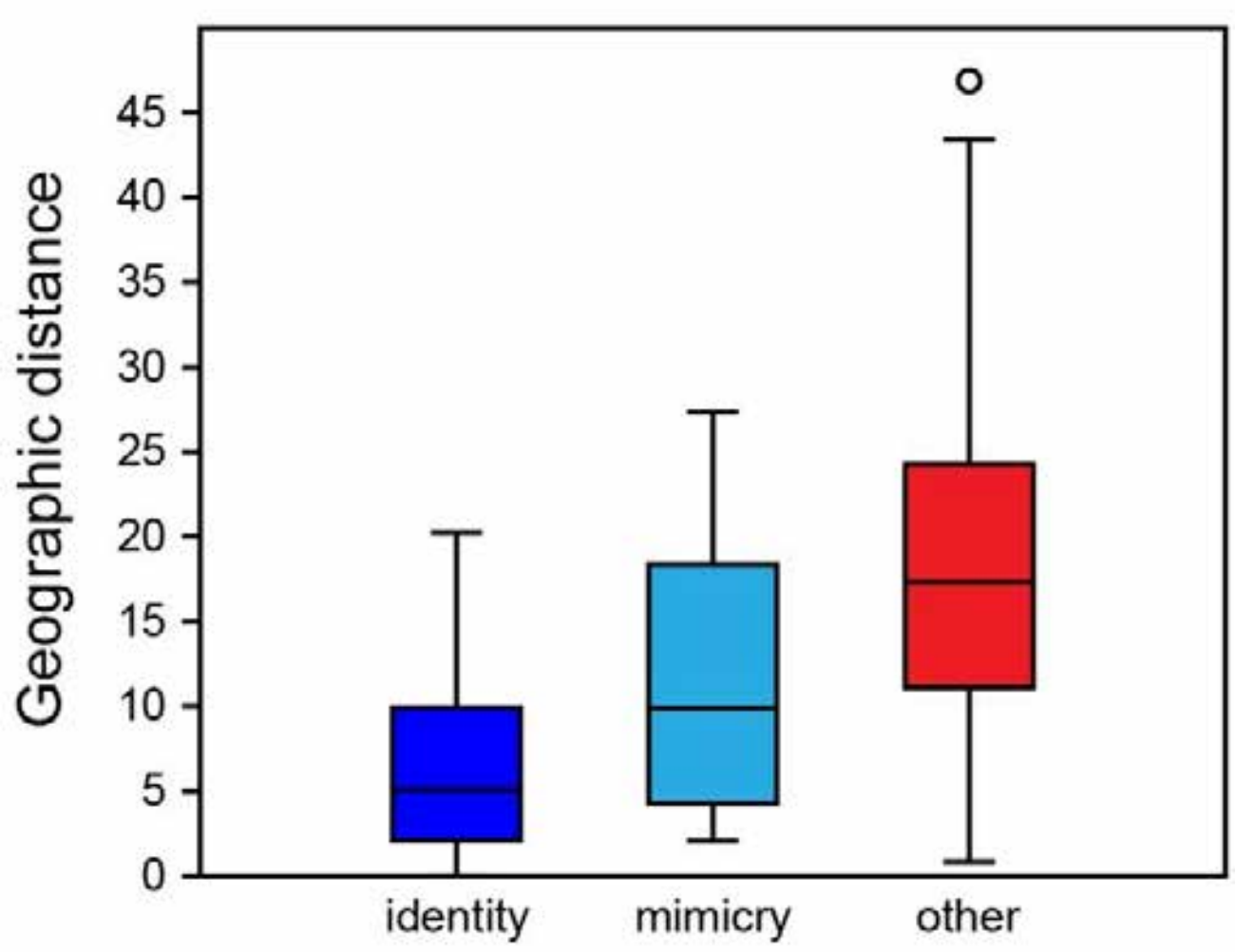

**Fig. S5. Average pairwise Euclidean geographic distances between subspecies of *H. erato* and *H. melpomene*.** Box plot of mean pairwise geographic distances (Table S6): within subspecies (identity), between co-mimic subspecies (mimicry) and between all other subspecies (other). Sample sizes: 38, 19 and 684 subspecies pairs, respectively. Boxes show 25-75% quartiles; horizontal lines, medians; whiskers, inner fence within 1.5 × box height; circles outliers within 3 × box height.

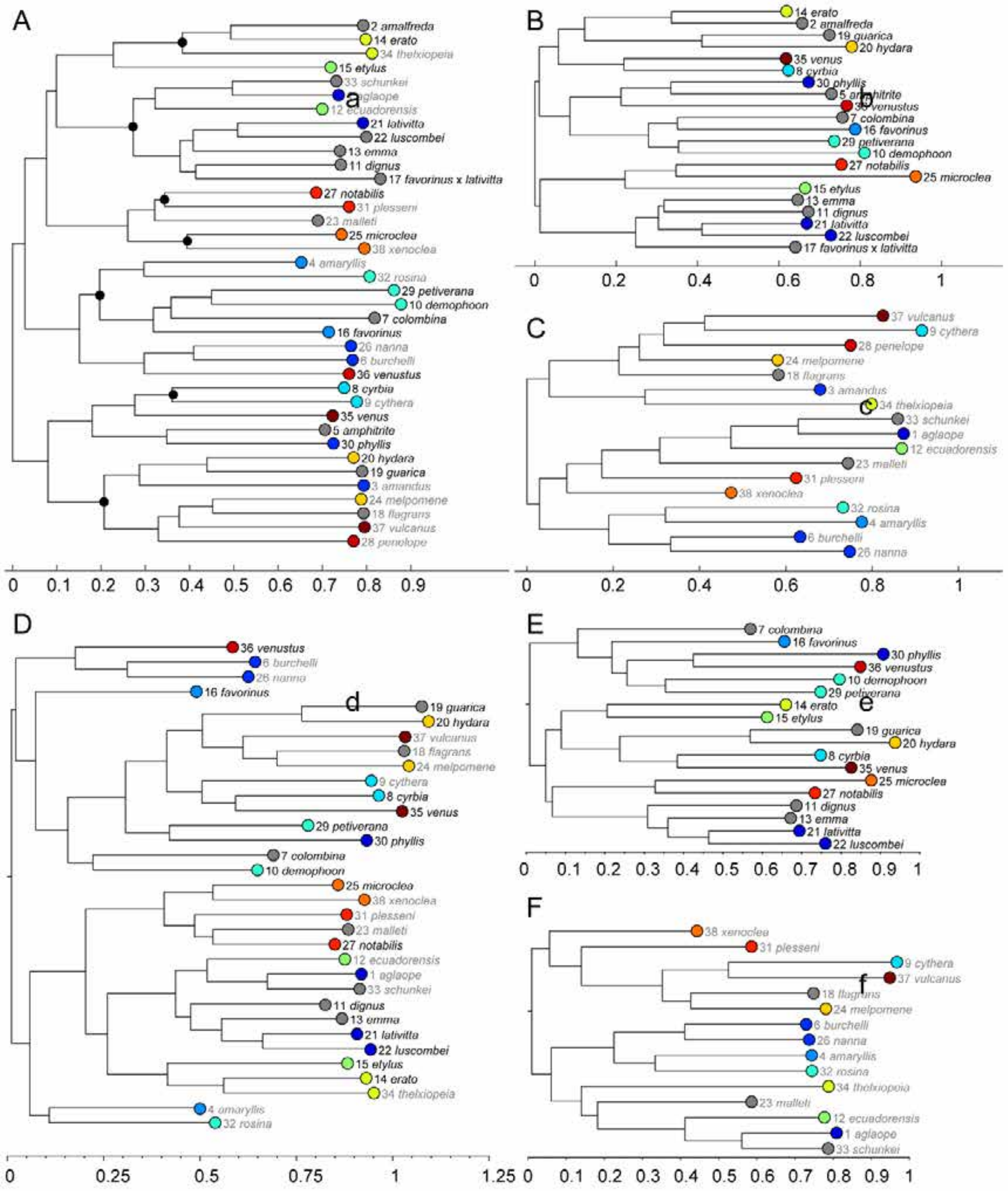

**Fig. S6. Neighbor-joining trees of phenotypic distance between subspecies of *H. erato* and *H. melpomene*. (A, D)** All subspecies. **(B, E)** Only subspecies of *H. erato*. **(C, F)** Only subspecies of *H. melpomene*. Subspecies label colour indicates species (black *H. erato*, grey *H. melpomene*). Leaf node colours correspond to mimicry groups of Fig. 1. Subspecies numbers correspond to Table S1. Black internal nodes (**A**) show independent clades containing interspecies co-mimics. **(D-F)** Phylogenies reconstructed from average subspecies distances calculated after exclusion of hybrid specimens (Table S2).

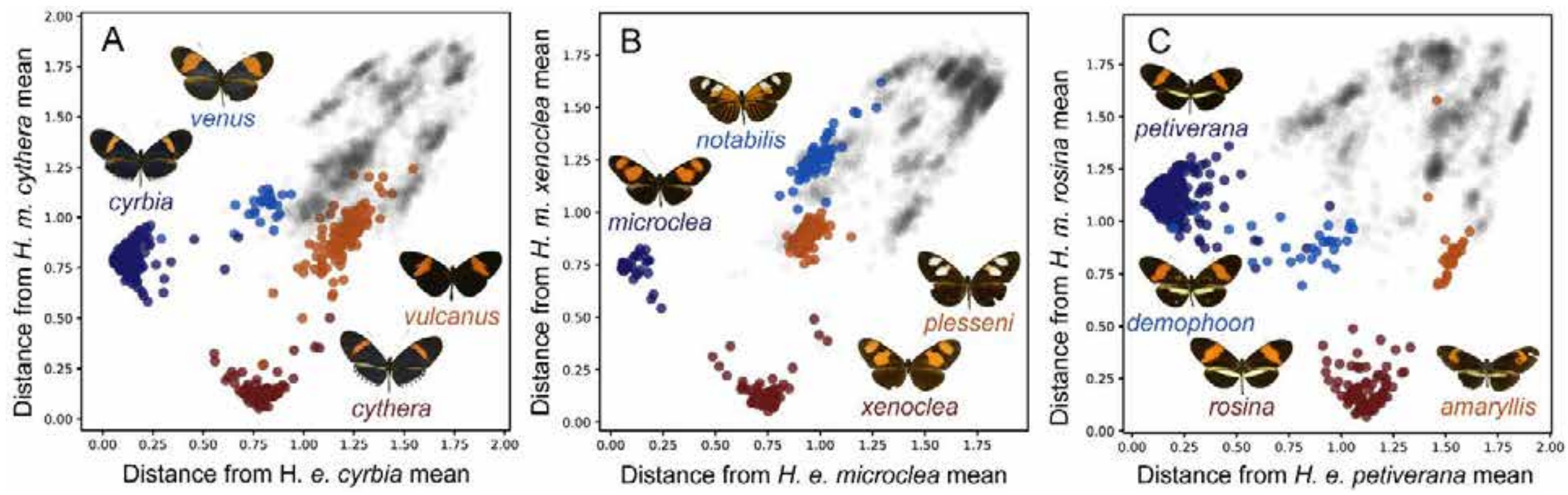

**Fig. S7. Comparative analyses of the extent of phenotypic convergence in mimicry.** The locations of six focal subspecies **(A-C)** dark blue, *H. erato*: *cyrbia*, *microclea*, *petiverana*; dark red, *H. melpomene*: *cythera*, *xenoclea*, *rosina*) in phenotypic space are compared alongside their six nearest conspecific subspecies **(A-C)** *H. erato*, light blue; *H. melpomene*, light red). Subspecies are illustrated by dorsal photographs of the butterfly closest to the mean location for the subspecies. Grey points indicate images of the other subspecies in the dataset. Axes show the squared distance from the mean location of the focal co-mimic, summed across all 64 spatial embedding axes. As polarised towards the focal taxa, these comparative analyses indicate cases of mutual convergence **(A-B)** as well as implied divergence by *H. erato* where the two species ranges cease to overlap **(c)** in Central-North American *H. erato petiverana* (*23*). In **(A)** the generally less abundant species *H. melpomene* has converged further towards *H. erato*, in line with the frequency dependent fitness benefits predicted in Müllerian mimicry. In another case, **(B)** *H. erato microclea* shows greater convergence on its co-mimic *H. melpomene xenoclea* than *vice versa*. Subspecies mean convergent distance from the focal co-mimic of the other species (distance = conspecific – focal conspecific) and corresponding Mann Whitney *p* values for *H. erato* and *H. melpomene* respectively: **(A)** distance = 0.26, *p* = 1.0195E-15, distance = 0.41, *p* = 5.1718E-31; **(B)** distance = 0.52, *p* = 8.2445E-16, distance = 0.20, *p* = 2.1749E-22; **(C)** distance = -0.22, *p* = 3.2368E-16, distance = 0.41, *p* = 1.1133E-19. Values with reversed evolutionary polarities (focal conspecifics, a-c: *venus*, *vulcanus*; *notabilis*, *plesseni*; *demophoon*, *amaryllis*): **(A)** distance = 0.23, *p* = 1.3176E-14, distance = 0.08, *p* = 3.8215E-12; **(B)** distance = 0.19, *p* = 4.4761E-11, distance = 0.51, *p* = 5.003E-27; **(C)** distance = 0.23, *p* = 6.1463E-17, distance = -0.42, *p* = 1.1133E-19. Statistical values with hybrids excluded (and standard evolutionary polarities) for *H. erato* and *H. melpomene* respectively: **(A)** distance = 0.29, *p* = 2.37E-15, distance = 0.40, *p* = 1.35E-10; **(B)** distance = 0.51 , *p* = 1.44E-11, distance = 0.22, *p* = 4.70E-16; **c,** distance = -0.21, *p* = 6.20E-16, distance = 0.41, *p* = 3.24E-10.

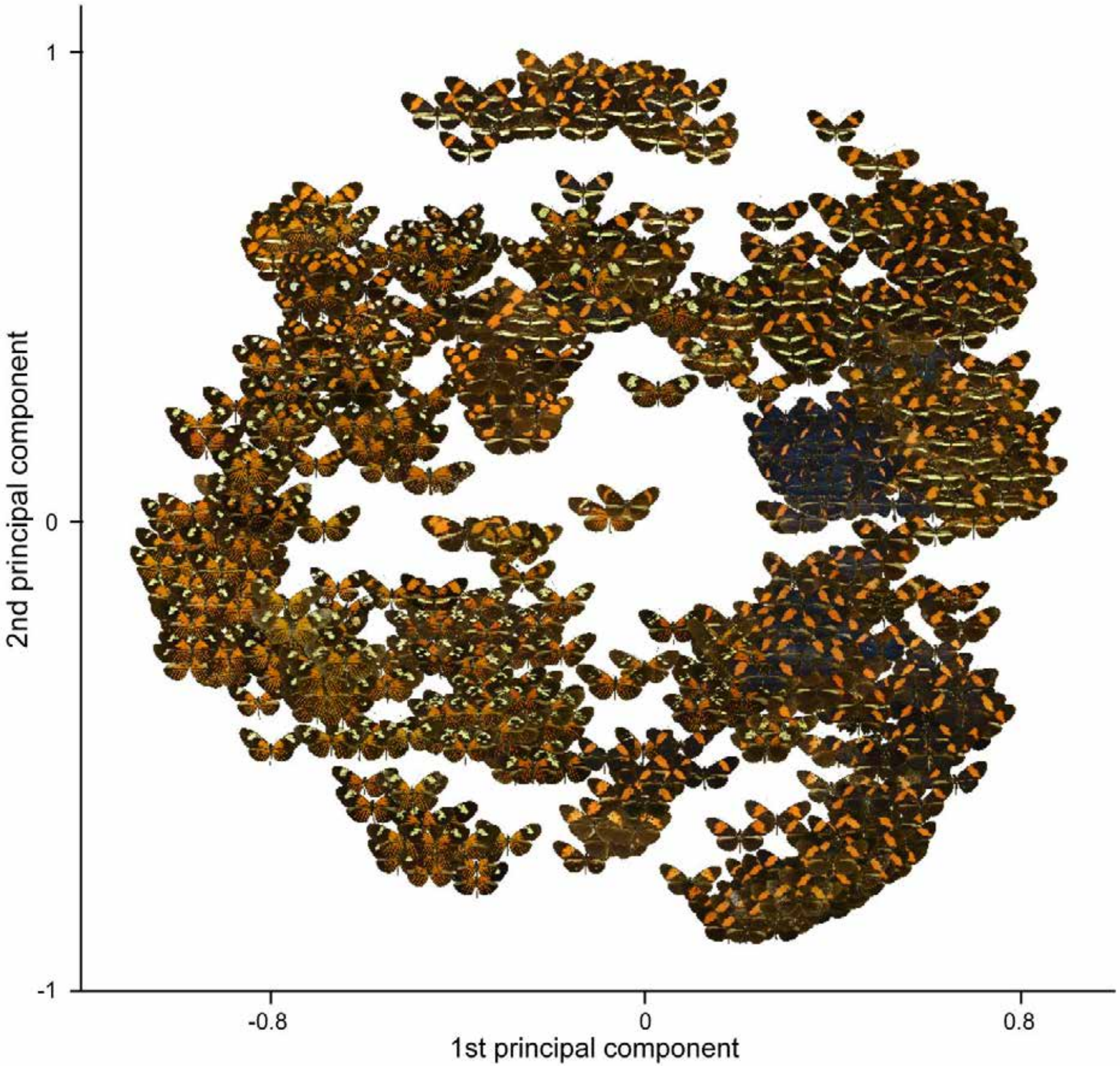

**Fig. S8. Principal component visualization of *Heliconius* butterflies.** Dorsal photographs of 1234 *Heliconius* butterflies visualised in the space of PCA scores calculated from the deep learning spatial embedding coordinates as described for Fig. 2.